\documentclass[12pt,a4paper]{article}
\usepackage[T1]{fontenc}
\usepackage[english]{babel}
\usepackage{graphicx}
\usepackage{amssymb}
\usepackage{amsmath}
\usepackage{caption}
\usepackage{subcaption}
\usepackage{indentfirst}
\usepackage[utf8]{inputenc} 
\usepackage{cite}
\usepackage{fullpage}
\usepackage{authblk}
\usepackage{multicol}
\usepackage{float}
\usepackage{xcolor}
\usepackage{tabularx}
\usepackage{program}
\usepackage[colorlinks=true,linkcolor={green},citecolor={blue},urlcolor={red}]{hyperref}
\usepackage{sectsty} %Стиль названия Раздела

\providecommand{\keywords}[1]
{
	\small	
	\textbf{Key words:} #1
}

\usepackage{titlesec}
\titlelabel{\thetitle.\quad}

\subsectionfont{\normalfont\itshape}

\newcolumntype{Y}{>{\raggedleft\arraybackslash}X}% raggedleft column X

\title{\textbf{Stellar Wind of Components of Detached Binary Systems}}

\author{A. V. Tutukov$^1$, A. V. Sobolev$^1$\thanks{e-mail: asobolev@inasan.ru} \\%
	
\textit{\small $^1$ Institute of Astronomy of the Russian Academy of sciences, Moscow, Russia}\\% 
}
\date{}

\begin{document}
\maketitle

\begin{abstract}
The paper is devoted to the consideration of the role of the donor stellar wind in the matter exchange between the components of detached binary systems. A classification of close binary systems with
interacting components is proposed. A list of potential donors and accretors of such systems, including X-ray binary and symbiotic stars, is given. Analytical tasks have been completed to evaluate the conditions and efficiency of interaction through the stellar wind, a criterion was found for maintaining the self-induced stellar wind of X-ray binaries, and a condition for the formation of an accretion disk during accretion of stellar wind matter by a compact accretor. Three-dimensional gas dynamic models of component interaction are constructed for the five initial velocities of the stellar wind using the example of Sco X-1 type systems. The simulation results are illustrated by pictures of streamlines, temperature distribution, and wind gas densities in the orbital and frontal planes. Model focusing of the donor wind flow by the accretor is confirmed by the observed phase X-ray light curve of Vela X-1.
\end{abstract}


\section{Introduction}{\label{sec-1}} %"*" выключает нумерацию Раздела

The vast majority of stars in the Galaxy and the Universe are members of multiple systems, reflecting the hierarchical nature of star formation during the collapse of inhomogeneous, rotating gas clouds \cite{Masevich1988,Cherepaschuk2013, BZB2013}. Approximately half of all stars belong to close binary systems whose components fill their Roche lobes over the course of evolution, which significantly enriches
the picture of their observable manifestations. Intensive observations of binary stars across all ranges of the electromagnetic spectrum in recent decades have greatly expanded our understanding of the evolution
of previously known binary systems and have also made it possible to discover new classes of close binaries at various stages of their evolution. This has turned the study of the physics and evolution of close binaries into a central theme of astrophysics over the past half-
century --- this line of study was awarded by several Nobel Prizes.

A key role in studying the physics and evolution of close binary systems is played by the concept of the Roche lobe, which confines the gravitational influence of the components of a binary star. Systems in
which at least one component can fill its Roche lobe are called close binaries, while the others are considered wide binaries. At the same time, the distance between components, which can significantly increase
during evolution, can reach $\sim2000R_\odot$. According to statistics, the combination of this factors allows about half of young binary stars to be classified as close binaries. The distribution of binary stars by their initial intercomponent distance $A$ at $1 \lesssim \log(A/R_\odot) \leqslant 6$ is described by the function \cite{Masevich1988} 

\begin{equation}\label{eq-1}
dN \thickapprox 0.2 ~d\log (A/R_\odot),
\end{equation}	 
where  $R_\odot$ is the solar radius, and $N$ is the number of
stars.

This function has proven to be a reliable basis for developing scenario-based programs that allow us to accurately estimate the number of binary stars at different evolutionary stages and the frequency of various events in their lives \cite{Tutukov1992,Lipunov1994}.

\begin{figure}[H]
	\centering
	\begin{minipage}[h]{0.45\linewidth}
		\centering{\includegraphics[width=0.9\textwidth]{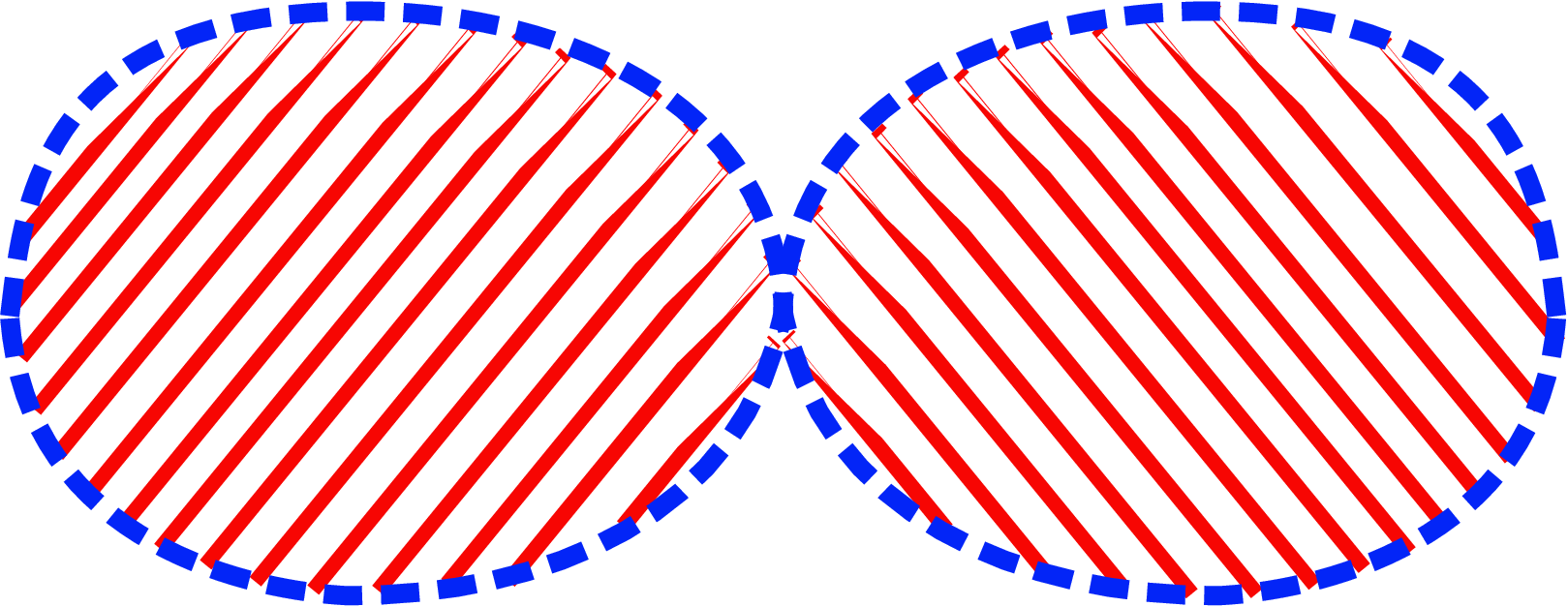} \\ (a)}
	\end{minipage}
	\hfill
	\begin{minipage}[h]{0.45\linewidth}
		\centering{\includegraphics[width=0.9\textwidth]{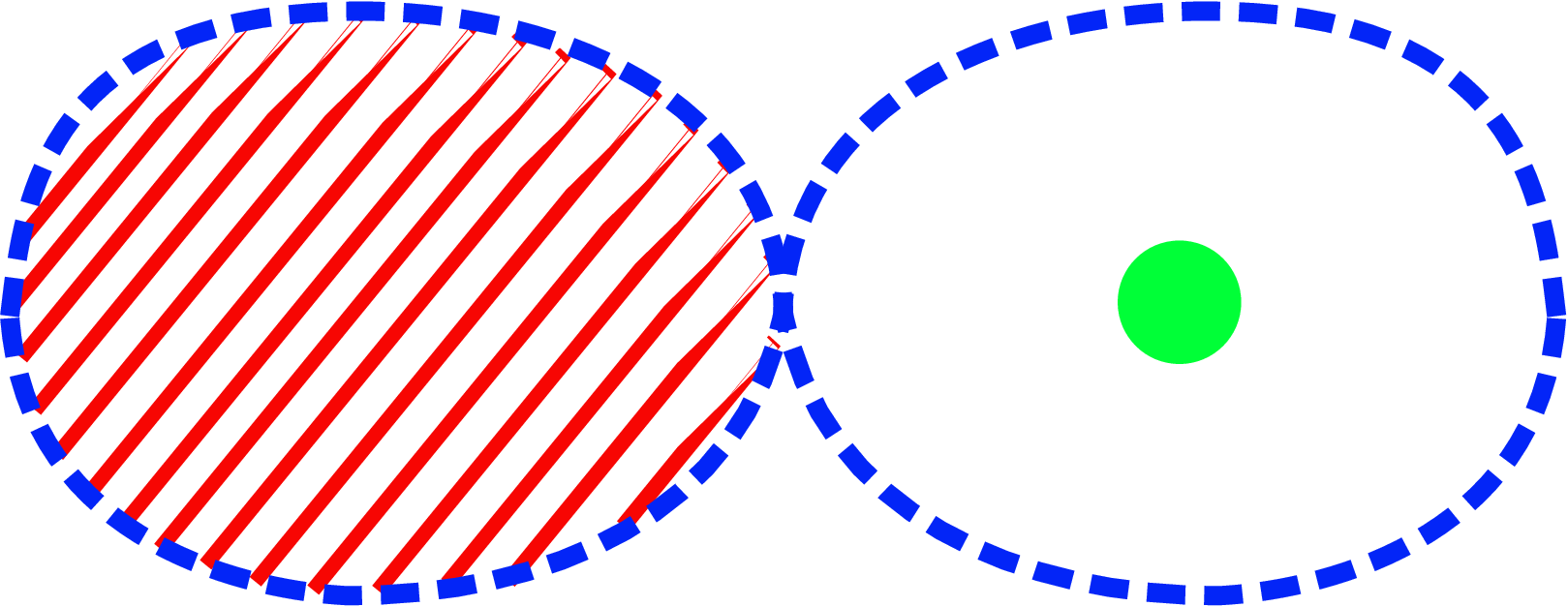} \\ (b)}
	\end{minipage}
	\hfill
	\begin{minipage}[h]{0.45\linewidth}	
		\centering{\includegraphics[width=0.9\textwidth]{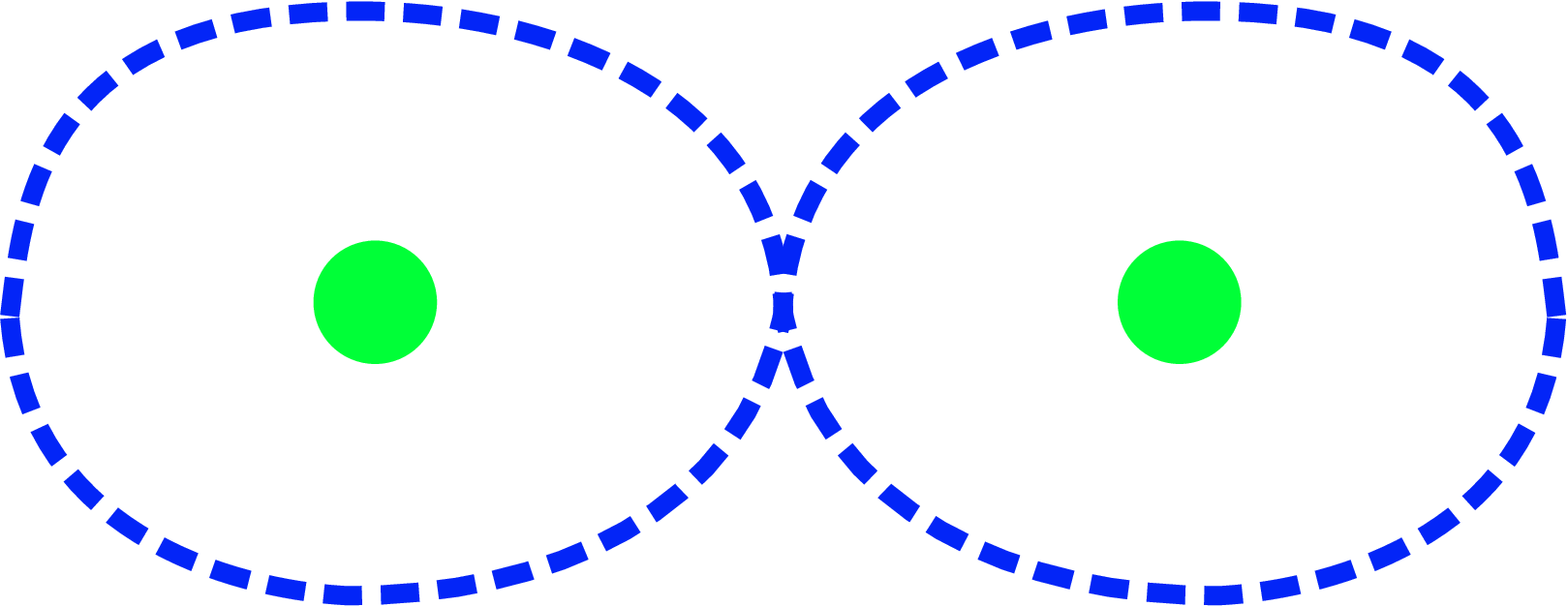} \\ (c)}
	\end{minipage}
	\hfill
	\begin{minipage}[h]{0.45\linewidth}
		\centering{\includegraphics[width=0.9\textwidth]{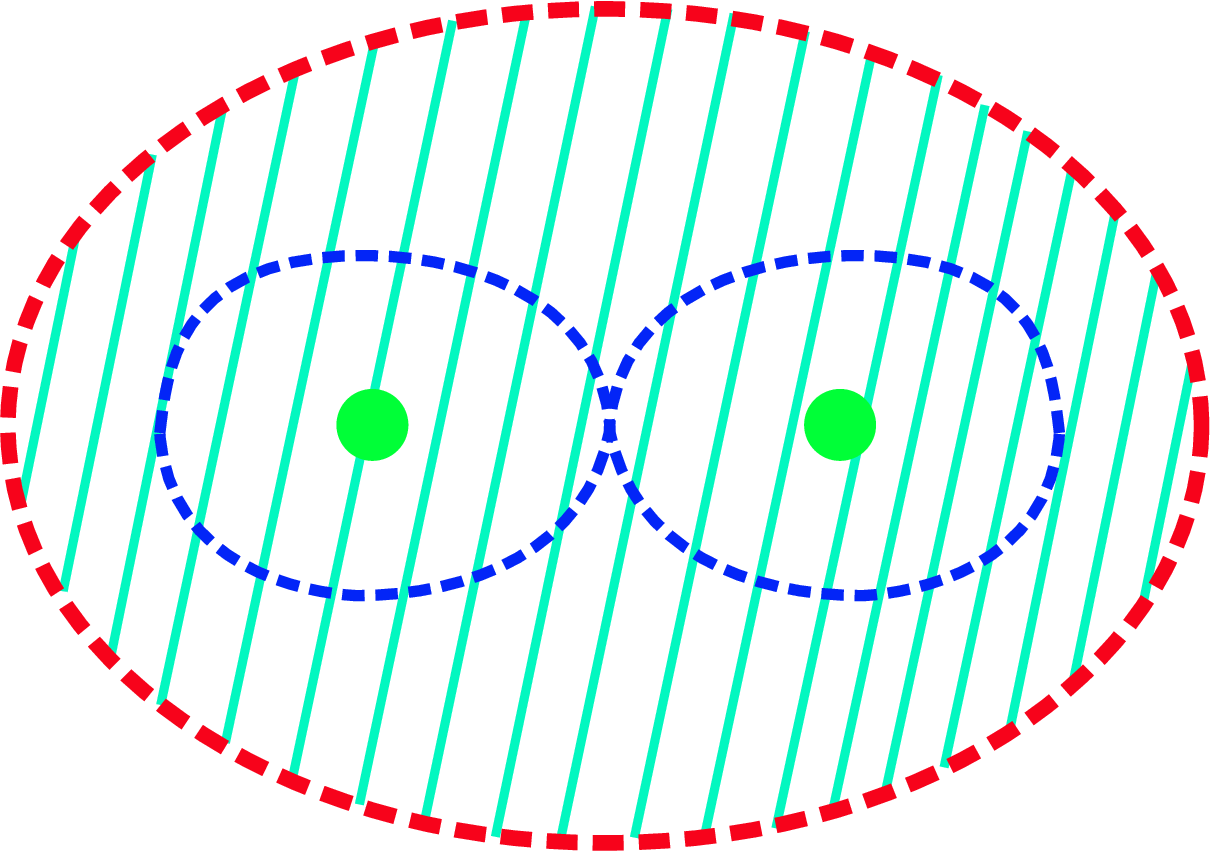} \\ (d)}
	\end{minipage}
	\caption{Classification of binary stars: contact (a), semi-detached (b), detached (c), and with a common envelope (d).}
	\label{fig-1}
\end{figure}

The assumption of synchronous axial rotation of
the donor filling its Roche lobe makes it possible to
estimate the size of the lobe for each component and
to distinguish the following classes of close binaries
(Fig. \ref{fig-1}):
\begin{enumerate}
   \item 
    \textit{Contact systems} (Fig. \ref{fig-1},a) --- systems in which both components fill their Roche lobes;
	\item
	 \textit{Semi-detached systems} (Fig. \ref{fig-1},b) --- systems in which only one component fills its Roche lobe. Well-known examples include systems like $\beta$ Persei (Algol);
   \item
	\textit{Detached systems} (Fig. \ref{fig-1},c) --- systems in which
	neither component fills its Roche lobe. Due to the long-held belief that no mass exchange occurs in such systems, they did not receive much attention;
	\item
	\textit{Systems with a common envelope} (Fig. \ref{fig-1},d). These form as a result of excessive mass transfer when the accretor cannot absorb all the matter flowing from the donor, and part of this matter escapes into a common envelope. Active gas loss from the donor contributes to the formation of extended envelopes, making these systems bright infrared sources \cite{Oskinova2018}
\end{enumerate}	

\begin{figure}[H]
	\centering
	\begin{minipage}[h]{0.45\linewidth}
		\centering{\includegraphics[width=0.9\textwidth]{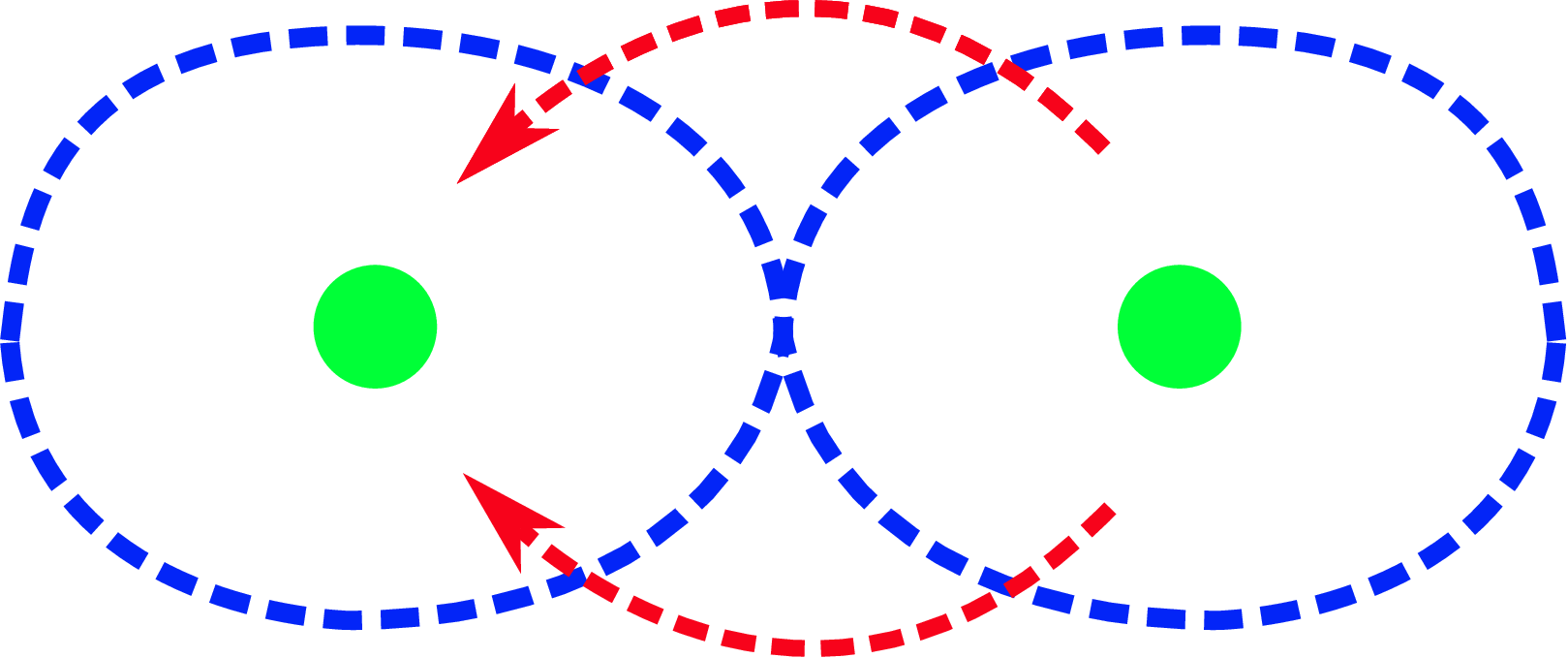} \\ (a)}
	\end{minipage}
	\hfill
	\begin{minipage}[h]{0.45\linewidth}
		\centering{\includegraphics[width=0.9\textwidth]{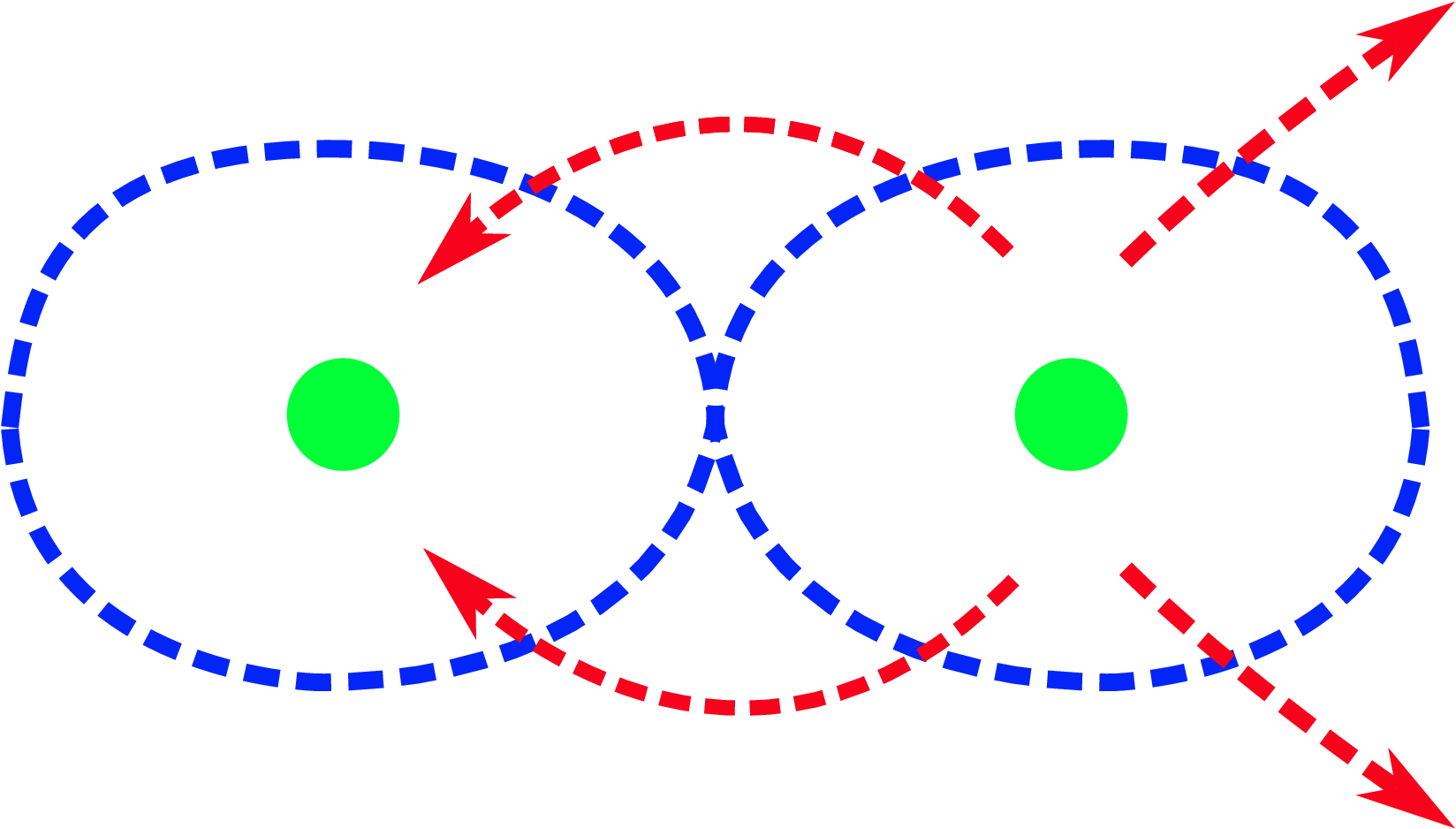} \\ (b)}
	\end{minipage}
	\hfill
	\begin{minipage}[h]{0.45\linewidth}	
		\centering{\includegraphics[width=0.9\textwidth]{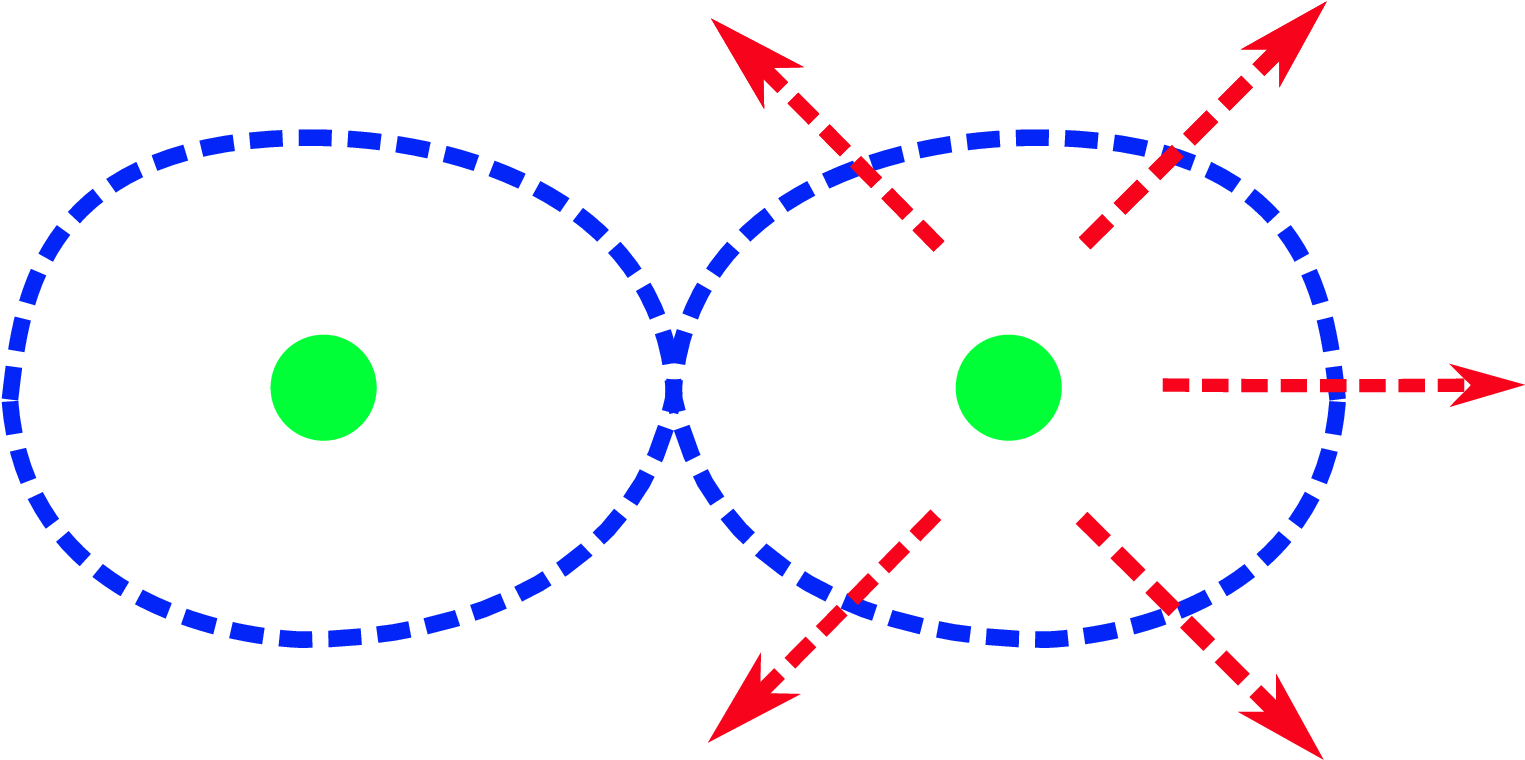} \\ (c)}
	\end{minipage}
	\caption{Classification of detached interacting binaries: quasi-conservative (a), semi-open (b), and open (c).}
	\label{fig-2}
\end{figure}

The classification of close binaries changed fundamentally after the development of the model of symbiotic stars, in which a red (super)giant, without filling its Roche lobe, possesses a strong stellar wind that allows a compact accretor—a degenerate white
dwarf—to capture part of the wind’s material \cite{Tutukov1976}.
Accretion onto the white dwarf gives rise to a series of outburst phenomena that draw observers’ attention to these systems. Thus, the presence of a stellar wind from one component radically alters the behavior of detached binaries, which requires them to be included
in the class of interacting binaries (Fig. \ref{fig-2}).

Since the gravitational structure of a binary star contains a region between the inner Roche lobe including the Lagrange point $\rm{L}_1$ and the common lobe including the external Lagrange points $\rm{L}_2$ and $\rm{L}_3$, the system maintains a quasi-conservative mass exchange
structure between components at significant stellar wind speeds (Fig.  \ref{fig-2},a). An increase in the initial stellar wind speed beyond the parabolic speed for the system leads to partial mass loss through the vicinity of the external point $\rm{L}_2$. The mass transfer structure then becomes semi-open (Fig. \ref{fig-2},b). A further increase in wind speed results in most of the wind mass being lost
not only through the external Lagrange points but also through the rest of the inner Roche lobe boundary, leaving only a small fraction to fall onto the accretor. The captured material then forms either an accretion disk or an accretion column around the star, depending on the angular momentum of the incoming gas and the strength of the compact object’s magnetic field. This creates an open type of interacting system (Fig. \ref{fig-2},c).

The presence of a stellar wind from one of the components changes the generally accepted view of close binaries. Filling the Roche lobe ceases to be a necessary criterion for the onset of mass exchange between components. As a result, detached systems can become effectively interacting, as demonstrated by the example of symbiotic stars. A candidate for nearby symbiotic stars is Betelgeuse ($\alpha$ Orionis), which has a companion with an orbital period of about 6 years and a mass lower than the Sun’s, most likely a degenerate
white dwarf \cite{MacLeod2024}.

The role of stellar wind in shaping gas flow between components and in the evolution of detached binaries has been repeatedly addressed in theoretical studies. In particular, it has been found that accretion of a red giant’s wind by a degenerate dwarf determines the activity of the dwarf companion in symbiotic stars \cite{Tutukov1976}. The physics of colliding winds in Wolf-Rayet binaries has been considered in \cite{Hill2002,Romeo2007}, and their three-dimensional gas dynamics was first studied for O+O and WR+O pairs in \cite{Parkin2008}. It is worth noting that symbiotic stars remain popular objects for modeling gas flows. Several works have focused on studying the stellar wind structure in such binaries \cite{Bisikalo1994,Bisikalo2006,Skopal2015}. An important result of the 2D modeling conducted in \cite{Bisikalo2006} was the discovery of a transition in the interaction regime between the donor’s wind and the accretor’s gravitational field. It was shown that this transition occurs at wind speeds on the order of the binary system’s orbital speed. The present paper discusses a possible cause of this change.

The role of stellar wind in the evolution of massive binaries where the accretor is a neutron star has been studied through numerical simulations in  \cite{Pittard2006,Taani2022}. Three-dimensional gas dynamical models of the stellar wind have reproduced light curves of X-ray binaries comparable to observations \cite{Pittard2010}. Such models are becoming reliable tools for studying the physics of various binary types. Modeling the profiles of spectral emission lines of gas surrounding the components and the X-ray light curves also provides significant opportunities to explore stellar wind dynamics.

This paper examines some issues related to the physics and evolution of systems with detached but interacting components. In these systems, mass transfer mainly occurs through the wind of one component, which is fully or partially captured by its companion. Such interaction is particularly effective when the accretor is a compact object --- a neutron star or black hole --- since matter falling onto these objects generates a powerful stream of X-ray radiation.

\section{MASS TRANSFER IN DETACHED BINARY	\newline SYSTEMS}{\label{sec-2}}

\subsection{Components of Detached Interacting Binary Systems (DIBS)}{\label{subsec-2.1}}

Let us consider the currently known potential components of DIBS, drawing on experience from studies of the evolution of single and close binary stars \cite{Masevich1988, Cherepaschuk2020, Tutukov2023, Ransome2024}. The role of donors in these systems can be played by all astrophysical objects with sufficiently intense stellar winds: massive Of and Bf stars, as well as Wolf–Rayet stars. This list must also include Ae and Be stars surrounded by decretion gas disks \cite{Tutukov1969}. In addition, red (super)giants and main-sequence stars with masses $(0.3-1.5)M_\odot$ and convective envelopes \cite{Iben1984} can serve as donors in such systems. This type of donor
is especially effective when its companion is a neutron star or black hole, since this helps to induce a powerful self-sustaining stellar wind \cite{Iben1997}.

Interestingly, the source of the wind in dense stellar nuclei of galaxies can be the gas disk of an accreting supermassive black hole (SMBH). A special and still little-studied role in wind generation is played by stars located near active galactic nuclei. Observations show
that gas accretion by an SMBH from its surrounding disk generates a wind with an intensity of approximately $\dot M = 10^{-2} - 10^2 M_\odot$/year and speed $\sim v_{\text{w}} = 10^3$ km/s for a characteristic SMBH mass $M_{\text{bh}} = 10^5 - 10^7 M_\odot$ \cite{Baron2019}. A large number of stars in the galactic nucleus with an active SMBH appear immersed in this wind field and are able to accrete it's matter. Neutron stars and black holes near an active
galactic nucleus, by accreting the nuclear wind, produce a number of interesting phenomena and objects. In this sense, such stars together with the galactic nucleus can be viewed as a kind of binary system
belonging to the class of interacting systems. A simple estimate shows that for a neutron star to achieve Eddington luminosity in the field of a nuclear wind with a mass loss rate $\dot M = 10^2 M_\odot/$year and speed $v_{\text{w}} = 10^3$ km/s, it must be located no farther than $\sim 10^{-3}$ pc from the SMBH. Considering that there are
a great number neutron stars and stellar-mass black holes in a galactic nucleus, we can conclude that they make a noticeable contribution to wind accretion and X-ray emission of the nucleus.

Wolf–Rayet stars are known as sources of powerful ($\dot M = 10^{-5} M_\odot/$year) and fast ($v_{\text{w}} = 1000 - 3000$ km/s) stellar winds. The wind power depends on the mass $M_{\text{WR}}$ of the helium Wolf–Rayet star: $\dot M_{\text{w}} = 3 \times 10^{-7} \left(\frac{M_{\text{WR}}}{M_\odot}\right)^{1.5}, M_\odot/$year \cite{Shaposhnikov2024,Rochowicz1995}. The presence of a powerful stellar wind makes Wolf–Rayet stars potential Eddington-rate donors in actively interacting detached binary systems.

Observed donors in X-ray binaries include Be stars \cite{Karino2021}. The orbital periods of such systems, derived from observational analyses, range from several days to hundreds of days, and their X-ray luminosities reach $10^{39}$ erg/s. The excessively high rotation velocity of Be stars may be due to either angular momentum transfer
from the contracting stellar core to the envelope during evolution \cite{Tutukov1969,Tutukov2007}, or it may result from prior mass accretion from a companion in a close binary system. Existing three-dimensional models of Be star envelopes already allow detailed study of this process and show good agreement with observations \cite{Harmanec2002}. Modeling the evolution of massive main-sequence
stars has demonstrated that over this period, the star may lose several percent of its initial mass \cite{Tutukov1969,Tutukov2007}.
This leads to the Be star shedding gas in the form of a decretion disk. The estimated mass loss rate of a Be star with mass $M_{\text{s}} = 4 - 30 M_\odot$ gives the following result:

\begin{equation}\label{eq-2}
  \dot M_{\text{w}} = 3\times 10^{-10} \left(\frac{M_{\text{s}}}{M_\odot}\right)^2, M_\odot/\text{year} 	
\end{equation}

Such decretion rates are sufficient to feed accretors in
X-ray binaries with luminosities up to $L = 10^{36} - 10^{38}$ erg/s.

Non-stationary mass transfer in X-ray binaries with Be stars as donors is driven by the interaction of the accretor’s X-ray radiation with the Be star’s decretion disk. Accretion onto the compact star can, as
often observed in cataclysmic variables, be non-stationary in nature \cite{Niwano2024}. In this case, donor matter likely accumulates in the disk up to a critical mass before the phase of mass transfer onto the accretor begins. Since accretion is accompanied by a powerful flow of X-rays
from the disk surrounding the accretor, the outer regions of the Be donor’s decretion disk begin to evaporate actively, creating a wind from this disk itself, thereby reducing its size. This circumstance can lead to a decrease in the mass transfer rate between components or even halt it altogether. The further evolution of the system involves the expansion of the Be star’s decretion disk and the start of a new cycle of gas accumulation in the accretor’s disk up to a critical mass. In this way, non-stationary mass transfer can develop in X-ray binaries with Be stars as donors. Other causes of non-stationary transfer may include the inclination of the accretor’s orbital plane relative to the decretion disk plane and a large orbital eccentricity.

Let us identify potential candidates for accretors in such systems. These may include low-mass main-sequence stars as well as helium, carbon–oxygen, and oxygen–neon degenerate white dwarfs. In so-called
X-ray binaries, the accretors can be neutron stars and stellar-mass black holes. Overall, considering these two groups of donors and accretors, there arises a large variety of objects encompassing several dozen types of stellar systems whose parameters can determine the observable properties of the majority of DIBS.

In this study, we present estimates of some parameters and properties of DIBS, along with numerical 3D modeling of gas flows in them based on the characteristics of the X-ray binary Sco X-1.

\subsection{Possible Causes of Stellar Wind from Donors}{\label{subsec-2.2}}

The phenomenon of stellar wind is likely typical for most stars. Establishing this fact is limited by the observational selection methods used. Nevertheless, the stars in which stellar wind has been detected through observations are quite diverse. Let us consider
some examples below and identify probable causes of their winds.
Bright emission lines observed in OB stars were the first clear evidence of an intense stellar wind in main-sequence stars. These lines are caused by light scattering in the lines of heavy elements. The acceleration zone of the radiatively driven stellar wind of OB stars lies at about $2 - 3$ stellar radii, and the terminal velocity determined by the surface potential is $1000 - 3000$ km/s \cite{Krticka2001}. Estimating the mass-loss rate of hot massive stars remains an active area of observational and theoretical study. A modern theoretical estimate for stars of solar chemical composition
can be expressed as $\log \dot M [M_\odot/\text{year}] \approx -9.13+2.1 \log (\frac{M}{M_\odot})$ \cite{Vink2018}. It should be noted that, for the mass-luminosity ratio of massive main-sequence
stars $\frac{L}{L_\odot} \approx 10^2 (\frac{M}{M_\odot})^2$ , this estimate quantitatively reduces to the condition of equality between the momentum fluxes of the stellar wind and the radiation, or $\dot M = L/(v \cdot c)$ , where $v$ is the terminal wind
speed and $c$ is the speed of light.

The intense wind of Wolf–Rayet stars is also probably radiative in nature. The wind of A stars and early F stars is weak due to their comparatively low luminosity but becomes stronger for late F stars as well as G and M stars. The solar wind is an example of the stellar wind of cool main-sequence stars that have convective envelopes. The cause of this wind may be turbulence in magnetized plasma driven by convection in the stellar envelope \cite{Shoda2021}. In turn, the magnetic field of the wind in cool main-sequence stars causes their
rotation to slow down over time \cite{Skumanich1972}. In this case, the
equatorial rotation velocity $v_{\text{e}}$ can be estimated from
the following expression:

\begin{equation}\label{eq-3}
	v_{\text{e}} = \frac{1.5\times10^5}{\sqrt{t}}, \text{km/s},	
\end{equation}

where $t$ is the age of the star in years.

For the Sun, the mass-loss rate due to the wind is small, about $10^{-14} M_\odot/$year, but it can be significantly higher for rapidly rotating young stars and components of cataclysmic binaries with solar-mass companions ($0.3 - 1.5 M_\odot$) due to tidal interaction between fast-rotating components \cite{Iben1984}. Together with H II regions around massive stars, such a wind could contribute to clearing young star clusters of gas \cite{Tutukov1978}.

Red giants and supergiants represent another family of stars with powerful stellar winds. At the stage of planetary nebula formation, the wind intensity can reach $\sim 10^{-1} M_\odot/$year. Observational estimates of the stellar wind intensity in massive red supergiants, the precursors of supernovae, range from $10^{-4}$ to $10^{-1} M_\odot/$year \cite{Ransome2024}. Red supergiants in symbiotic binaries whose donors do not fill their Roche lobes were
the first known examples of detached interacting binaries \cite{Tutukov1976}. Accretion by a degenerate dwarf of even a small fraction of the gas lost by such a donor leads to a series of non-stationary accretion and nuclear phenomena that attract observational interest.

Finally, in binary systems with compact accretors --- neutron stars and stellar-mass black holes --- there is the possibility of an irradiation-induced donor
stellar wind \cite{Davidson1973}. The cause is the large energy release
during accretion of matter at a specific energy output of $\sim 0.1 c^2$ onto the compact object. Accretion of 1 g matter releases about $\sim 10^{20}$ of energy. In extremely close binaries with comparable component masses, the donor absorbs around $4\times 10^{18}$ erg. Meanwhile, to lose one gram of mass, a main-sequence donor needs
about $10^{15}$ erg. Thus, strong irradiation can in principle trigger a self-sustaining stellar wind even in a donor without filling the Roche lobe.

A special case arises from the interaction of compact stars in a galactic nucleus with the fast wind of an actively accreting SMBH. The presence of a hot (with a temperature $T = 3\times10^6$ K) wind from such a black hole in the nucleus of galaxy M82 was established by
the authors of \cite{Bu2024,Boetcher2024}. A wind with a speed of
$\sim 500$ km/s can be generated either by the accretion disk of the SMBH or by a star-formation burst in the nucleus accompanied by type II 1b massive supernovae \cite{Loose1982}. In this case, the mass-loss rate of the nucleus can reach $10 M_\odot/$year. Accretion of SMBH wind gas by neutron stars and black holes in the galactic nucleus
can explain part of the X-ray emission from active galactic nuclei. Observations have detected extended X-ray sources in such nuclei \cite{Gupta2024,Kimbrell2024}.

\subsection{Condition for the Formation of an Accretion Disk
	during Donor Wind Accretion}{\label{subsec-2.3}}

For the donor wind matter captured by the gravity of the accretor, two scenarios are possible: either direct infall onto the accretor’s surface, or, if there is sufficient angular momentum, the formation of a gas torus around it. In the presence of sufficient viscosity, the
gas torus eventually evolves into an accretion disk, observable by its X-ray emission.

For analytical estimates of the conditions of donor wind interaction with the accretor, two classes of stellar winds can be distinguished --- hot and cold. We will call the wind <<hot>> if the speed of sound in its gas exceeds the orbital velocities of the components. If the speed of sound is lower than the orbital velocity, the wind is <<cold>>. Let us find the condition for the donor’s wind to cool to a temperature that provides a low speed of sound in its gas. The concentration of wind matter at the level of the accretor’s orbit is

\begin{equation}\label{eq-4}
	n_{\text{a}} = \frac{\dot M_{\text{w}}}{4\pi v_{\text{w}} A^2 \mu},	
\end{equation}
where $\dot M_{\text{w}}$ is gas loss rate due to the donor’s wind, $A$ is the intercomponent distance, and $\mu$ is the molecular weight of the gas. The cooling time of hot gas, according to the cooling function $\Lambda = 10^{-22}n^2$, erg/$(\text{cm}^3 \times \text{c})$ \cite{Semenov2003}:
\begin{equation}\label{eq-5}
	t_{\text{cool}} = \frac{10^6 T}{n},	\text{c},
\end{equation}

where $T$ is the gas temperature, n is the gas concentration. The time for the wind to traverse the intercomponent distance $A$ is defined by the simple relation
\begin{equation}\label{eq-6}
	t_{\text{orb}} = \frac{A}{v_{\text{w}}}.	
\end{equation}

Now assuming $t_{\text{orb}} = t_{\text{cool}}$ we find an estimate of
the wind intensity $\dot M_{\text{w}}$, which separates hot and cold
donor winds:
\begin{equation}\label{eq-7}
	\dot M_{\text{w}} = 10^{-10} \left(\frac{v_{\text{w}}}{300 ~[\text{km/s}]}\right)^4 \left(\frac{A}{R_\odot}\right), M_\odot/\text{year}.	
\end{equation}

\begin{figure}[H]
 \centering{\includegraphics[width=0.9\textwidth]{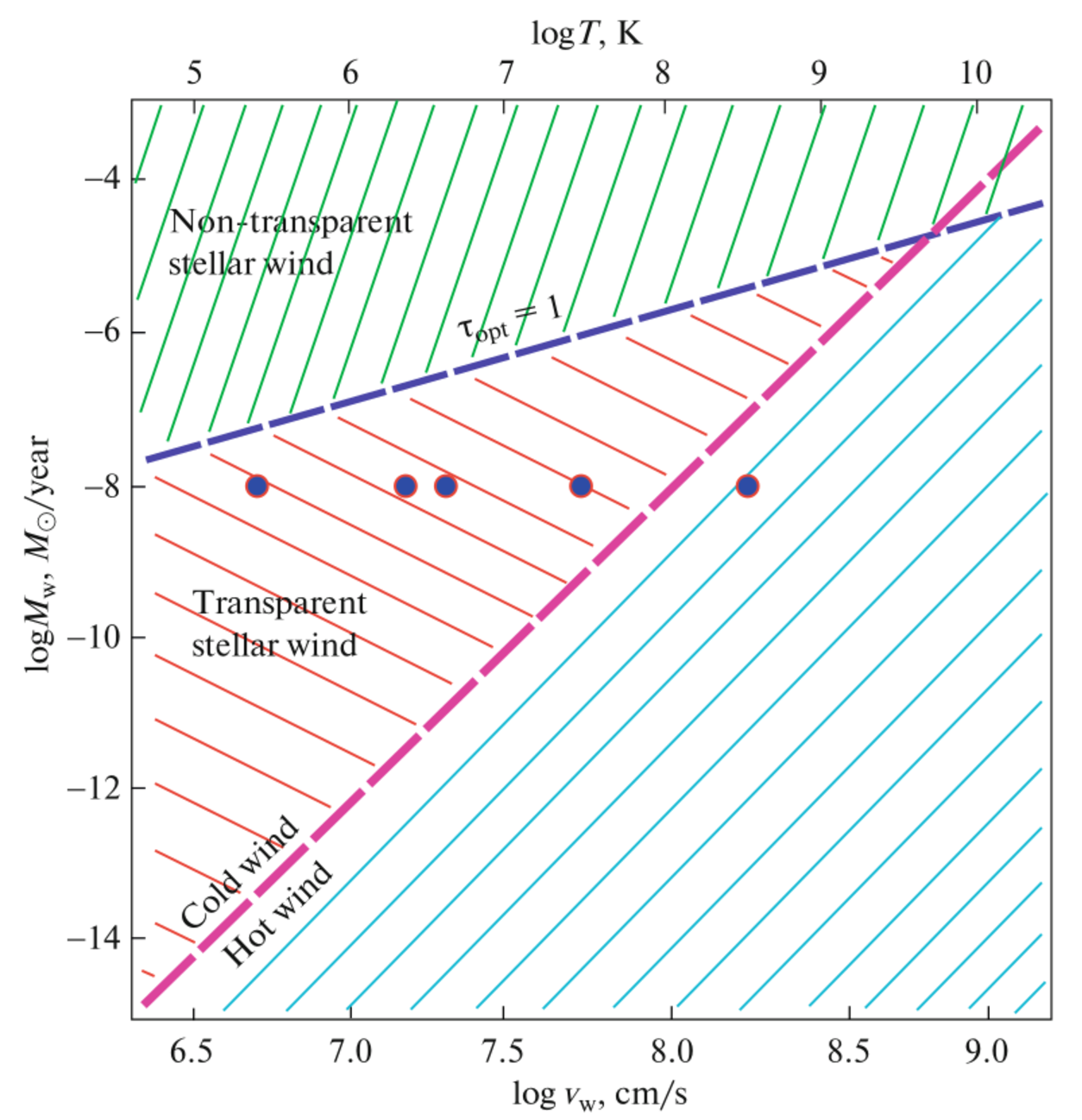}}
\caption{Classification of donor stellar winds in close binaries. The semi-major axis A is taken equal to the solar radius. The plot accounts for cooling of wind gas due to radiation. The red dashed line shows the dependence defined by (Eq. \ref{eq-7}); the blue dashed line indicates the transparency condition (Eq. \ref{eq-8}). Blue dots mark calculation variants of the binary system modeled in this study.}
    \label{fig-3}	
\end{figure}	 

It is evident that the boundary intensity of the wind strongly depends on its speed, since speed determines the wind density at the level of the accretor’s orbit and thus the gas cooling rate.

For volumetric cooling of the stellar wind, it must be transparent within the donor’s orbit. The transparency condition, within our simplified model, can be written as: 
\begin{equation}\label{eq-8}
	\dot M_{\text{w}} < 10^{-6} \left(\frac{v_{\text{w}}}{300 ~[\text{km/s}]}\right)^4 \left(\frac{A}{R_\odot}\right), M_\odot/\text{year}.	
\end{equation}

Graphically, conditions (Eq. \ref{eq-7}) and (Eq. \ref{eq-8}) are shown in
Fig. \ref{fig-3}. Despite the simplified approach to estimating essential parameters, the figure helps orient us in understanding the nature of the donor wind in close binary systems.

Fig. \ref{fig-3} shows that modeling intense winds with $\dot M_{\text{w}} \gtrsim 10^{-6} M_\odot/$year requires solving the radiative transfer equation in the wind flow. Modeling wind
dynamics with speed $v_{\text{w}} \lesssim 300$ km/s and $10^{-12} \leqslant \dot M_{\text{w}} \leqslant 10^{-7} M_\odot/$year requires accounting for gas cooling, which leads to large contrasts in gas density and temperature within the accretor’s orbit. A wind with
high gas temperature and low intensity is transparent; thus, its gas can be considered nearly adiabatic within a close binary system.

Now let us consider an analytical approximation describing the conditions of accretor interaction with the donor wind gas. The radius of capture of cold wind matter by the accretor is
\begin{equation}\label{eq-9}
	r_{\text{cap}} \approx \frac{G M_{\text{a}}}{v_{\text{w}}^2} \approx 10^{26}\left(\frac{M_{\text{a}}}{M_\odot}\right)\left(\frac{1}{v_{\text{w}}^2}\right), \text{cm},	
\end{equation}
where $G$ is the gravitational constant, and $M_{\text{a}}$ is the
accretor’s mass.

Let us consider the condition for accretion disk formation in the case of a cold wind. Suppose the donor’s angular velocity is
\begin{equation}\label{eq-10}
	\Omega_{\text{d}} = \alpha \left[G(M_{\text{d}} + M_{\text{a}})\right]^{\frac{1}{2}} A^{-\frac{3}{2}},	
\end{equation}
where $\alpha \approx 1$ is the condition of synchronous axial
rotation of the donor with the orbital motion. Given the radius of capture of the donor’s wind matter by the accretor $r_{\text{cap}}$, the condition for disk formation is that the angular momentum of the captured matter exceeds the angular momentum of a maximally compact disk with radius $r_{\text{disc}}$. After straightforward transformations, we find the condition for the formation of an accretion-decretion disk around the accretor
\begin{equation}\label{eq-11}
	v_{\text{w}} < \frac{\alpha^{\frac{1}{4}} G^{\frac{1}{2}} M_{\text{a}}^{\frac{3}{8}} (M_{\text{d}} + M_{\text{a}})^{\frac{1}{8}}}{A^{\frac{3}{8}} R_{\text{a}}^{\frac{1}{8}}}.	
\end{equation}

The wind speed constraint (Eq. \ref{eq-11}) can be rewritten in
dimensionless form using the expression for orbital velocity:
\begin{equation}\label{eq-12}
	v_{\text{orb}} = \sqrt{\frac{G(M_{\text{d}} + M_{\text{a}})}{A}}.
\end{equation}
For the parameters of Sco X-1 given in Table \ref{tab-1},
$v_{\text{orb}} = 2 \times 10^7$ cm/s. Then,
\begin{equation}\label{eq-13}
	\frac{v_{\text{w}}}{v_{\text{orb}}} < \alpha^{\frac{1}{4}} \left(\frac{M_{\text{a}}}{M_{\text{d}} + M_{\text{a}}}\right)^{\frac{3}{8}} \left(\frac{A}{R_{\text{a}}}\right)^{\frac{1}{8}}.	
\end{equation}

This expression is convenient for analysis. For most solid-body rotating binary systems with donors possessing stellar winds, the critical wind speed for accretion disk formation is on the order of the orbital speed. In the Sco X-1 system, as in other X-ray binaries with compact components, the critical speed for disk formation is several times higher than the orbital speed (Eq. \ref{eq-13}) due to the large system size relative to the accretor’s radius. Considering the simplicity of our analytical model, it is difficult to obtain a reliable numerical estimate of the critical speed. However, the model
allows establishing the existence of two interaction regimes between the wind and the accretor.

Let us estimate the initial accretor disk size $r_{\text{disc,0}}$,
corresponding to the angular momentum of the captured matter, assuming its angular velocity $\Omega_{\text{w}} = \Omega_{\text{b}} \frac{r_{\text{cap}}}{A}$ (Eq. \ref{eq-9}), (Eq. \ref{eq-11}), $\Omega_{\text{b}} = \frac{G^{1/2} \left(M_{\text{d}} + M_{\text{a}}\right)^{1/2}}{A^{3/2}}$. After simple transformations, we find
\begin{equation}\label{eq-14}
  r_{\text{disc,0}} = \frac{G^6 M_{\text{a}}^5 \left(M_{\text{d}} + M_{\text{a}}\right)}{A^5 v_{\text{w}}^{12}}.	
\end{equation} 
At the accretor’s orbital velocity
\begin{equation}\label{eq-15}
	v_{\text{acc}} = \frac{G^{\frac{1}{2}} M_{\text{d}}}{\left(M_{\text{d}} + M_{\text{a}}\right)^{\frac{1}{2}} A^{\frac{1}{2}}}
\end{equation} 
the initial disk size is
\begin{equation}\label{eq-16}
  r_{\text{disc,0}} =	r_{\text{cap}} \left(\frac{M_{\text{a}}}{M_{\text{d}}}\right)^4 \left(\frac{M_{\text{d}} + M_{\text{a}}}{M_{\text{d}}}\right)^6 
  \left(\frac{v_{\text{acc}}}{v_{\text{w}}}\right)^{10}.
\end{equation} 

It is clear that the initial disk radius strongly depends on wind speed. Numerical viscosity in the model brings the accretion disk size, over several orbital periods, to the characteristic value of the model disk (Fig. \ref{fig-6}).

The model disk around the accretor gradually expands over time due to numerical viscosity. The characteristic expansion time can be estimated using the relation derived by linearizing the diffusion equation: $\tau_{\text{diff}} \approx \frac{r_{\text{disc}}^2}{l v_{\text{w}}}$, where $r_{\text{disc}} \approx 1 R_{\odot}$, $v \sim 0.1 v_{\text{k}}$, $v_{\text{k}}$ is the Keplerian velocity of gas in the disk, following from the relative disk thickness (Fig. \ref{fig-10}), and $l = 1.5 \times 10^7$ cm is the characteristic size of the computational grid cell. Substituting these values into the formula gives $\tau_{\text{diff}} = 1 P_{\text{orb}}$. It should be noted that accretion disks formed by donor wind accretion typically arise around compact objects --- white dwarfs, neutron stars, and black holes --- when the wind speed satisfies condition (Eq. \ref{eq-13}).

\subsection{Conditions for Sustaining a Stellar Wind Induced by X-Ray Irradiation of the Donor}{\label{subsec-2.4}}

Let us consider the condition for sustaining an induced stellar wind in low-mass X-ray binaries (LMXB) with neutron stars and black holes as accretors. The energy release during wind accretion onto a compact object is quite large, on the order of $10^{20}$ erg/s, and absorption of just a small fraction of this energy by the donor --- a low-mass main-sequence star --- is sufficient to sustain its stellar wind.

Suppose the donor, under irradiation from a nearby accretor, has a stellar wind with speed $v_{\text{d}} = \alpha \sqrt{\frac{G M_{\text{d}}}{R_{\text{d}}}}$. In this case, mass infall onto the accretor occurs from a region of radius given by expression (Eq. \ref{eq-9}). Then the accretion rate is
\begin{equation}\label{eq-17}
	\dot M_{\text{a}} = \frac{G^2 M_{\text{a}}^2 \dot M_{\text{w}}}{4 v_{\text{d}}^4 A^2},	
\end{equation} 
and the X-ray luminosity of the accretor, assuming spherically symmetric accretion, is
\begin{equation}\label{eq-18}
	L_{\text{x}} = 2.5 \times 10^{19} \left(\frac{\dot M_{\text{w}}}{\alpha^4}\right) \left(\frac{r_{\text{cap}}}{A}\right)^2, \text{erg/s}.	
\end{equation}  
The donor absorbs a part of this energy equal to $L_{\text{abs}} = \left(\frac{r_{\text{cap}}}{2A}\right)^2$, with efficiency $\beta$, which sustains its stellar wind. As a result, to sustain the induced stellar wind, the following condition must be met:
\begin{equation}\label{eq-19}
	v_{\text{d}} < 2.5 \times 10^{9} \left(\frac{\beta^{\frac{1}{2}}}{\alpha}\right) \left(\frac{M_{\text{d}}}{M_{\text{a}}}\right) \left(\frac{r_{\text{cap}}}{A}\right)^2, \text{cm/s}.	
\end{equation}    

Assuming for simplified numerical estimates
$v_{\text{d}} \simeq 5 \times 10^7$ cm/s, $M_{\text{d}} \approx M_{\text{a}}$, $\alpha = 1$ and $r_{\text{cap}} = r_{\text{R}}$, where $r_{\text{R}}$ is the Roche lobe radius of the accretor, we find that to sustain the induced wind it is necessary to have $\beta >10^{-2}$. The choice of values for parameters $\alpha$ and $\beta$ is determined by the reliability of the numerical model of the stellar wind. At the same time, analysis of equation (Eq. \ref{eq-19}) shows that filling the Roche lobe is not a necessary condition for maintaining the donor’s
wind. This circumstance, as noted in the Introduction, allows us to extend the concept of interacting binaries to include detached systems with relatively close components.

In reality, the problem of stellar wind generation by X-ray irradiation from an accreting compact object --- a neutron star or black hole—requires the development of a complex three-dimensional gas-dynamical model that takes into account heating of the donor’s surface by the accretor’s X-ray radiation as well as radiative losses from the hot wind matter. However, an estimate of the condition for exciting a stellar wind in an X-ray binary, accounting for radiative energy losses, can be made within a simple analytical model of the system’s energetics to identify the main parameters of this problem. As initial conditions, we set the donor’s mass $M_{\text{d}}$ and radius $R_{\text{d}}$, the accretor’s mass $M_{\text{a}}$, and the
orbital semi-major axis $A$. From the condition of continuity, the donor’s mass-loss rate due to the wind is $\dot M_{\text{w}} = 4 \pi \rho v_{\text{d}} A^2$, where $\rho$ is the wind matter density.
We find an expression for the wind matter concentration $n_{\text{w}}$ at the accretor’s orbit:
\begin{equation}\label{eq-20}
	n_{\text{w}} = 10^9 \frac{\dot M_{\text{w,-10}}}{v_{\text{d,8}} (\frac{A}{R_\odot})^2}, \text{cm}^{-3},	
\end{equation}    
where $\dot M_{\text{w,-10}}$ is the donor’s mass-loss rate (in units of $10^{-10} M_\odot/$year), and $v_{\text{d,8}}$ is the wind speed (in units of $10^8$ cm/s).

To estimate the intensity of the emission lines of the hot corona of an X-ray binary system, an estimate of the emission measure is necessary:
\begin{equation}\label{eq-21}
	ME = n_{\text{w}}^2 A^3 \simeq 10^{51} \left(\frac{\dot M_{-10}}{v_{\text{w,8}}}\right)^2 \left(\frac{A}{R_\odot}\right)^3, \text{cm}^{-3}.	
\end{equation}

To estimate the efficiency of radiative energy losses from the hot wind matter, we estimate its optical depth due to Thomson scattering:
\begin{equation}\label{eq-22}
	\tau_{\text{rad}} = \varkappa \rho A \simeq 4 \times 10^{-5} \left(\frac{\dot M_{-10} R_\odot}{v_{\text{w,8}} A}\right),	
\end{equation}
where $\varkappa$ is the absorption coefficient. Obviously, it is small over a wide range of wind intensities at $\dot M \lesssim 10^{-6} M_\odot/$year. This makes it possible to estimate the radiative losses from the hot wind matter, knowing its density distribution and cooling rate $\Lambda \sim 10^{-22} n_{\text{w}}^2$, erg/(cm$^3 \cdot$ s) \cite{Kholtygin2002}:
\begin{equation}\label{eq-23}
	\frac{L_{\text{R}}}{L_{\odot}} \approx 2 \times 10^{-5} \left(\frac{\dot M_{-10}}{v_{\text{w,8}}}\right)^2 \left(\frac{R_\odot}{R_{\text{d}}}\right).	
\end{equation}

Evaporation of donor wind matter and its acceleration to parabolic velocity requires absorption of energy at the rate of
\begin{equation}\label{eq-24}
	\frac{L_{\text{w}}}{L_{\odot}} \approx 2 \times 10^{-3} \dot M_{-10} \left(\frac{M_{\text{d}}}{M_\odot}\right) \left(\frac{R_\odot}{R_{\text{d}}}\right).	
\end{equation} 
The accretor’s X-ray luminosity is determined by wind matter infalling within a circle of radius $r_{\text{cap}} \sim G^2 M_{\text{a}}^2 / v_{\text{w}}^4$:
\begin{equation}\label{eq-25}
	\frac{L_{\text{x}}}{L_{\odot}} \approx \left(\frac{\dot M_{-10}}{v_{\text{w,8}}^4}\right) \left(\frac{M_{\text{a}}}{M_\odot}\right)^2 \left(\frac{R_\odot}{A}\right)^2.	
\end{equation} 
Now, from the energy balance condition
\begin{equation}\label{eq-26}
	L_{\text{x}} = L_{\text{R}} + L_{\text{w}}	
\end{equation}  
we find the condition for existence and intensity of the induced stellar wind from the donor in the binary:
\begin{equation}\label{eq-27}
	\dot M_{-10} = 5 \times 10^4 v_{\text{w,8}}^2 \left(\frac{R_{\text{d}}}{R_\odot}\right) \left[\left(\frac{M_{\text{a}}}{M_\odot}\right)^2 \left(\frac{R_\odot}{A}\right)^2 \left(\frac{1}{v_{\text{w,8}}^4}\right) - 2 \times 10^{-3} \left(\frac{M_{\text{d}}}{M_\odot}\right) \left(\frac{R_\odot}{R_{\text{d}}}\right) \right], M_\odot/\text{year}.	
\end{equation} 
Then, the condition for sustaining a self-induced stellar wind is that the expression in square brackets in (Eq. \ref{eq-27}) is positive, or:
\begin{equation}\label{eq-28}
	\frac{A}{R_\odot} \lesssim \left(\frac{20}{v_{\text{w,8}}^2}\right) \left(\frac{M_{\text{a}}}{M_\odot}\right) \left(\frac{R_{\text{d}}}{R_\odot}\right)^{\frac{1}{2}} \left(\frac{M_\odot}{M_{\text{d}}}\right)^{\frac{1}{2}}.	
\end{equation} 
This condition shows that X-ray binaries with solar-mass components can have an induced stellar wind in detached systems when $2.5 \lesssim \frac{A}{R_\odot} \lesssim 20$, if $v_{\text{w,8}} \simeq 1$.

Neglecting the comparatively small contribution of $L_{\text{w}}$ in (Eq. \ref{eq-27}), we estimate the donor’s stellar wind intensity:
\begin{equation}\label{eq-29}
	\dot M_{\text{w}} = \left(\frac{5 \times 10^{-6}}{v_{\text{w,8}}^2}\right) \left(\frac{R_{\text{d}}}{R_\odot}\right) \left(\frac{M_{\text{a}}}{M_\odot}\right)^2 \left(\frac{R_\odot}{A}\right)^2, M_\odot/\text{year}	
\end{equation} 
and, taking (Eq. \ref{eq-25}) into account, the accretor’s X-ray luminosity:
\begin{equation}\label{eq-30}
	\frac{L_{\text{x}}}{L_{\odot}} = \left(\frac{5 \times 10^4}{v_{\text{w,8}}^6}\right) \left(\frac{M_{\text{a}}}{M_\odot}\right)^4 \left(\frac{R_{\text{d}}}{R_\odot}\right) \left(\frac{R_\odot}{A}\right)^4.	
\end{equation}

Equation (Eq. \ref{eq-30}) is illustrative in many respects. Its dimensionless form clearly shows not only the main factors determining the system’s X-ray luminosity fed by the donor’s induced wind, but also their relative importance. In particular, this luminosity strongly
depends on the donor wind velocity near the accretor.

In addition, it must be noted that a realistic model of wind generated by X-ray irradiation does not yet exist. The known parametric expression for wind speed as a function of distance from the donor,
$v_{\text{w}} = 200 (\frac{R}{R_\odot} - 1)$, km/s \cite{Cranmer1999} at $\frac{R}{R_\odot} \lesssim 3.5$ and $v_{\text{w}} = v_{\text{w,0}} \left(1 - \frac{R_{\text{s}}}{R}\right)^{\beta}$, where $R$ is the distance from star with the radius $R_{\text{s}}$, $\beta \approx 1$ \cite{Antokhin2022} implies a large wind acceleration zone nearly three times the stellar radius. Clearly, the <<uncertainty>> in the donor wind speed, which strongly determines the system’s X-ray luminosity (see (Eq. \ref{eq-30})), complicates realistic estimates
using this equation. However, this equation provides a reliable estimate of speed $v_{\text{w,8}}$ for the Sco X-1 system: $v_{\text{w,8}} \simeq 0.36$. More importantly, it also indicates the
need for a robust stellar wind model to estimate the X-ray luminosity of such systems.

Equation (Eq. \ref{eq-30}) also shows the possibility of achieving high, nearly Eddington-level, X-ray luminosity for a neutron star of mass $m_{\text{NS}} \simeq 1.4 M_\odot$ and a black hole of mass $m_{\text{BH}} \simeq 10 M_\odot$ as a result of self-induced stellar
wind from a nearby solar-mass donor. Noticeable is the strong dependence of luminosity on component separation. This means that the system’s X-ray luminosity is maximal only for donors nearly filling or filling their Roche lobe.

At the same time, this condition indicates the brevity of the bright X-ray source stage in the life of a binary with massive donors. Filling the Roche lobe leads to a common envelope for the system. It is noteworthy that the intensity of the wind matter flow accreted by the compact companion of the solar-mass donor (see (Eq. \ref{eq-29})) is comparable to the accretion rate in semi-detached X-ray binaries driven by angular momentum loss through the donor’s magnetic stellar
wind \cite{Iben1984}.

The modeling shows the presence of a shock wave on the order of the size of the binary system, reflecting the accretor’s gravitational focusing effect on the stellar wind (Fig. \ref{fig-10},\ref{fig-12} and \ref{fig-14}). It forms from part ($\alpha$) of the wind matter. Taking the donor corona’s radiative losses from (Eq. \ref{eq-23}) and the accretor’s X-ray luminosity from (Eq. \ref{eq-25}), we can compute the ratio of radiative loss energy to accretor X-ray luminosity:
\begin{equation}\label{eq-31}
	\frac{L_{\text{R}}}{L_{\text{x}}} = 2 \times 10^{-5} \dot M_{-10} \left(\frac{v_{\text{w,8}} M_\odot}{M_{\text{a}}}\right)^2 \left(\frac{A}{R_{\odot}}\right).	
\end{equation}  

This relation shows the main factors determining the relative brightness of the shock wave in objects accreting wind matter. To estimate the observability of the shock wave, it is necessary to consider the gas’s radiative properties.

Of course, the formalism adopted for estimating the conditions of induced stellar wind generation in close binaries is simple. It is based on two channels for expending X-ray energy absorbed by the donor:
acceleration of the donor’s envelope matter and radiative losses from the wind matter. Each channel is complex due to wind inhomogeneity and the difficulties in constructing models of wind heated by the accretion
disk’s radiation. Therefore, the aim of this work is not to build a self-consistent model of such systems, but to provide an estimate of the energetics of the two main processes --- coronal evaporation of the donor and radiative cooling of its corona.

The extensive family of LMXB has been actively and comprehensively studied over the past 60 years since the discovery of its brightest representative, Sco X-1. The total number of known Galactic LMXBs
today is 339 systems \cite{Fortin2024}. The orbital period of 50 of
them, with neutron stars as accretors, lies within $0.4 - 10^4$ h, while the neutron stars’ own spin periods reach $10^{-3}$ s \cite{Fortin2024}. Short-period neutron accretors occur only in systems with orbital periods shorter than $\sim 10$ h, indicating that the driving forces of their evolution coincide with the causes of evolution of cataclysmic systems \cite{Iben1997}. X-ray luminosities of LMXBs correspond to accretion rates of $10^{-15} - 10^{-8} M_\odot$year \cite{Fortin2024}. The question of Roche lobe filling by donors in such systems remains open due to the possibility of an
induced wind generated by the accretor’s X-ray irradiation \cite{Iben1997}.

The above mentioned analytical estimates of the conditions for sustaining an induced stellar wind demonstrate a strong dependence of the system’s luminosity on input parameters and the wind model. Currently, the solar wind model is well developed, since its physical characteristics are directly measured in the Solar System by satellites, allowing modeling results to be compared with observations. For stars of other types and spectral classes, gas-dynamic models of winds are built assuming similarity to the solar wind model, as there is no reliable information on the physical processes of wind generation and flow for these objects.

In this study, we construct the wind model using the solar wind approximation, without accounting for certain physical processes related to wind propagation within the computational domain. In particular, the model does not include possible wind acceleration
mechanisms far from the donor. Even for the Sun, estimates of this acceleration differ greatly among researchers. Since the donor in our model is a red dwarf rather than the Sun, the obtained numerical
results are mainly illustrative and should be regarded as a demonstration of a possible wind pattern in detached systems.

\section{NUMERICAL MODELING OF MASS EXCHANGE IN DETACHED BINARY SYSTEMS}{\label{sec-3}}

\subsection{Object of Modeling}{\label{subsec-3.1}}

Three-dimensional modeling of the gas dynamics of winds in detached binary stars of various classes is a traditional problem in computational astrophysics. One popular problems of this type is studying the interaction of winds from the components \cite{Folini2000,Pittard2010,Madura2013}. Another example is modeling the gas flow structure in symbiotic binaries \cite{Mitsumoto2005,Saladino2019,Gawryszczak2003}, where the giant’s stellar wind is partially captured by a degenerate companion \cite{Tutukov1976}.

To illustrate the process of mass exchange in detached binaries via the donor’s stellar wind, we carried out a series of three-dimensional numerical simulations of the gas flow structure. As shown in Subsection \ref{subsec-2.2}, the systems considered can include various
types of stars as donor and accretor. To present comparable results, we used a generalized configuration of a binary system for the simulations, based on the characteristics of the X-ray binary Sco X-1.

Sco X-1 is among the first X-ray sources known to astrophysicists. It has been actively and comprehensively studied for several decades, which has made it possible to determine its parameters with sufficient
accuracy. Current estimates of this binary’s parameters are given in \cite{Karino2021,Antokhin2022,Cherepaschuk2022}. It should be noted that the donors in about a dozen similar systems with orbital periods of several hours are mostly main-sequence stars \cite{Campana2018}, and thus Sco X-1 is representative of an entire family of LMXB. The neutron stars in these systems have very short spin periods, $P_{\text{spin}} = 2 \times 10^{-3} - 4 \times 10^{-3}$ s, indicating a long X-ray phase, $10^{7}$ years, and efficient accretion, $\sim 0.1 M_\odot$.

Table \ref{tab-1} lists the main characteristics of Sco X-1 that we used in our calculations. To simplify the gas dynamics modeling of flows in this system, we used not a neutron star but a white dwarf with typical size $R_{\text{a}} = 0.01 R_{\odot}$ as the accretor. Since we mainly investigate the dynamics of the donor’s stellar wind, this simplification of the model is acceptable. The donor is represented by a red dwarf with radius $R_{\text{d}} = 0.15 R_{\odot}$,
which does not fill its Roche lobe. The distance between the components was taken as $A = 4 R_{\odot}$.

\begin{table*}[h!]
	\caption{Main parameters of the X-ray binary Sco X-1.} 
	\label{tab-1}
	\centering
	\setlength{\arrayrulewidth}{1.0pt}
	\begin{tabular}
		{|
			!{}>{\centering\arraybackslash}m{5.0cm}|
			!{}>{\centering\arraybackslash}m{2.8cm}|
			!{}>{\centering\arraybackslash}m{3.0cm}|
			!{}>{\centering\arraybackslash}m{2.8cm}|
	    }	
		\hline
		\textbf{Parameter} &
		\textbf{Notation} &
		\textbf{Value} &
		\textbf{Unit of measurement} \\ 
		\hline
		Accretor mass & $M_{\text{a}}$ & 1.4 & $M_\odot$ \\ 
		\hline
		Accretor radius & $R_{\text{a}}$ & $2 \times 10^{-5}$ & $R_\odot$ \\
		\hline 
		Donor mass &  $M_{\text{d}}$ & 0.4 & $M_\odot$ \\
		\hline 
		Donor radius & $R_{\text{a}}$ & 0.4 & $R_\odot$ \\
		\hline 
		Donor’s Roche lobe radius & $R_{\text{Roch,d}}$ & 1.26 & $R_\odot$ \\
		\hline
		Intercomponent distance & A & 4.8 & $R_\odot$ \\	
		\hline
		Orbital period & $P_{\text{orb}}$ & 19 & h \\
		\hline
		Mass transfer rate &  $\dot M$ & $1 \times 10^{-8}$ & $M_\odot/$year \\
		\hline		
	\end{tabular}
\end{table*}

\subsection{Numerical Model}{\label{subsec-3.2}}

The three-dimensional numerical simulations for this task were performed using an MHD code that we had previously applied to model cataclysmic variables --- polars and intermediate polars --- which determined the above modifications of model parameters relative to the Sco X-1 system \cite{Zhilkin2019}. This code allows simulation of the gas dynamics of binary systems, including systems with magnetic fields of varying strengths. A detailed description of the physical, mathematical, and numerical models of the code can be found in \cite{Zhilkin2019, Balsara1998}.

Here, we outline the initial and boundary conditions used in our numerical model. Since the inner boundary of the computational domain follows the donor’s Roche lobe, we present the initial conditions
in this zone. The normal wind speed $v_n$ was set equal to the initial wind speed $v_{\text{w}}$, proportional to the square root of the effective temperature of the donor’s corona $T_\text{c}$ (Eq. \ref{eq-32}) below. The accretor is defined as a sphere of radius $R_{\text{a}}$ with conditions of free inflow applied at the boundary. The magnetic field at the outer boundary of the computational domain was set to $\boldsymbol{b}_b = 0$. For the boundary velocity of the matter $\boldsymbol{v}_b$ free outflow conditions were imposed: when the velocity is directed outward, the conditions $\partial \boldsymbol{v}_b/\partial \boldsymbol{n} = 0$ were used, where $\boldsymbol{n}$ is the normal to the donor’s surface. The initial conditions in the computational domain are as follows: density $\rho_0 = 10^{-8} \rho(\rm{L}_1)$ (for the $\rho(\rm{L}_1)$ values, see Table \ref{tab-2}), temperature $T_0 = T_{\text{c}}$, velocity $\boldsymbol{v}_0 = 0$ and magnetic field $\boldsymbol{b}_0 = 0$.

The applied MHD code lacks a necessary parameter --- the initial speed of the stellar wind at the donor’s surface --- because it had previously been used only for models in which the donor completely filled its Roche lobe, and mass outflow was determined by the mass-loss rate $\dot{M}$ through the vicinity of the inner Lagrangian point $\rm{L}_1$. To model a stellar wind with a given initial velocity, the donor’s temperature $T_{\text{d}}$, was used as an input parameter; however, since in our model the donor does not fill its Roche lobe, the $T_{\text{d}}$ value corresponds to the temperature at the base of the donor’s corona $T_{\text{c}}$, which determines the wind gas expansion rate.

To study the dynamics of the stellar wind, we calculated five models corresponding to different initial wind speeds $v_{\text{w,0}}$, determined by its initial temperature (Table \ref{tab-2}). The initial temperature values were chosen to yield velocities on the order of the parabolic velocity at the donor’s surface.

The dependence of the v w,0 value on the donor corona temperature $v_{\text{w,0}}$ is given by the expression
\begin{equation}\label{eq-32}
	v_{\text{w,0}} = \sqrt{\frac{3 k_{\text{B}} T_{\text{c}}}{m_{\text{p}}}}, 
\end{equation}
where $k_{\text{B}}$ is Boltzmann’s constant and mp is the proton
mass. The velocity vector is directed normal to the donor’s Roche lobe surface.

\begin{table*}[h!]
	\caption{Numerical models of donor wind calculations.} 
	\label{tab-2}
	\centering
	\setlength{\arrayrulewidth}{1.0pt}
	\begin{tabular}
		{|
			!{}>{\centering\arraybackslash}m{3.0cm}|
			!{}>{\centering\arraybackslash}m{3.5cm}|
			!{}>{\centering\arraybackslash}m{3.5cm}|
			!{}>{\centering\arraybackslash}m{3.5cm}|
		}	
		\hline
		\textbf{Model number} &
		\textbf{Donor corona temperature, $T_{\text{d}}$, K} &
		\textbf{Characteristic wind speed, $v_{\text{w0}}$, km/s} & 
		\textbf{Density of matter at Lagrange point $\rm{L}_1$, $\rho(\rm{L}_1)$, g/cm$^3$}\\ 
		\hline
		1 & $3 \times 10^5$ & 90 & $4 \times 10^{-11}$ \\
		\hline
		2 & $3 \times 10^6$ & 270 & $1 \times 10^{-12}$ \\
		\hline
		3 & $5.5 \times 10^6$ & 370 & $5 \times 10^{-13}$ \\ 
		\hline
	    4 & $3 \times 10^7$ & 860 & $4 \times 10^{-14}$ \\
		\hline 
		5 & $3 \times 10^8$ & 2700 & $1 \times 10^{-15}$ \\
		\hline		
	\end{tabular}
\end{table*}

The use of the MHD code assumes the presence of a magnetic field at the accretor. Since our task is to study the pure gas dynamics of the donor’s wind, we ran the simulations with a very small magnetic field
strength at the accretor, 1 Gs. Such field induction corresponds to an effective magnetosphere radius (Alfv\'en radius) of $1.5 \times 10^6$ cm, which is two orders of magnitude smaller than the accretor’s radius. This allows us to state that in our problem, the wind flow structure near the accretor is mainly determined by its gravitational potential.

\begin{figure}[H]
  \centering{\includegraphics[width=0.55\textwidth]{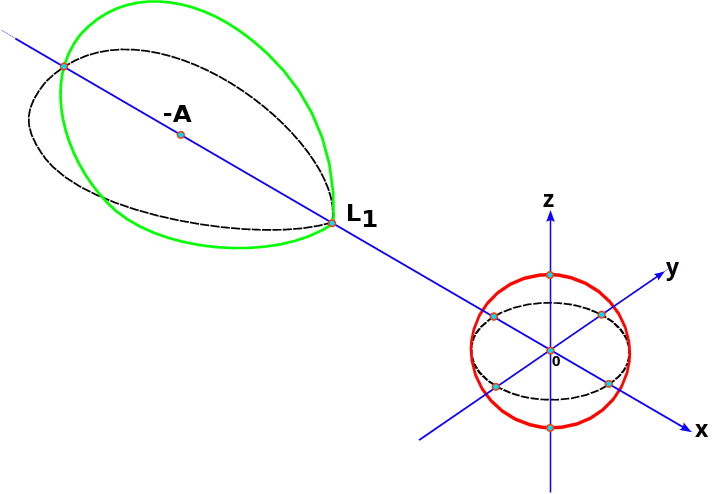}}
  \centering{\includegraphics[width=0.8\textwidth]{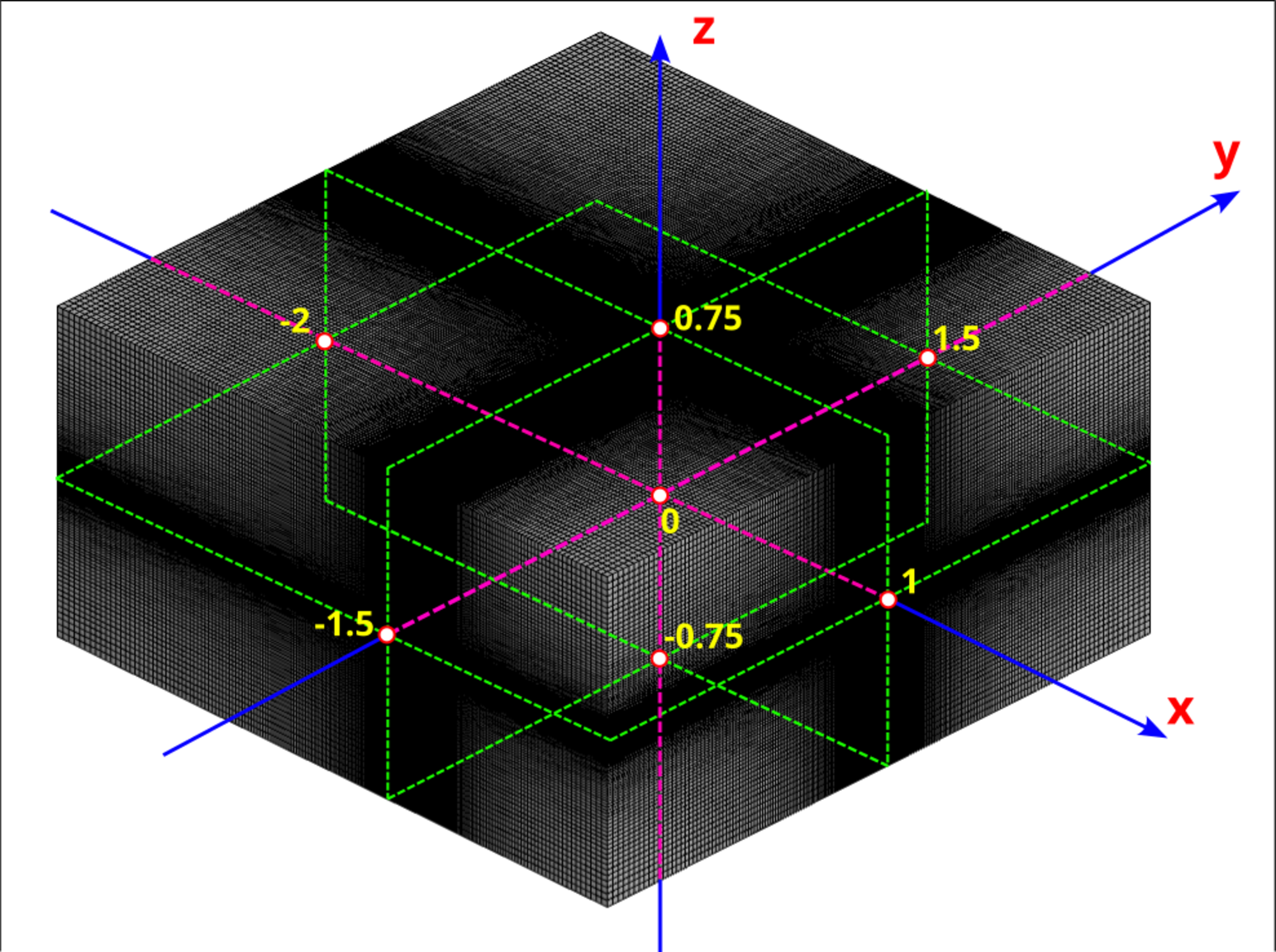}}
  \caption{Coordinate system and computational grid used in the model.}
  \label{fig-4}
\end{figure}

Figure \ref{fig-4} shows the Cartesian coordinate system used in the model and the grid over the computational domain. This grid contains 256 cells along the $x$ and $y$ axes and 128 cells along the $z$ axis. It should be noted that the grid is adaptive, with a decreasing step
size toward the origin along all three axes. This makes it possible to resolve the flow structure near the accretor in greater detail. The minimum grid step is $0.2 R_{\text{a}}$.

The computational domain fully includes the Roche lobes of both the donor and the accretor. The origin coincides with the center of the accretor star, and the donor center is offset from it by the intercomponent distance $A$. In the diagram, the donor is shown as its
Roche lobe (green line), and the accretor boundary (red line) coincides with the surface of the corresponding star. Since the model was built assuming a dipolar configuration of the accretor’s magnetic field, the figure also marks the position of the magnetic dipole axis $\mu$ and its defining spherical angles $\theta$ and $\phi$.

Since the computational domain is bounded not only by the external sides of the grid but also by the stellar surfaces (for the donor --- its Roche lobe), the internal volume of the binary components is excluded from the simulation. Therefore, outflow of matter from the donor is determined by the initial and boundary conditions of the numerical model. The wind matter captured by the accretor does not accumulate on its surface but leaves the computational domain into the
star’s interior volume. Moreover, due to the short simulation times --- on the order of several orbital periods --- we do not consider evolutionary changes in the binary system but rather show its state at a specific moment in time in the simulation results.

\subsection{Results of Numerical Simulations}{\label{subsec-3.3}}

In presenting the results of three-dimensional numerical simulations of gas dynamics here, it is important to note the difficulty of representing them in the two available coordinates of a journal page. To convey the geometry of the gas flows, we used a series of 2D cross-sections (in the $XY$ and $XZ$ planes) of the model’s three-dimensional picture, in each of which the distributions of density and temperature across the computational domain were plotted.

Assessing the adequacy of three-dimensional gas-dynamic modeling of the stellar wind in components of observed close X-ray binaries is complicated by the low gas density in the intercomponent space. This circumstance reduces the observational information available about the physics and kinematics of this gas, weakening its emission lines. As a result, the possibilities for directly comparing theoretical 3D models of the rarefied gas flow in a close binary system with observations are significantly limited. Currently, observers have access only to the optical emission of the donor, the X-ray emission of the accretion disk, and the bow shock cone formed as the accretor is enveloped by the donor’s high-velocity wind (see Figs. \ref{fig-16} and \ref{fig-18} below). The low-density wind gas that fills the space between the components and surrounds the system remains largely inaccessible for studying its physics and kinematics. The inability to directly compare three-dimensional gas-dynamic models with observations leaves them, in the absence of detailed physics of the stellar wind, as merely a set of potentially plausible flow patterns of gas kinematics in low-mass close binaries.

The three-dimensional numerical simulations were performed for the case of synchronous rotation of the accretor with the orbital motion of the binary system. The calculation results represent a stationary
solution corresponding to a fixed position of the binary system at a particular time point and the distribution of gas-dynamic parameters within the computational domain. A feature of the MHD code we used
is that after a certain period of computation, it establishes a quasi-stationary state of matter in the computational domain. This state is characterized by the constancy of the model gas mass, meaning the amount of gas generated within the system equals the amount
leaving it. However, within the quasi-stationary state, the gas-dynamic parameters still evolve, leading to some fluctuations in the flow structure while maintaining a constant total amount of matter. Nevertheless, such fluctuations do not cause significant
changes in the flow pattern, which allows us to speak of obtaining the required stationary solution to the problem. Typically, achieving a quasi-stationary state takes several (depending on the task) orbital periods of the binary system. In the model considered in this study, the time required to reach the state necessary for obtaining a stationary solution is 8 $P_{\text{orb}}$.
\begin{figure}[H]
	\begin{minipage}[h]{0.40\linewidth}	
	 \centering{\includegraphics[width=1.0\textwidth]{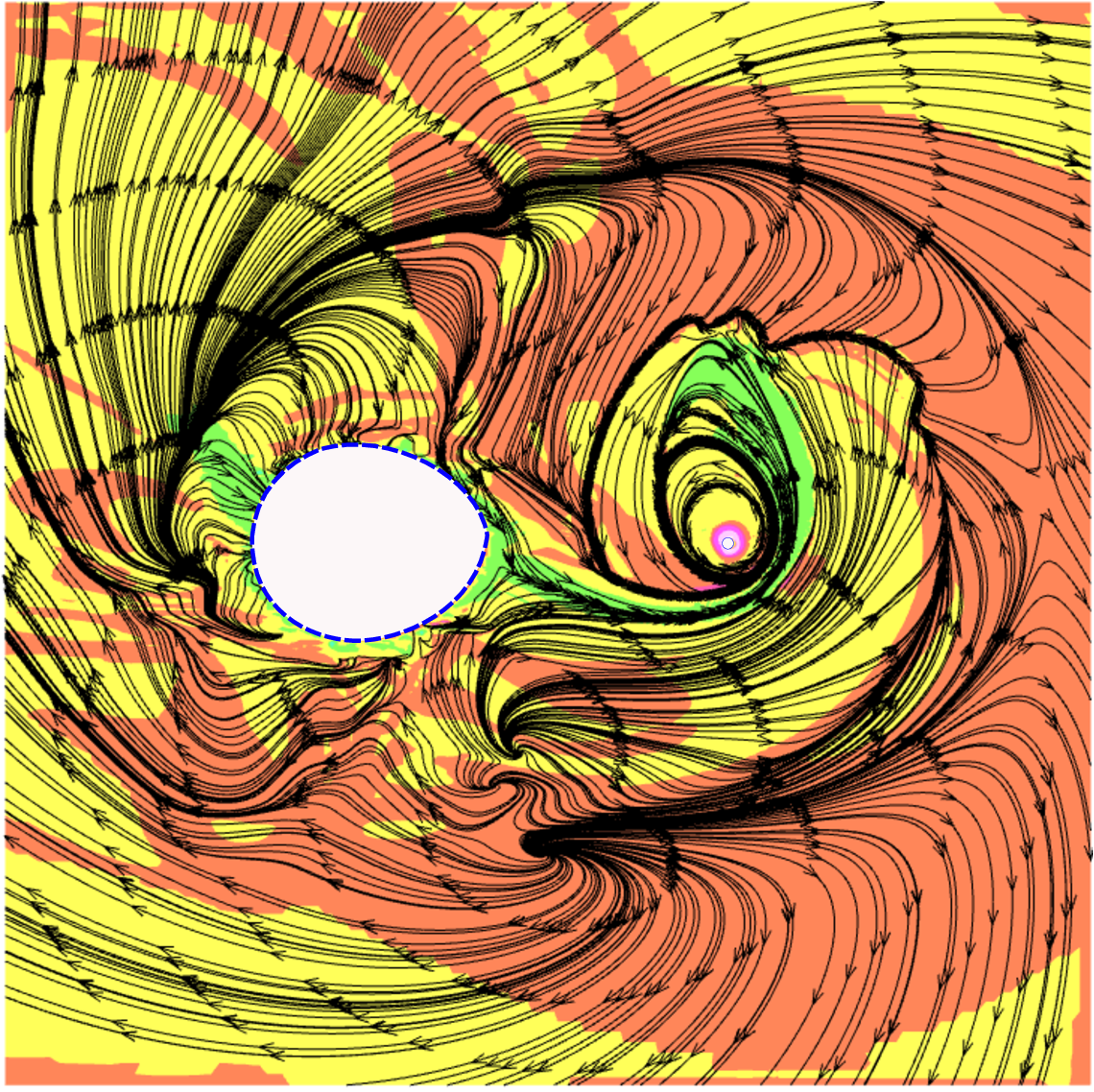} \\ $0.2 P_{\text{orb}}$}
	\end{minipage}
	\hfill
	\begin{minipage}[h]{0.55\linewidth}
	 \centering{\includegraphics[width=1.0\textwidth]{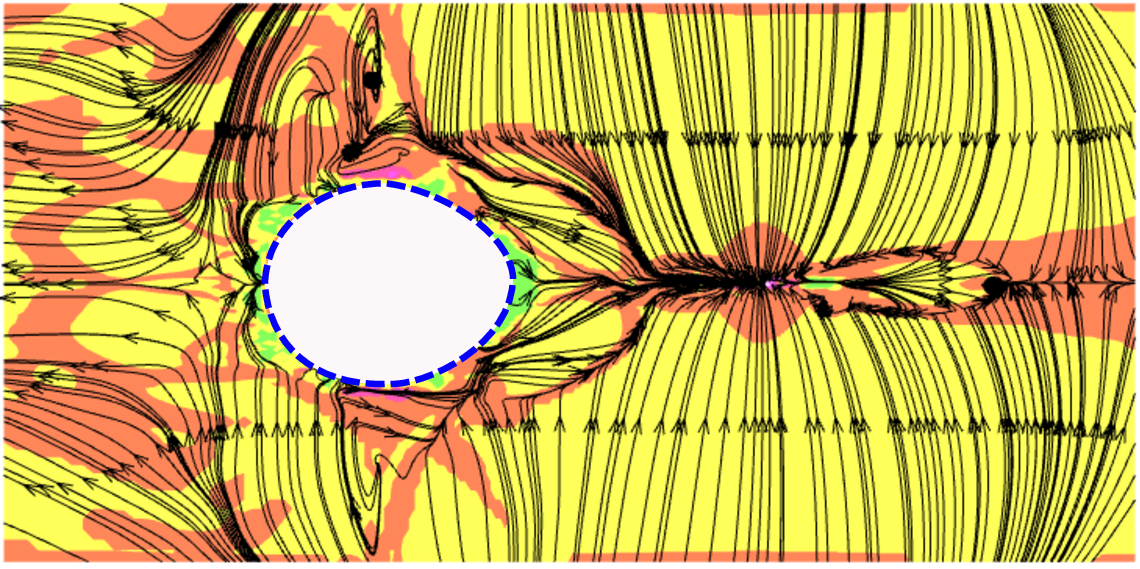} \\ $0.2 P_{\text{orb}}$}
	\end{minipage}
	\hfill
	\begin{minipage}[h]{0.45\linewidth}
	 \centering{\includegraphics[width=0.9\textwidth]{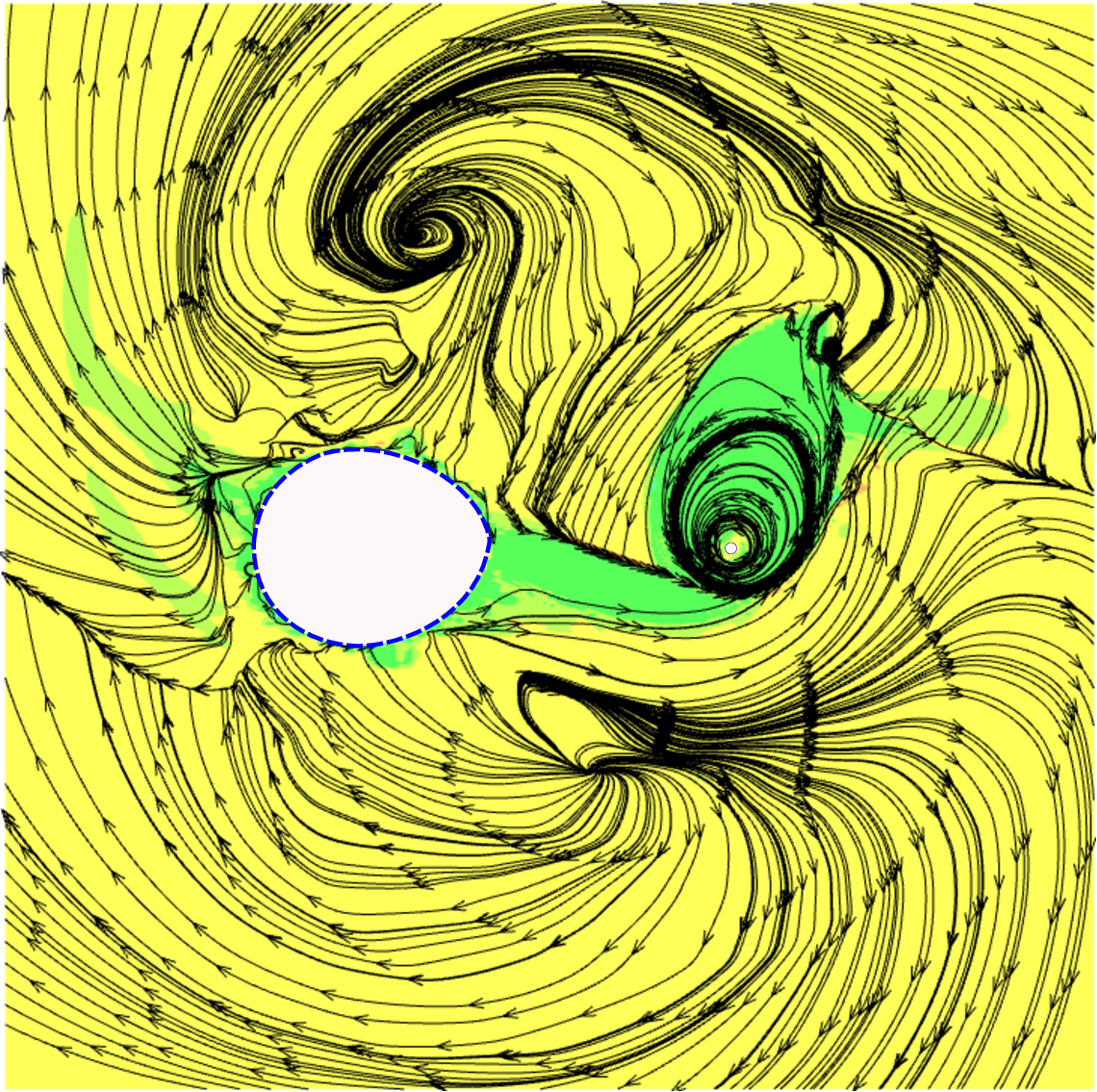} \\ $0.5 P_{\text{orb}}$}
	\end{minipage}
	\hfill
	\begin{minipage}[h]{0.50\linewidth}
	 \centering{\includegraphics[width=0.9\textwidth]{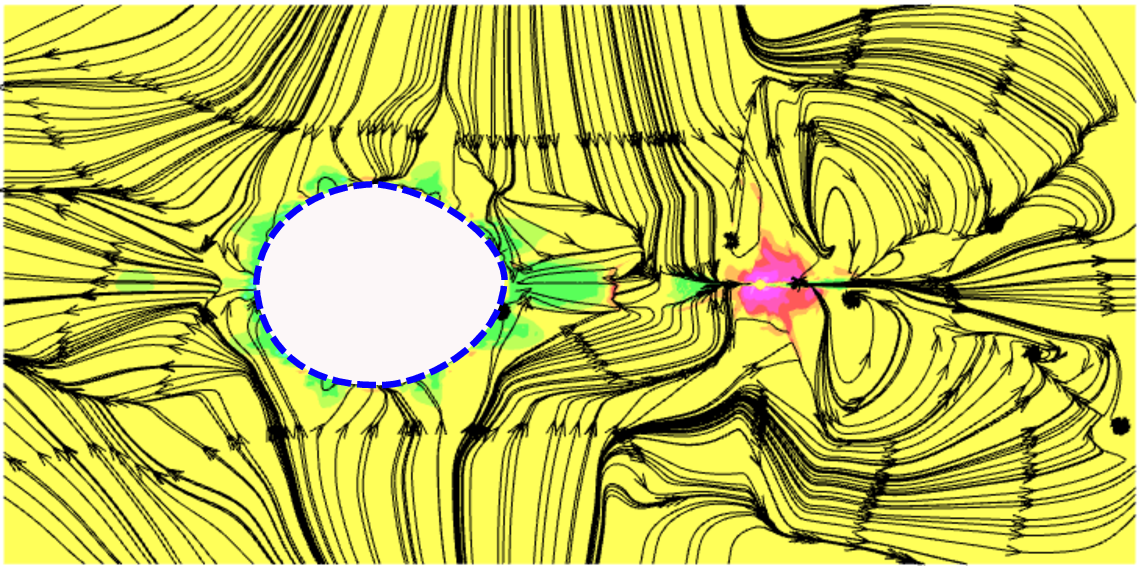} \\ $0.5 P_{\text{orb}}$}
	\end{minipage}
	\hfill
	\begin{minipage}[h]{0.45\linewidth}
	 \centering{\includegraphics[width=0.9\textwidth]{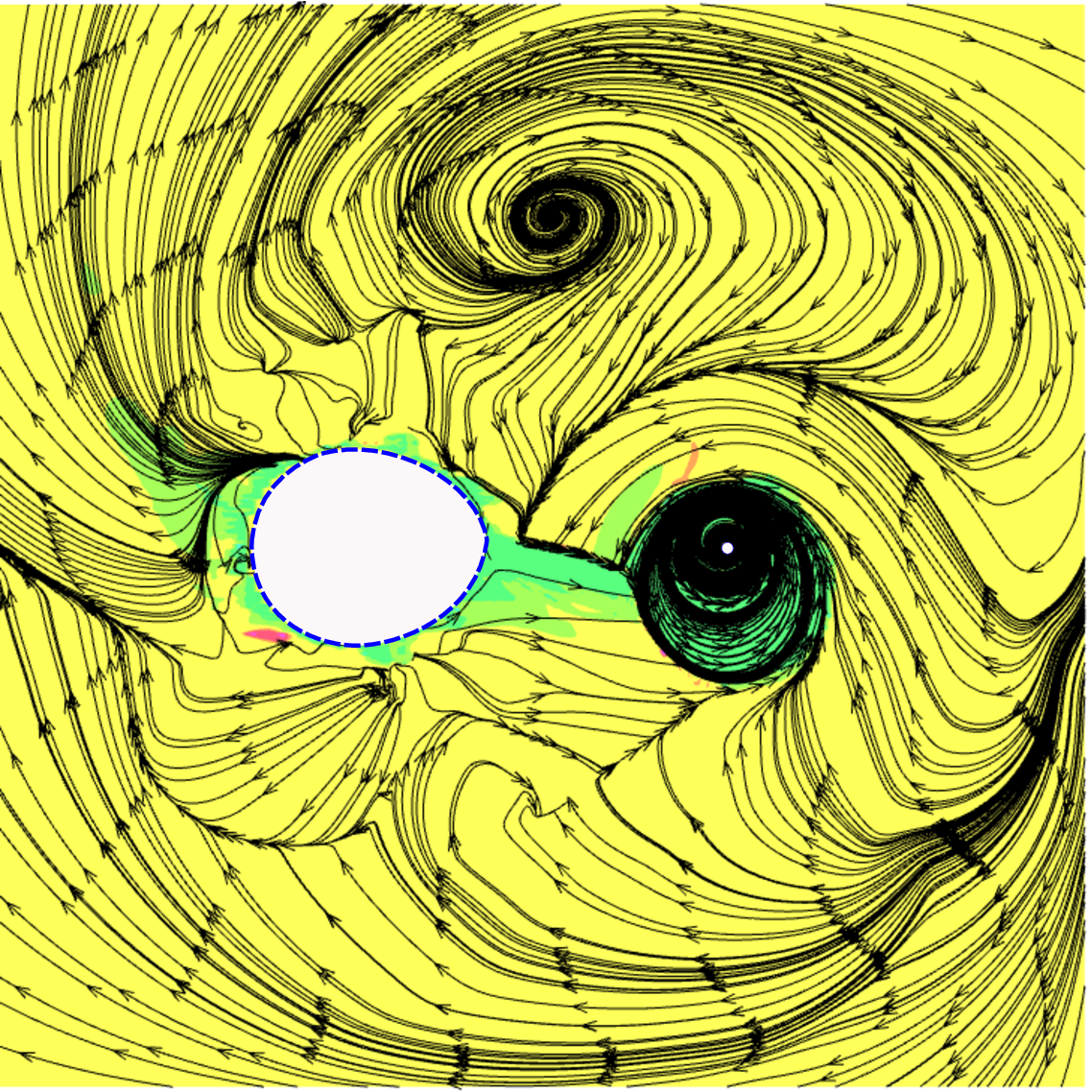} \\ $1.0 P_{\text{orb}}$}
	\end{minipage}
	\hfill
	\begin{minipage}[h]{0.50\linewidth}
	 \centering{\includegraphics[width=0.9\textwidth]{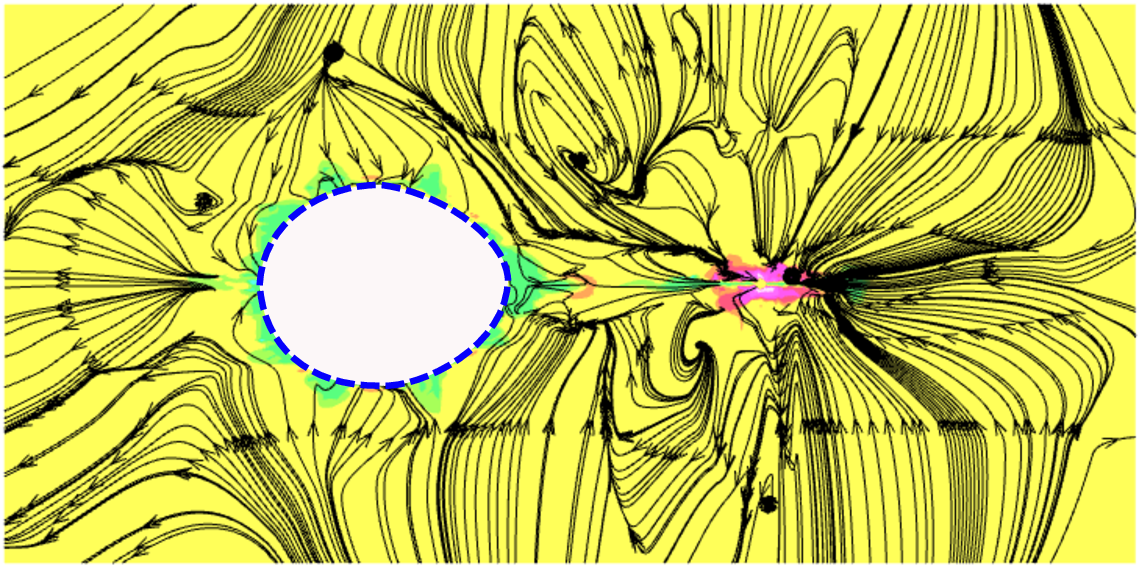} \\ $1.0 P_{\text{orb}}$}
	\end{minipage}
	\centering{\includegraphics[width=0.75\textwidth]{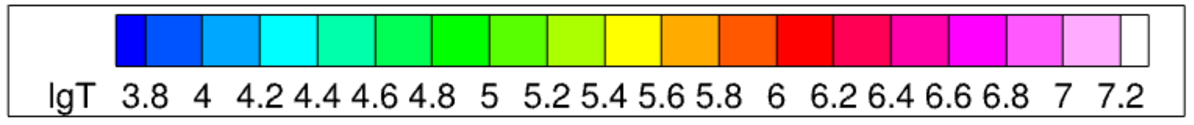}}
    \caption{Formation of a quasi-stationary state of gas flows in the binary system within the computational domain for a wind temperature $T_{\text{w}} = 3 \times 10^5$ K over the time interval $0.2 - 1.0 P_{\text{orb}}$.}
    \label{fig-5}
\end{figure}	
\begin{figure}[H]
	\begin{minipage}[h]{0.45\linewidth}
	 \centering{\includegraphics[width=0.9\textwidth]{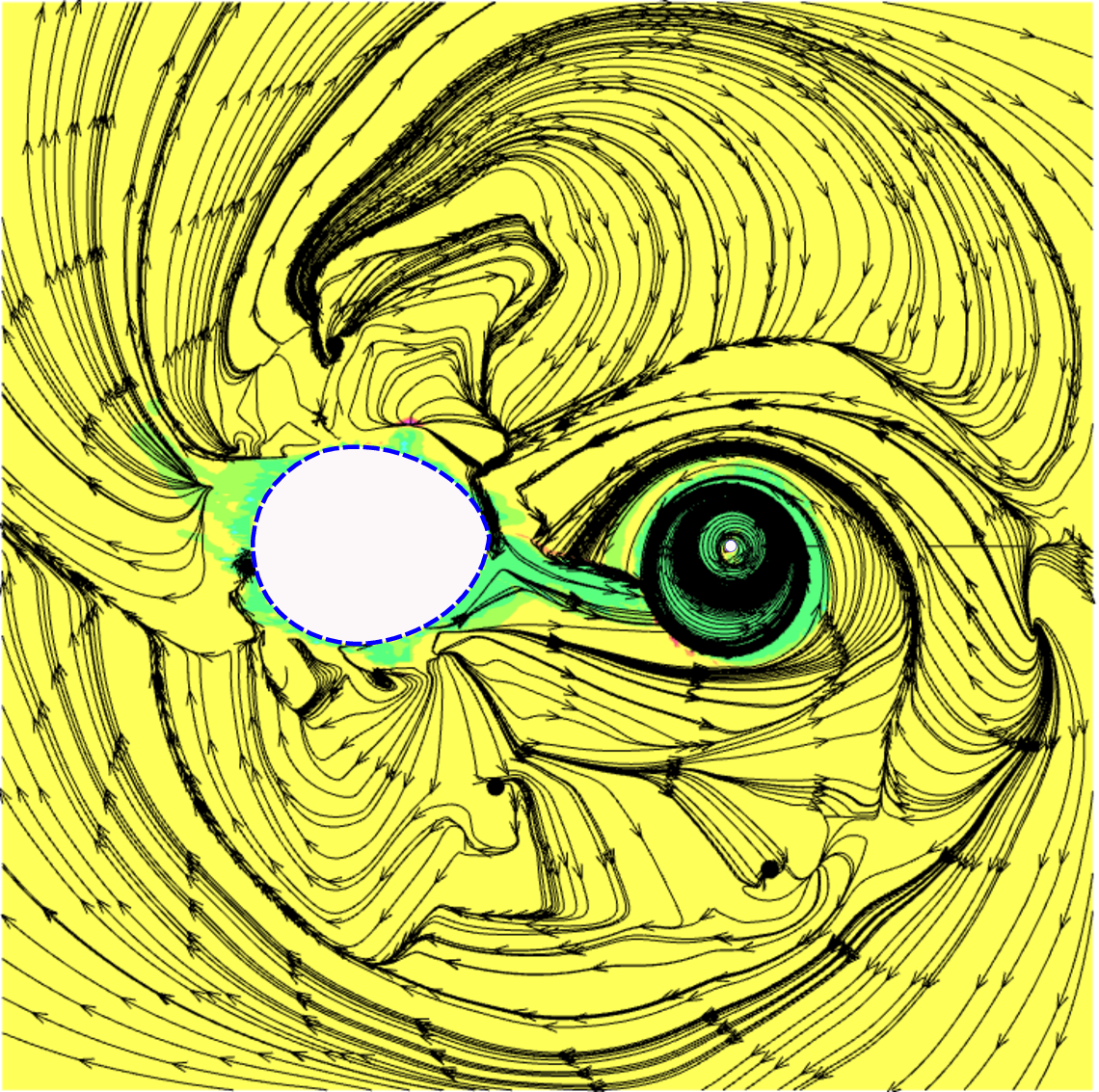} \\ $2.0 P_{\text{orb}}$}
	\end{minipage}
	\hfill
	\begin{minipage}[h]{0.50\linewidth}
	 \centering{\includegraphics[width=0.9\textwidth]{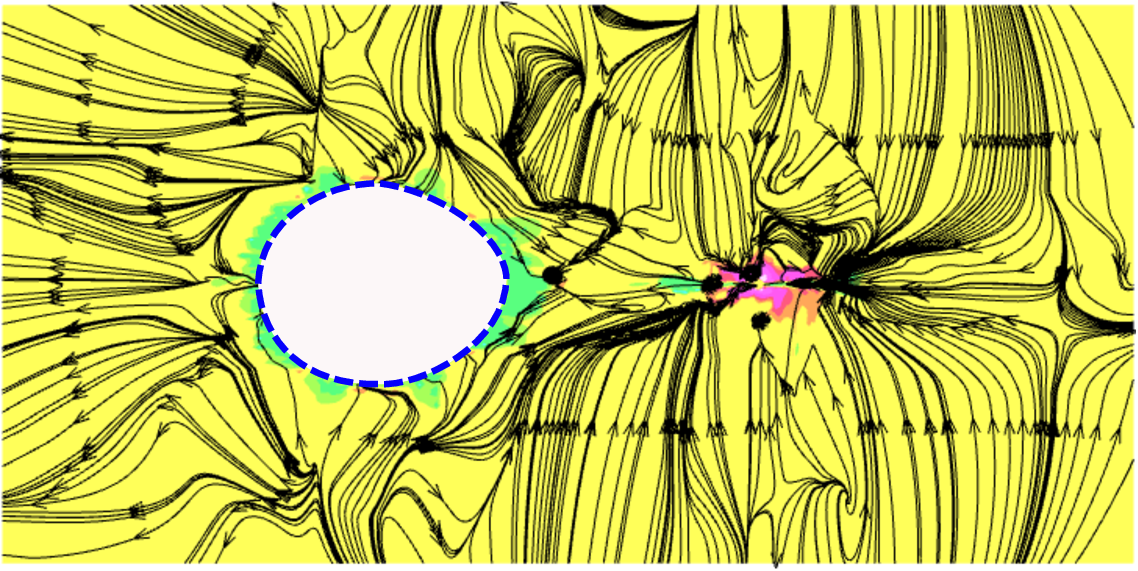} \\ $2.0 P_{\text{orb}}$}
	\end{minipage}
	\hfill
	\begin{minipage}[h]{0.45\linewidth}
	 \centering{\includegraphics[width=0.9\textwidth]{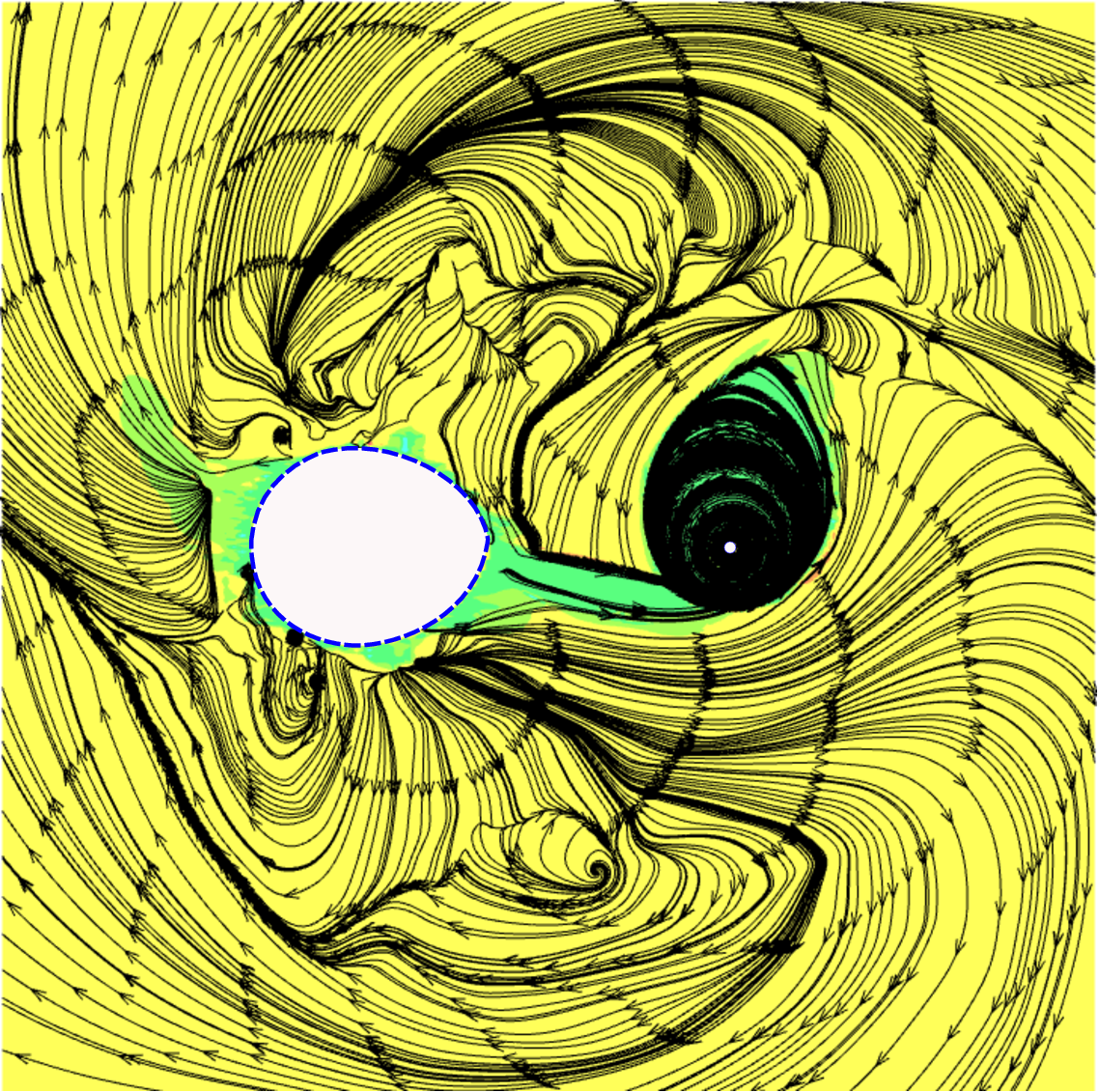} \\ $4.0 P_{\text{orb}}$}
	\end{minipage}
	\hfill
	\begin{minipage}[h]{0.50\linewidth}
	 \centering{\includegraphics[width=0.9\textwidth]{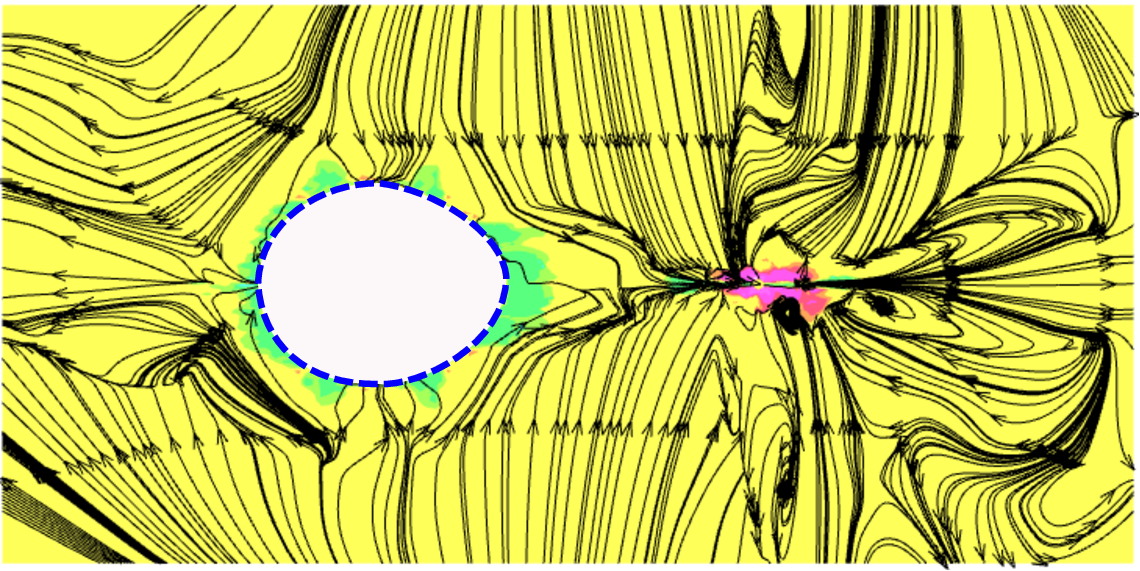} \\ $4.0 P_{\text{orb}}$}
	\end{minipage}
    \hfill
    \begin{minipage}[h]{0.45\linewidth}
	 \centering{\includegraphics[width=0.9\textwidth]{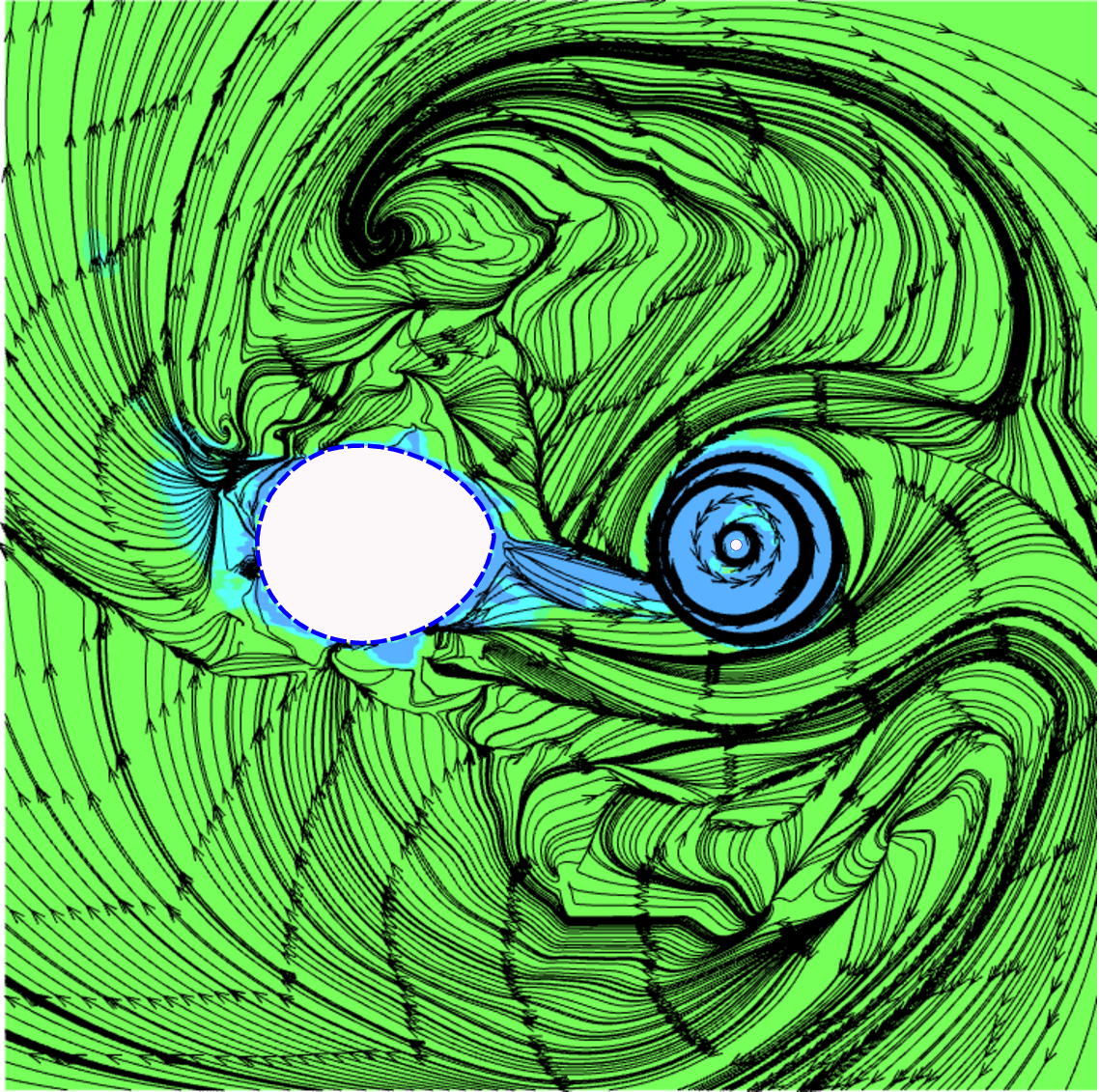} \\ $7.0 P_{\text{orb}}$}
     \end{minipage}
    \hfill
    \begin{minipage}[h]{0.50\linewidth}
	 \centering{\includegraphics[width=0.9\textwidth]{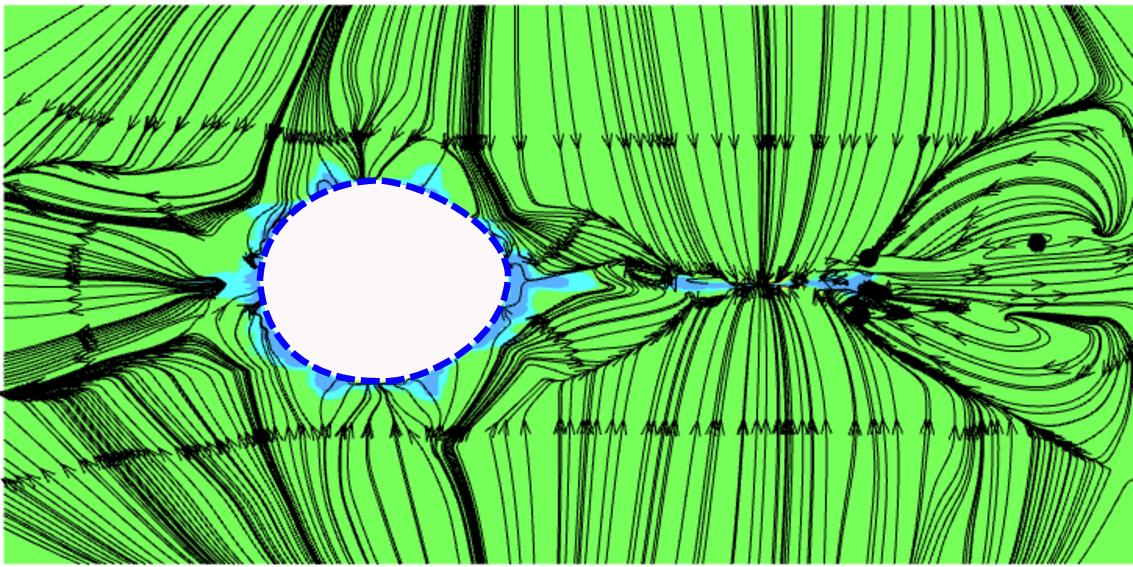} \\ $7.0 P_{\text{orb}}$}
    \end{minipage}
	\centering{\includegraphics[width=0.75\textwidth]{figs/Color_bar_105.eps}}
	\caption{Same as Fig. \ref{fig-5}, but for the time interval $2.0 - 7.0 P_{\text{orb}}$.}
	\label{fig-6}
\end{figure}
Figures \ref{fig-5} and \ref{fig-6} illustrate the process of establishing a quasi-stationary wind flow structure over seven orbital periods using Model 1 as an example. The temperature distributions shown in these figures are plotted in the $XY$ and $XZ$ planes and correspond to the most significant time phases of the process. The figures show that the donor’s wind fills the computational domain over the time $0.3 P_{\text{orb}}$.

Indeed, if we calculate the time it takes for the wind to travel from the source to the outer boundary of the computational domain using formula (Eq. \ref{eq-6}) and substituting $A = 4 R_\odot$, we obtain a minimum estimate of this time for Model 1, $t_{\text{orb}} \sim 0.25 P_{\text{orb}}$. 

The initial expansion of the wind gas from the donor’s Roche lobe is accompanied by an increase in the temperature of the circumstellar environment (according to the model’s initial conditions, the computational domain already contains gas with density $\rho_0 = 10^{-8}\rho({\rm{L}_1})$) due to the passage of a shock wave through it.

As can be seen from the figures, achieving a quasistationary flow structure in the computational domain takes about four orbital periods. Characteristic elements of the structure (the flow of matter from the donor’s Roche lobe, the accretion disk, and turbulent
vortices near the Lagrangian points $\rm{L}_4$ and $\rm{L}_5$) form
within the first orbital period. Variations in the gas flow structure at later stages are associated with changes in the wind’s motion, manifesting as emerging and disappearing regions of turbulence. In addition, during the establishment of the quasi-stationary flow pattern, the dynamics of the accretion disk formation are clearly visible: the early stages of this process are marked by significant motion of its center relative to the accretor.

For this model, we can observe the formation of an envelope around the donor’s Roche lobe, as well as a dense and broad stream emerging from the vicinity of the Lagrangian points $\rm{L}_1$ and $\rm{L}_3$. Due to the high density of the flow, its temperature drops tenfold compared to the surrounding environment. With the wind speed adopted in this model, this envelope subsequently prevents wind from escaping from areas of the Roche lobe other than these specified points. Thus, a
stable source of accretion disk feeding is formed.

It is particularly noteworthy to highlight the trajectory of matter in the accretion disk --- a spiral with variable step, where each branch corresponds to a specific value of angular momentum. As shown in \cite{Tutukov2016}, two types of structures can form in an accretion disk. Disk matter moves along a spiral from the outer edge to the
inner region if this motion is in the same direction as the orbital rotation of the perturbing body, which in our case is the donor. If these motions are in opposite directions, a ring structure appears. A likely cause of the ring structure is the resonance of the disk matter’s orbital motion relative to the donor’s orbital motion.

Such ring structures are common in the universe. They typically arise from interactions between extended disk-like objects of different natures: galaxies, stars with gas disks, and planets with dust disks
\cite{Brahic1974}. These structures have been studied many times.
For example, \cite{Chatterjee1981} examined the formation of ring
structures during collisions between disk and elliptical galaxies. The influence of close passages of other galaxies on the morphology of a galactic disk was studied in \cite{Tutukov2016}. Numerical modeling showed that a passage in the same direction as the disk’s rotation creates spiral density waves in it, while a passage in the opposite
direction leads to the formation of gas–star rings in the disk. A similar phenomenon is observed in the motion of trans-Neptunian objects and Neptune’s orbital rotation with a 2 : 3 resonance relationship \cite{Ipatov2000}.

\begin{figure}[H]
	\centering	
	\begin{minipage}[h]{0.48\linewidth}
		\centering{\includegraphics[width=0.9\textwidth]{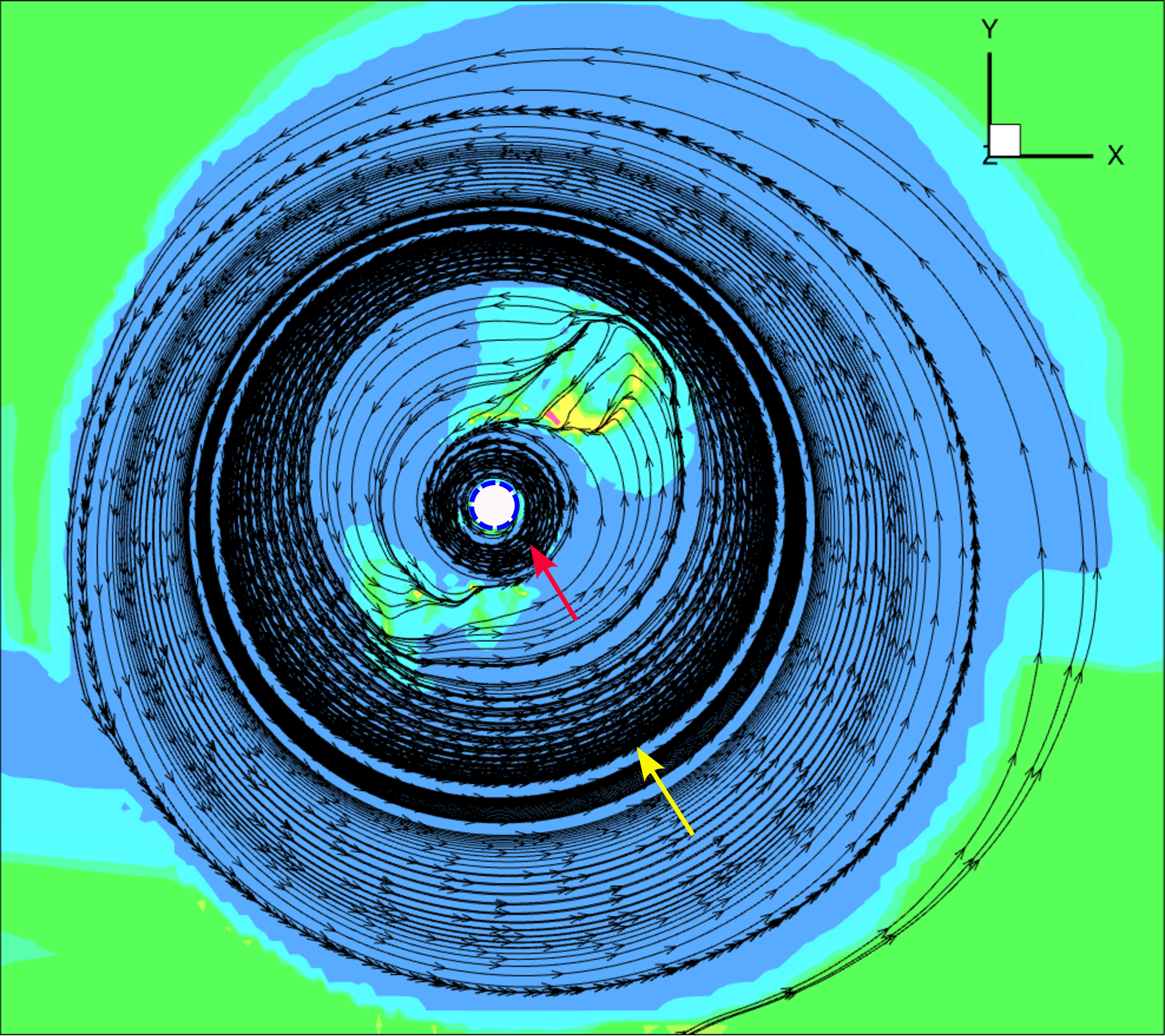}} \\ $z=0$
	\end{minipage}
	\hfill
	\begin{minipage}[h]{0.48\linewidth}
		\centering{\includegraphics[width=0.9\textwidth]{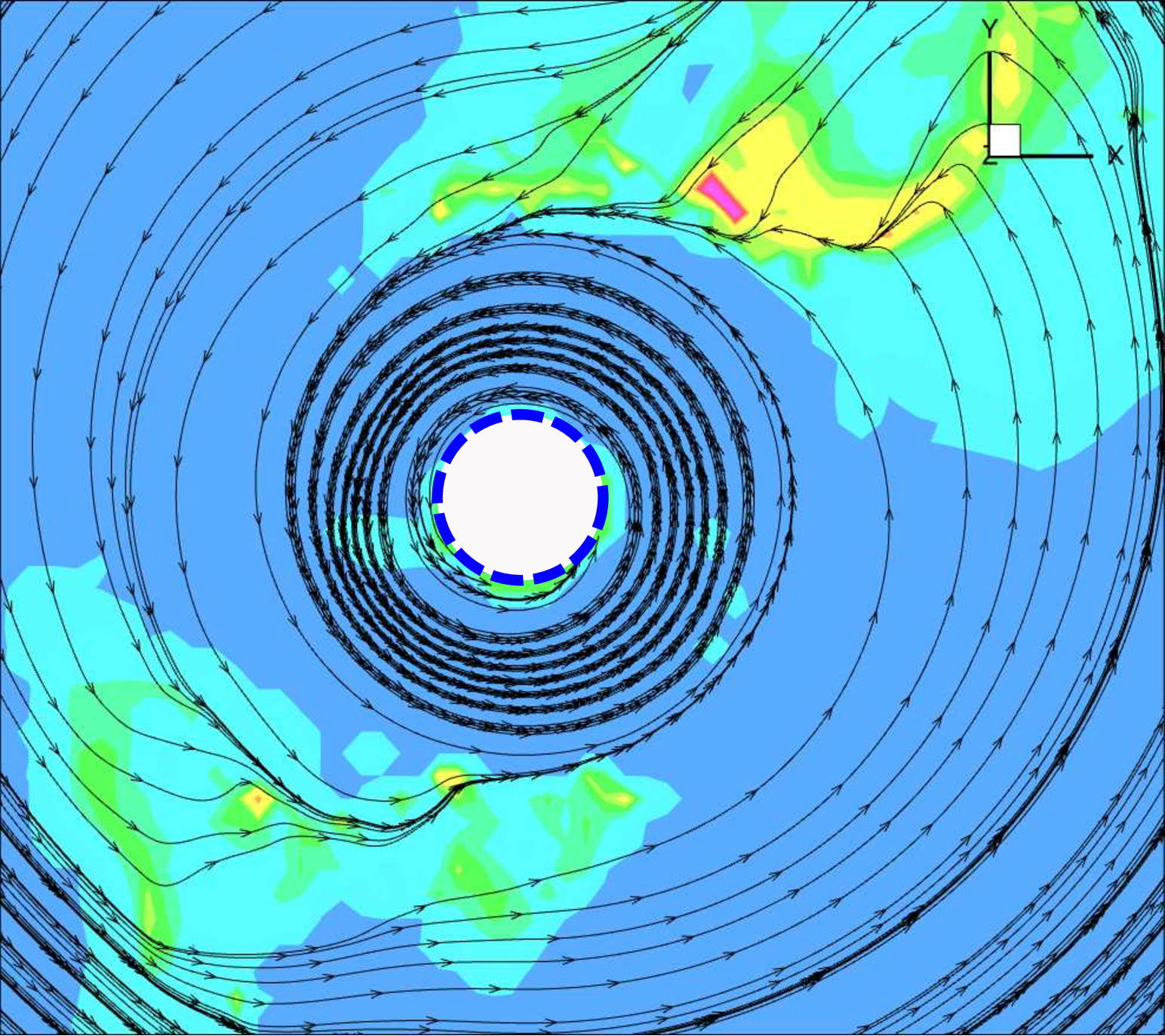}} \\ $z=0$ 
	\end{minipage}
	\hfill
	\begin{minipage}[h]{0.48\linewidth}
		\vspace{1cm}
		\centering{\includegraphics[width=0.9\textwidth]{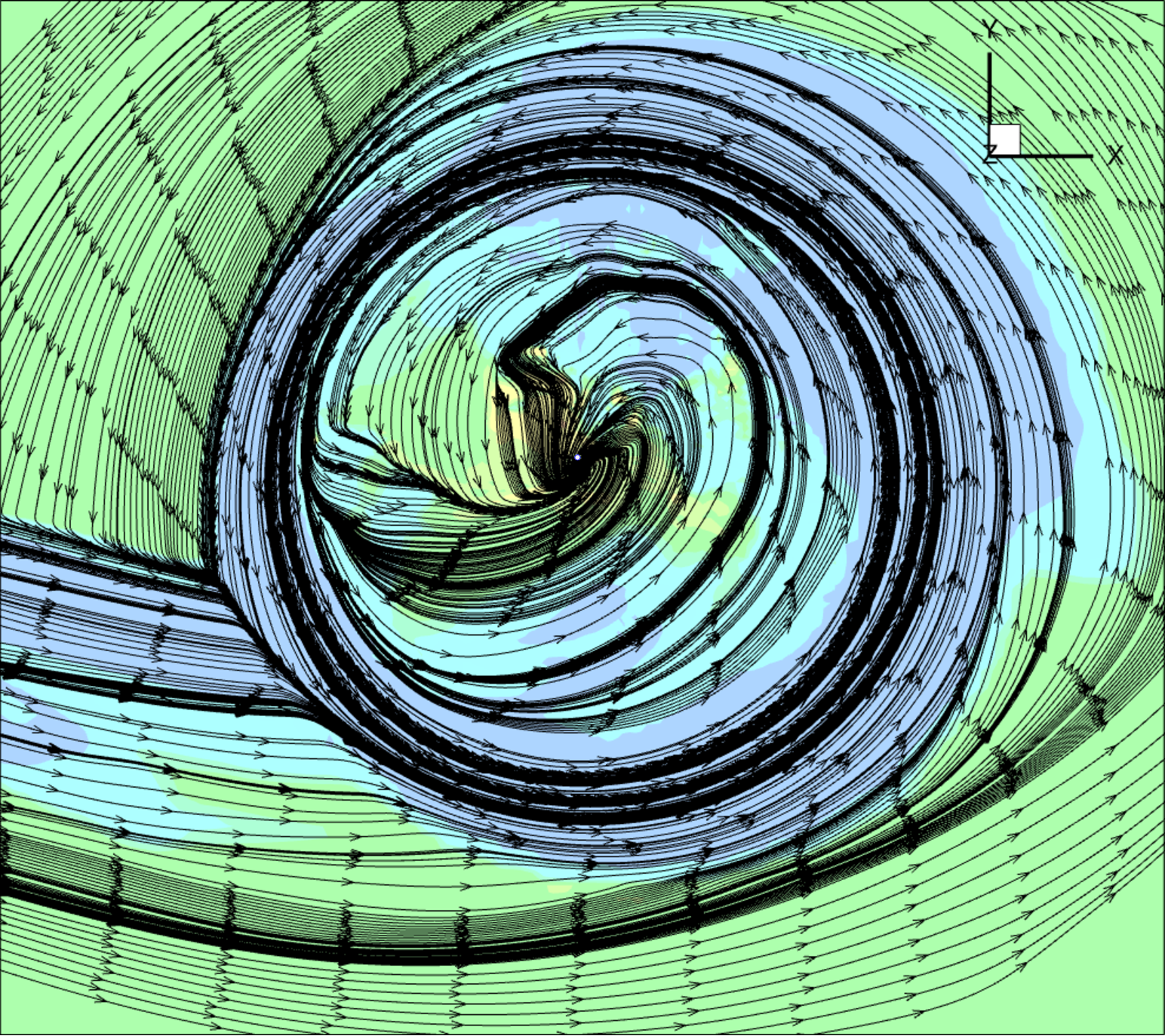}} \\ $z=-0.01$ 
	\end{minipage}
	\hfill
	\begin{minipage}[h]{0.48\linewidth}
		\vspace{1cm}
		\centering{\includegraphics[width=0.9\textwidth]{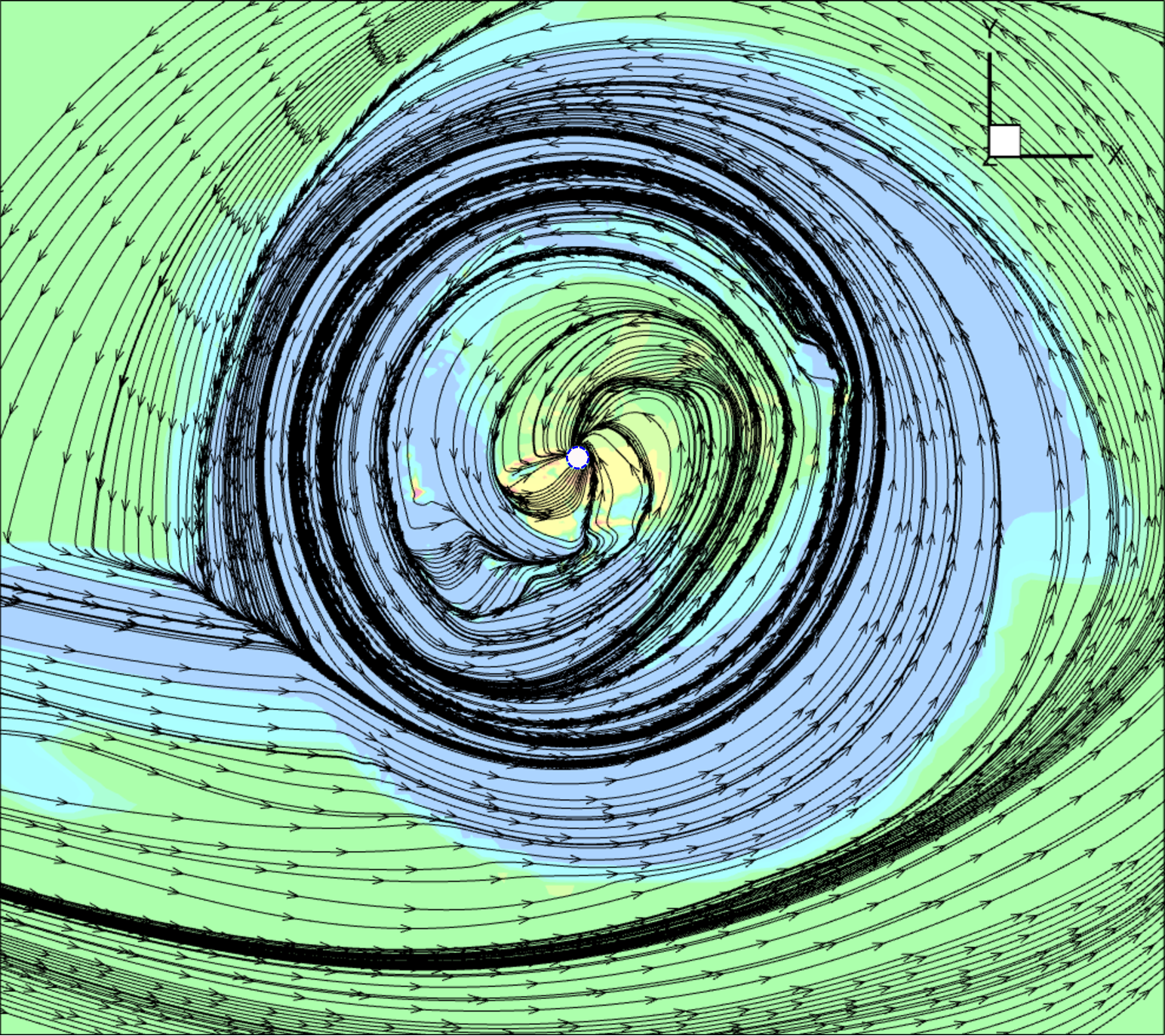}} \\ $z=+0.01$
	\end{minipage}
	\centering{\includegraphics[width=0.75\textwidth]{figs/Color_bar_105.eps}}
	\caption{Gas flow structure in the accretion disk for a wind temperature $T_{\text{w}} = 3 \times 10^5$ K at time of $8 P_{\text{orb}}$. The \textit{upper} panels show the overall flow pattern (left) and the flow structure near the accretor (right) in the orbital plane ($z = 0$). The \textit{lower} panels show the flow pattern in section planes shifted downward $z = -0.01$ (left) and upward $z = +0.01$ (right) relative to the orbital plane.}
	\label{fig-7}
\end{figure}

In Fig. \ref{fig-7}, the motion of donor matter in the accretion disk is shown for a wind gas temperature $T_{\text{w}} = 3 \times 10^5$ K at time $8 P_{\text{orb}}$. For the cross-section coinciding with the orbital plane ($z = 0$), the ring structure of the gas flow is clearly visible. At the same time, from the outer to the inner edge of the disk, a density gradient is observed, indicating the presence of
density waves. Two zones of highest density can be distinguished here: the middle part of the disk (marked with a yellow arrow) and the region near the accretor (marked with a red arrow). Between these zones, the gas structure has a spiral shape.

The top right panel of the figure shows a magnified view of the disk area near the accretor. This panel clearly shows two channels through which matter from the disk flows into the accretor’s circumstellar zone. In these channels, there is a noticeable increase in gas
temperature --- by about an order of magnitude --- caused by the formation of shock waves as the spiraling material collides with the channel gas. 

The lower panels of the figure show the flow structure in cross-sections offset along the $z$ axis relative to the orbital plane.
\begin{figure}[H]
	\centering
    \begin{minipage}[h]{0.48\linewidth}
	\vspace{1cm}
	\centering{\includegraphics[width=1.0\textwidth]{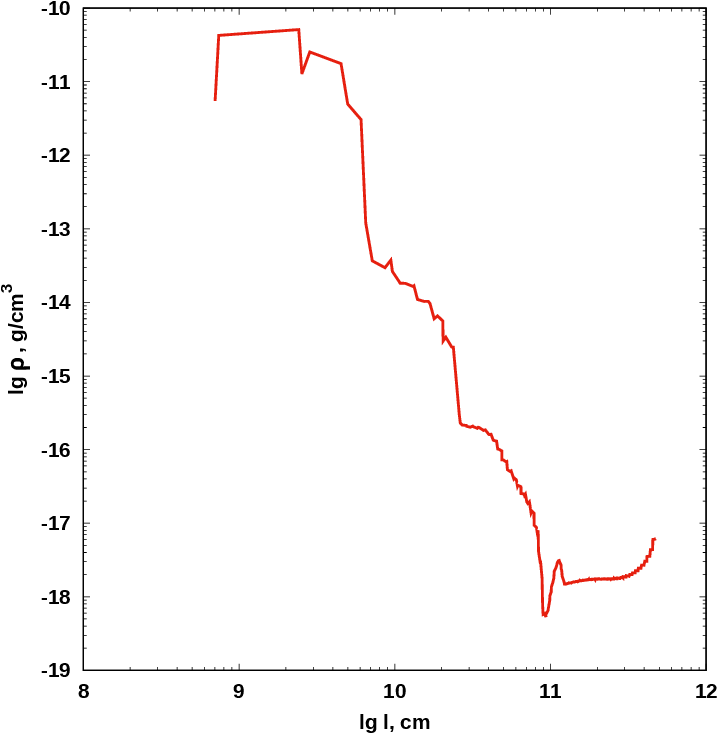}} \\ $z=0$ 
    \end{minipage}
    \hfill
    \begin{minipage}[h]{0.48\linewidth}
	\vspace{1cm}
	\centering{\includegraphics[width=1.0\textwidth]{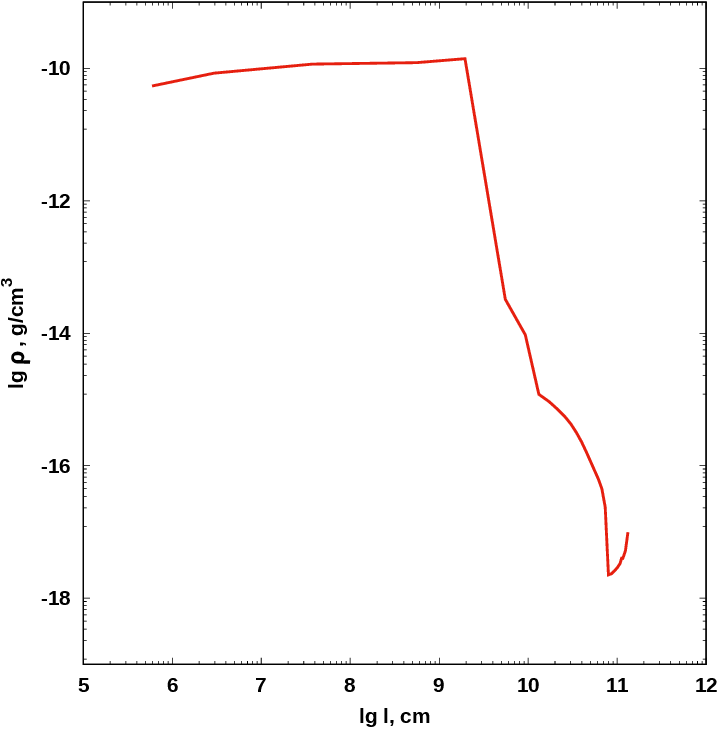}} \\ $z=+0.01$ 
    \end{minipage}
    \hfill
    \begin{minipage}[h]{0.48\linewidth}
	\vspace{1cm}
	\centering{\includegraphics[width=1.0\textwidth]{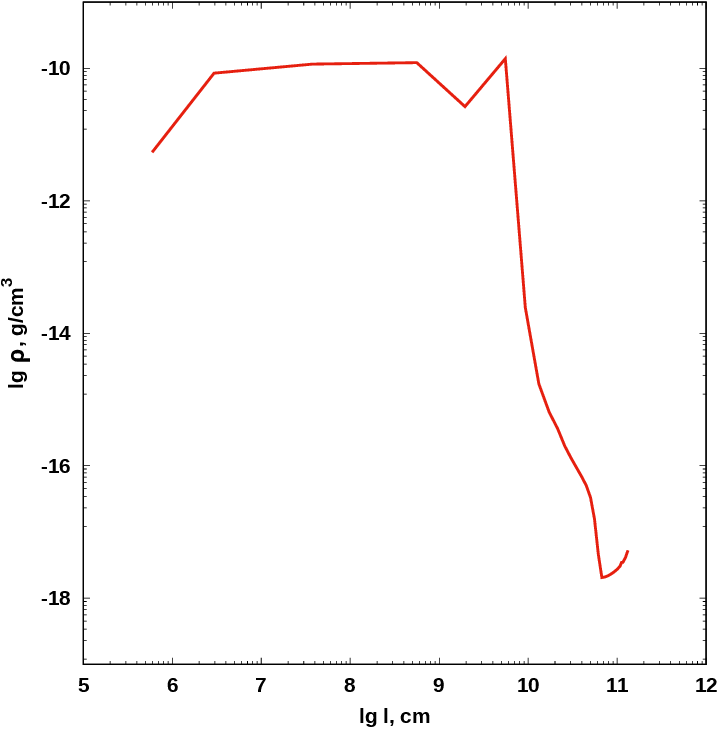}} \\ $z=-0.01$ 
    \end{minipage}
\caption{Radial density variation in the accretion disk for a wind temperature $T_{\text{w}} = 3 \times 10^5$ K at time of $8 P_{\text{orb}}$ in parallel planes shifted relative to the orbital plane along the $z$ axis by the indicated amount.}
\label{fig-8}
\end{figure}

The offset along the $z$ axis was chosen so that these sections are tangent to the surface of the white dwarf at its geographic poles. Recall that the unit of length in the model is the intercomponent distance $A$. Compared to the orbital plane, the spiral trajectory of gas motion is clearly expressed here. These sections, along with the section $z = 0$, allow us to understand the three-dimensional accretion picture of the flow onto the white dwarf. The overall gas flow from the vicinity of the Lagrange point $\rm{L}_1$ near the
accretor splits into several arms: its central part forms a thin accretion disk in the orbital plane, while the peripheral flows form a disk halo, which is a toroidal structure with an internal spiral pattern in the polar regions of the white dwarf, through which the main
accretion of matter occurs. 

Such a distribution of flows in the $XZ$ plane results in shockless interaction with the accretion disk structures and prevents the appearance of a pronounced <<hot line>>.

Figure \ref{fig-8} shows the radial density distribution of the
accretion disk from the accretor’s surface to its outer edge in the three previously indicated cross-sectional planes. As the line along which this distribution is considered, we chose the positive direction of the x axis. The zero value on the abscissa in the figure corresponds to the coordinate origin (the accretor’s center), and the density plot starts from the white dwarf’s surface, which corresponds to the point $8 \times 10^8$ cm ($x = 0.01$) for the equatorial plane ($z = 0$) and $5 \times 10^5$ cm ($x = 0.002$) for the planes passing through the accretor’s geographic poles. The latter value differs from zero because the minimum grid step is $0.2 R_{\text{a}}$.

Analysis of the figure shows that the density variation from the disk center to its outer edge in the orbital plane features a flat segment at the beginning of the plot, corresponding to the circumstellar region and marked with a red arrow in Fig. \ref{fig-7}. This is the densest part of the disk with an almost constant density. The ring of increased density (marked with a yellow arrow in Fig. \ref{fig-7}) is located at a distance of about one solar radius from the accretor’s surface ($7 \times 10^{10}$ cm). This region exhibits a noticeable, order-of-magnitude drop in density with the presence of resonant phenomena towards the disk’s outer edge. Since the plotted region
along the abscissa extends to the outer boundary of the computational domain, corresponding to the point $x = 1$, the figure shows the further behavior of the density up to this point. The density decrease is nearly linear almost to the periphery of the computational domain, rather than being limited only to the outer edge of the accretion disk located at the distance $\sim 2 R_{\odot}$.

The plots corresponding to the offsets $z = \pm 0.01$, show a similar pattern of density variation. Here, the zone of constant density extends from the white dwarf’s pole to the mark $1 R_\odot$. It corresponds to the area of spiral accretion of wind material onto the star’s surface. At greater distances from this zone, the density drops sharply and shows no significant parameter fluctuations. Thus, resonant phenomena develop only in the thin disk.

\begin{figure}[H]
  \centering
  \includegraphics[width=0.85\textwidth]{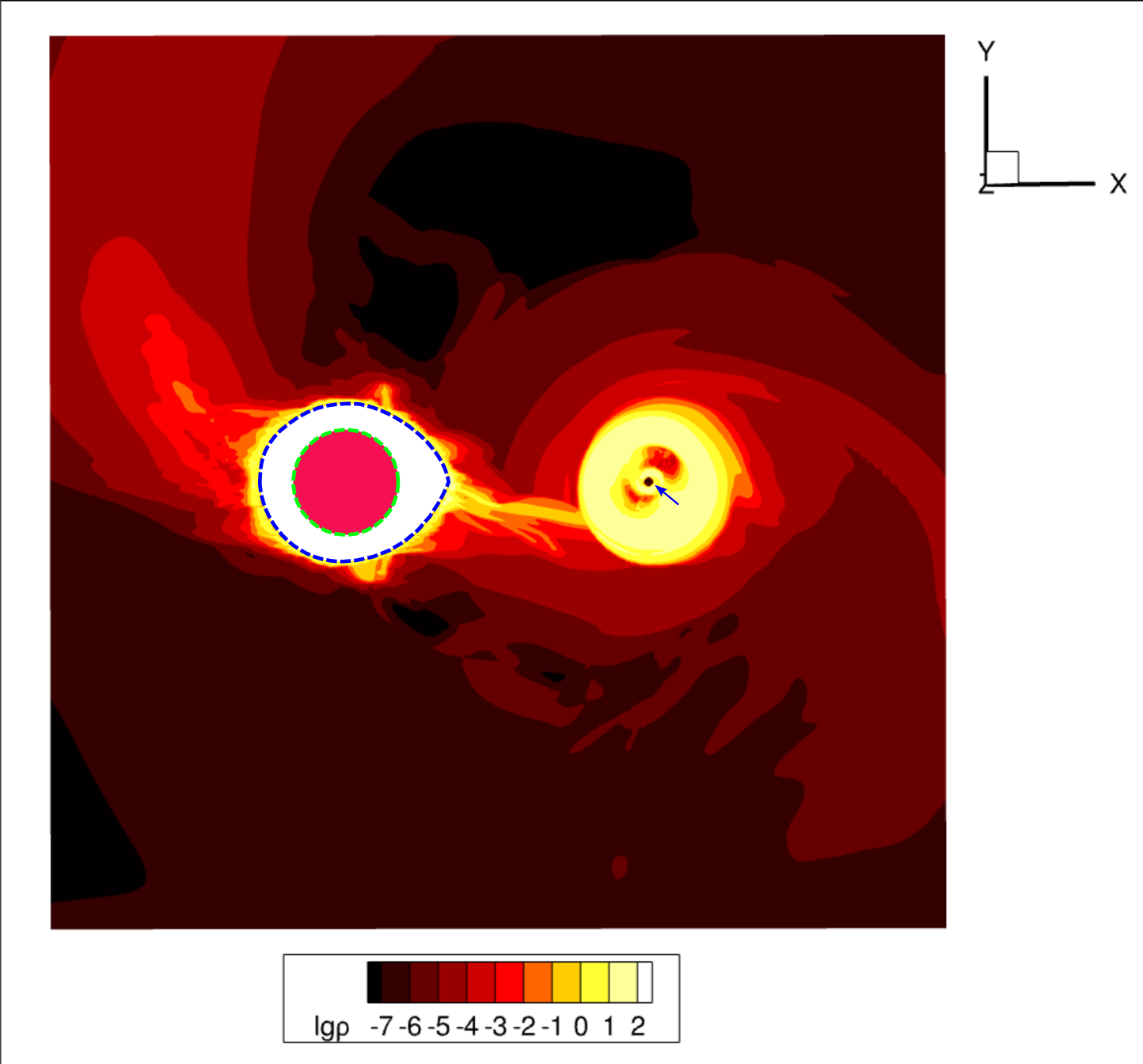}
  \includegraphics[width=0.85\textwidth]{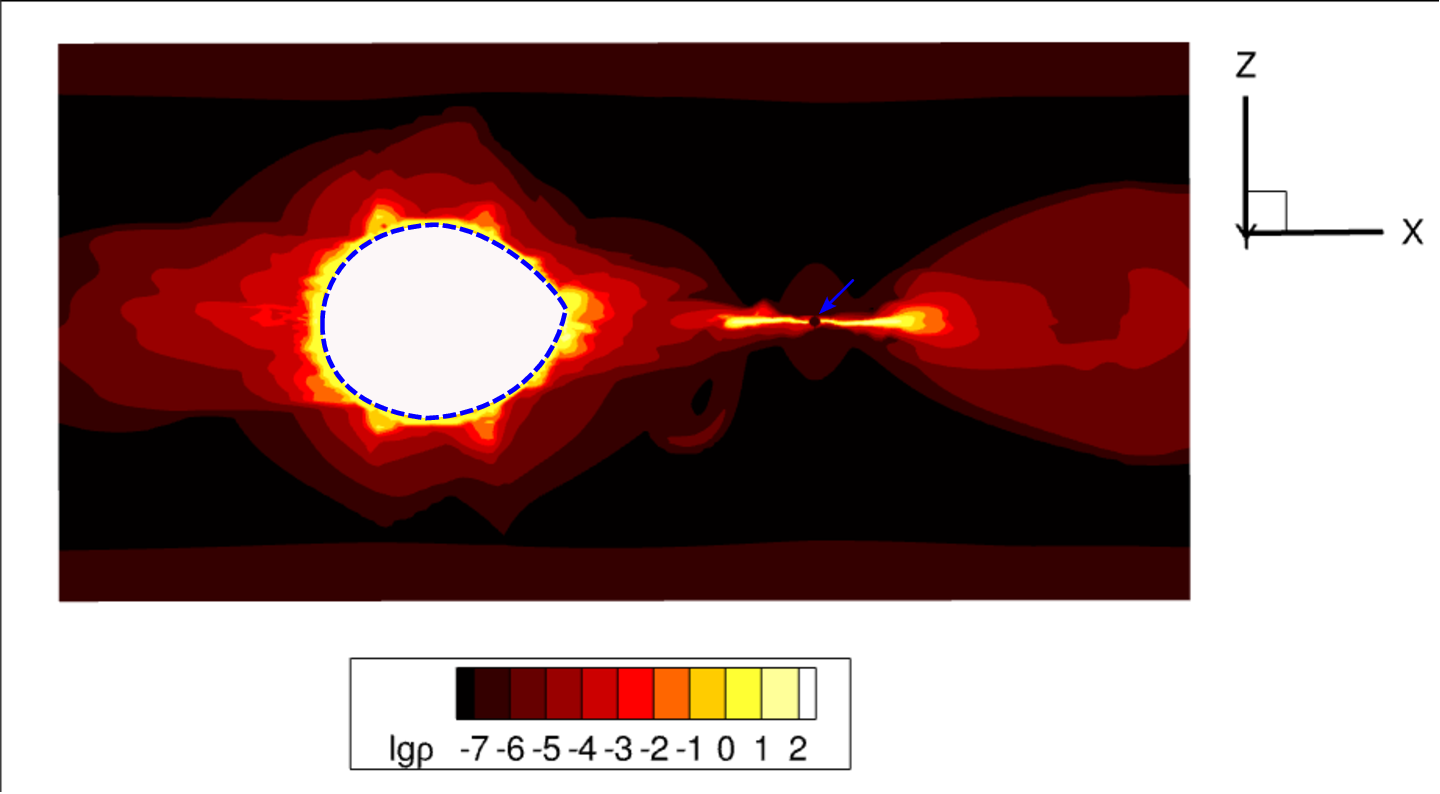}
  \caption{Numerical simulation results for a donor wind temperature $T_{\text{d}} = 3 \times 10^5$ K at time of $8 P_{\text{orb}}$ (with initial wind velocity $v_{\text{w}} = 90$ km/s). Shown is the density distribution in units of $\rho(\rm{L}_1)$. The upper panel presents a view of the system in the orbital plane $XY$, the lower panel --- in the frontal plane $XZ$. The donor’s Roche lobe is marked by a blue dashed line; the accretor’s position by a blue arrow. The red highlighted area inside the Roche lobe corresponds to the donor not filling it.}
  \label{fig-9}
\end{figure}

\begin{figure}[H]
	\centering
	\includegraphics[width=0.9\textwidth]{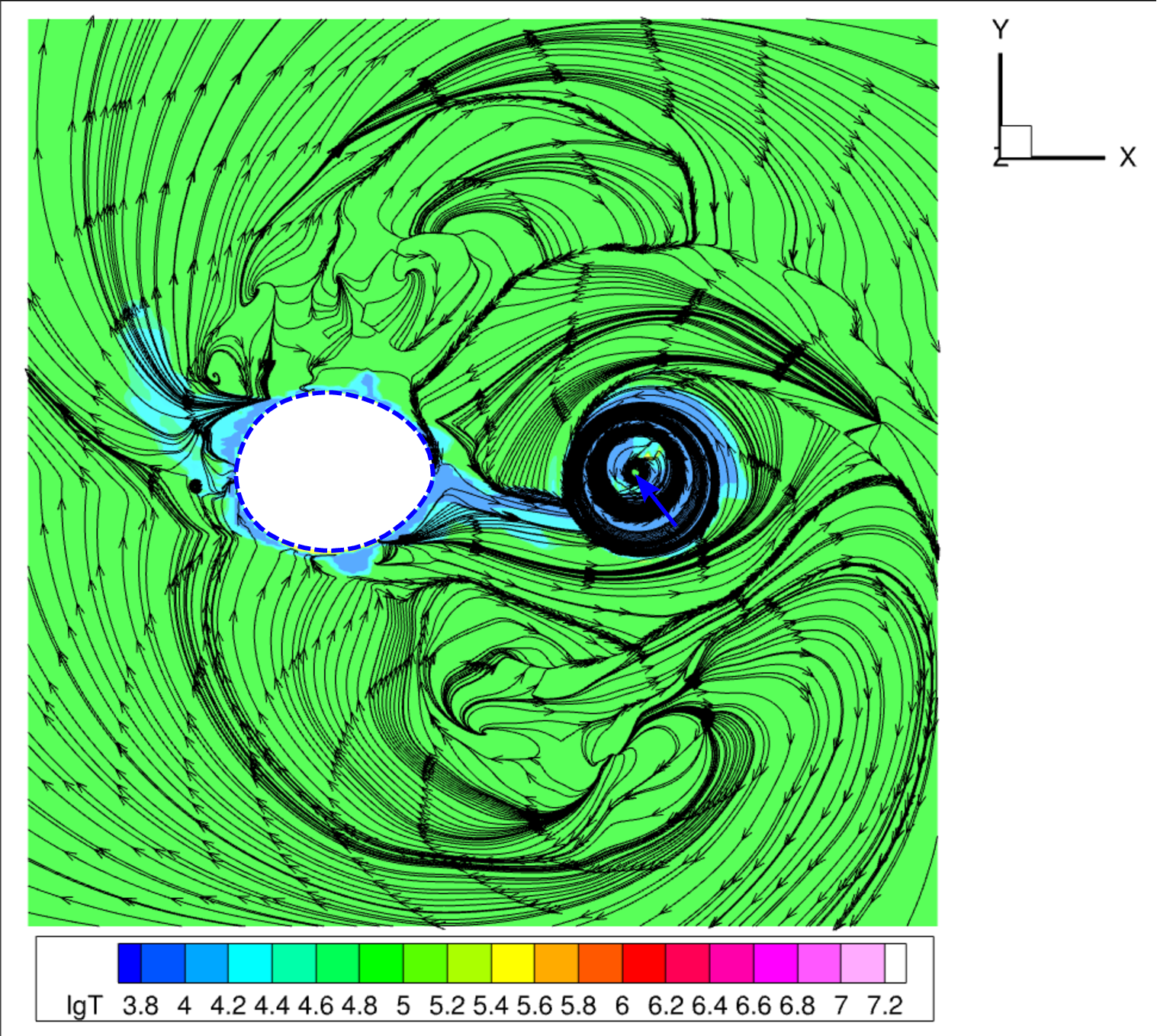}
	\includegraphics[width=0.9\textwidth]{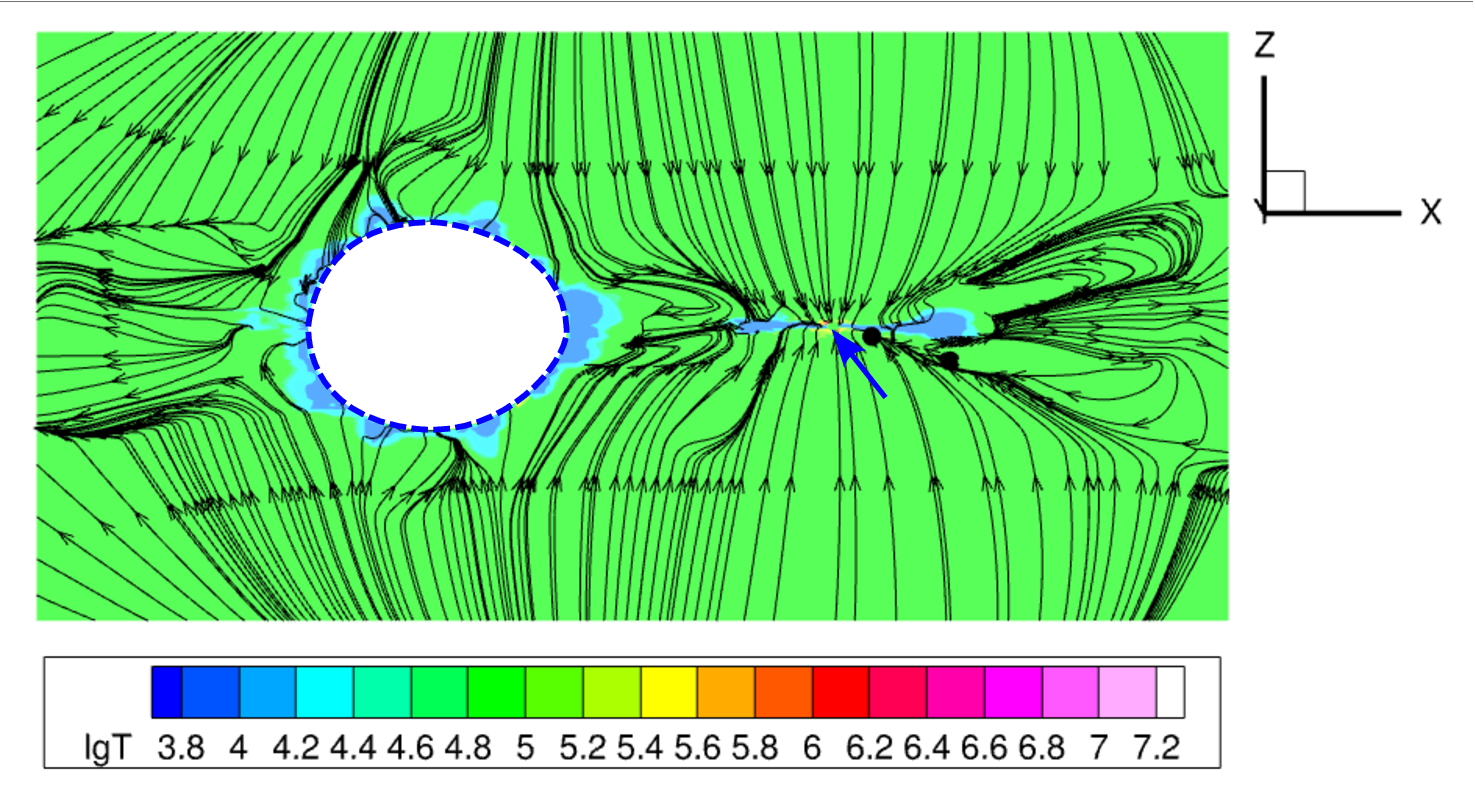}
	\caption{Numerical simulation results for a donor wind temperature $T_{\text{d}} = 3 \times 10^5$ K. Shown are the temperature distribution and streamlines of the donor wind material. The upper panel presents a view in the orbital plane $XY$, the lower panel --- in the plane $XZ$. The donor’s Roche lobe is marked by a blue dashed line; the accretor’s position by a blue arrow.}
	\label{fig-10}
\end{figure}

\begin{figure}[H]
	\centering
	\includegraphics[width=0.95\textwidth]{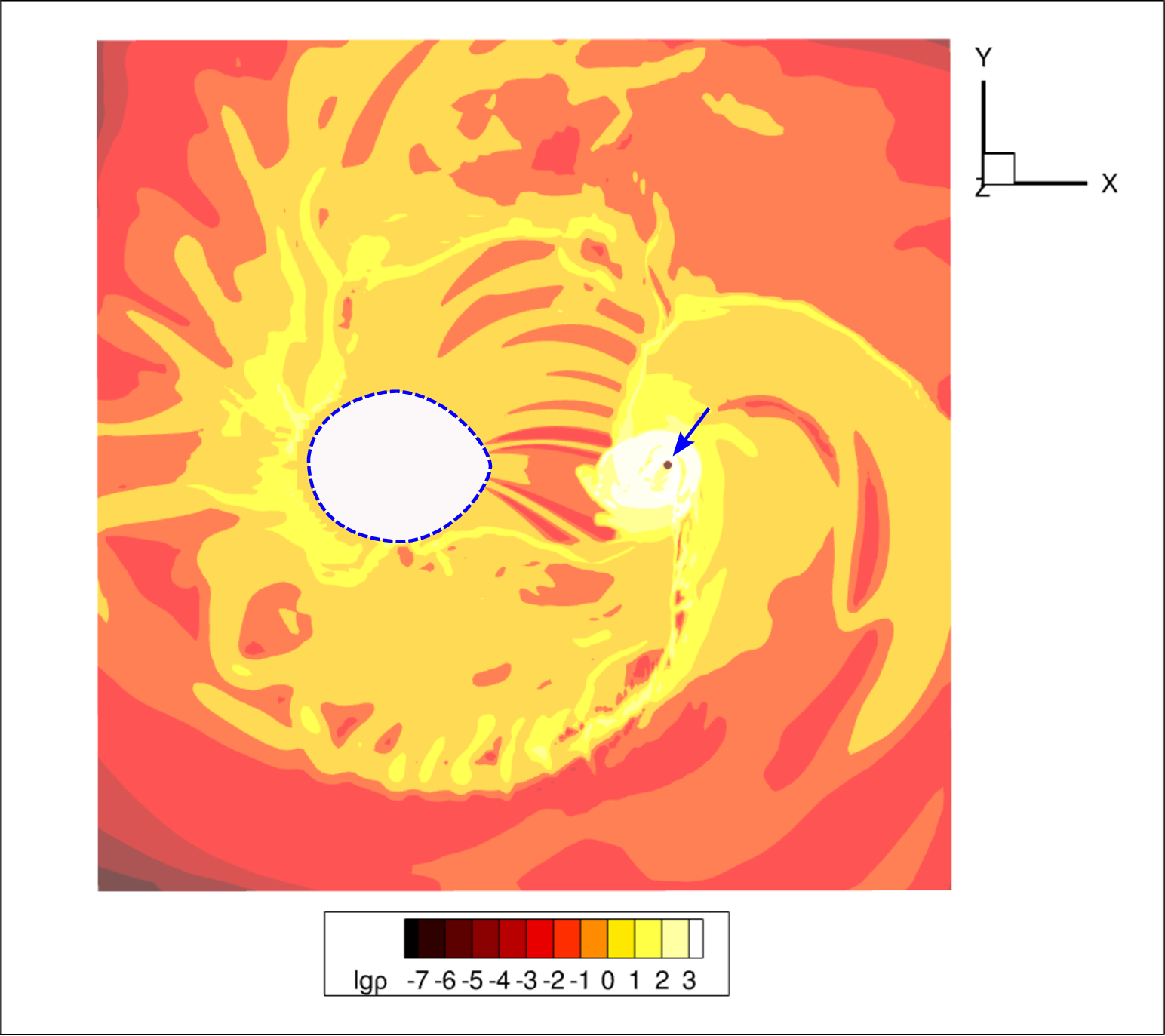}
	\includegraphics[width=0.95\textwidth]{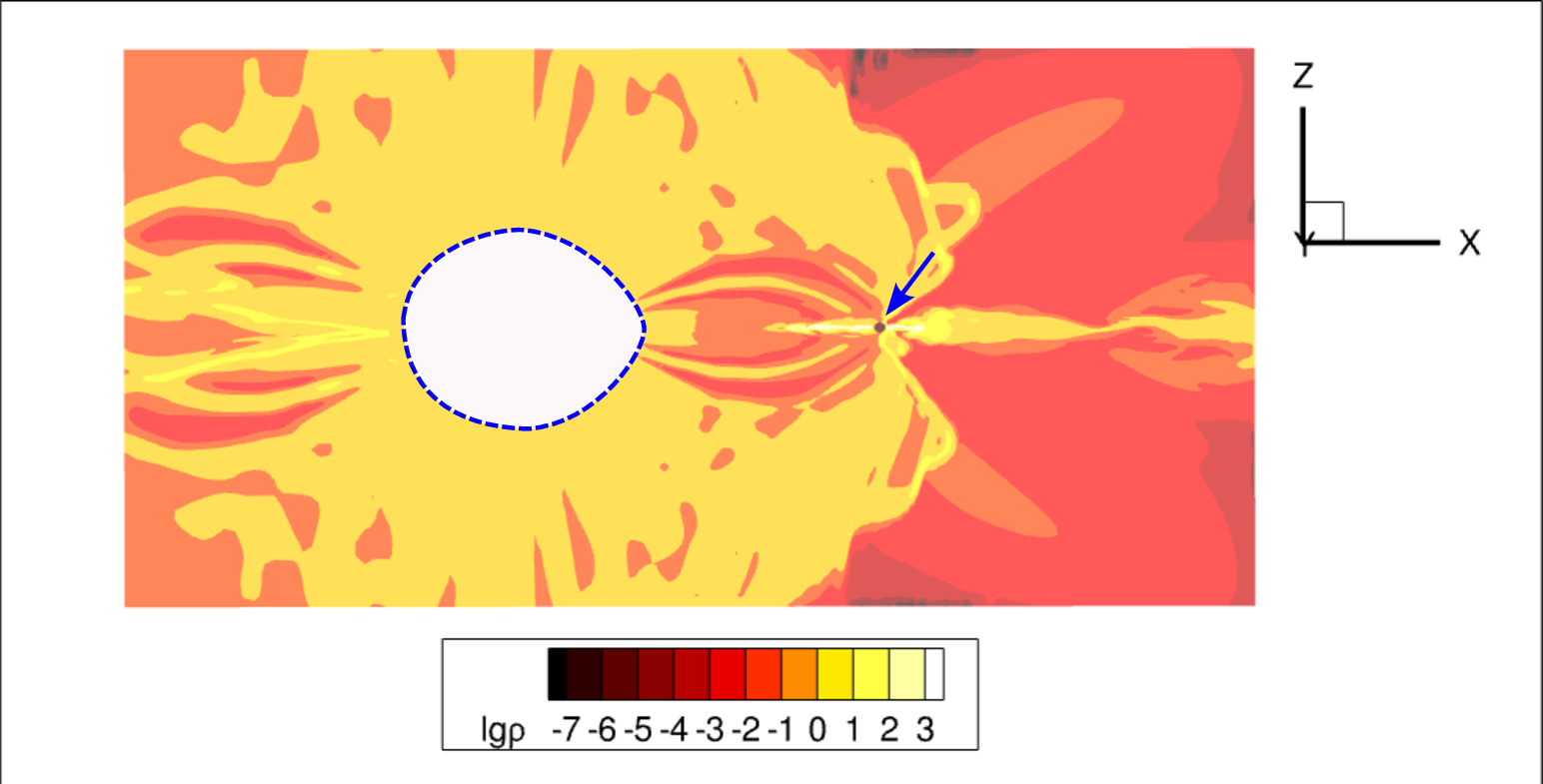}
	\caption{Same as in Fig. \ref{fig-9}, but for a donor wind temperature $T_{\text{d}} = 3 \times 10^6$ K (with initial wind speed $v_{\text{w}} = 270$ km/s).}
	\label{fig-11}
\end{figure}

\begin{figure}[H]
	\centering
	\includegraphics[width=0.95\textwidth]{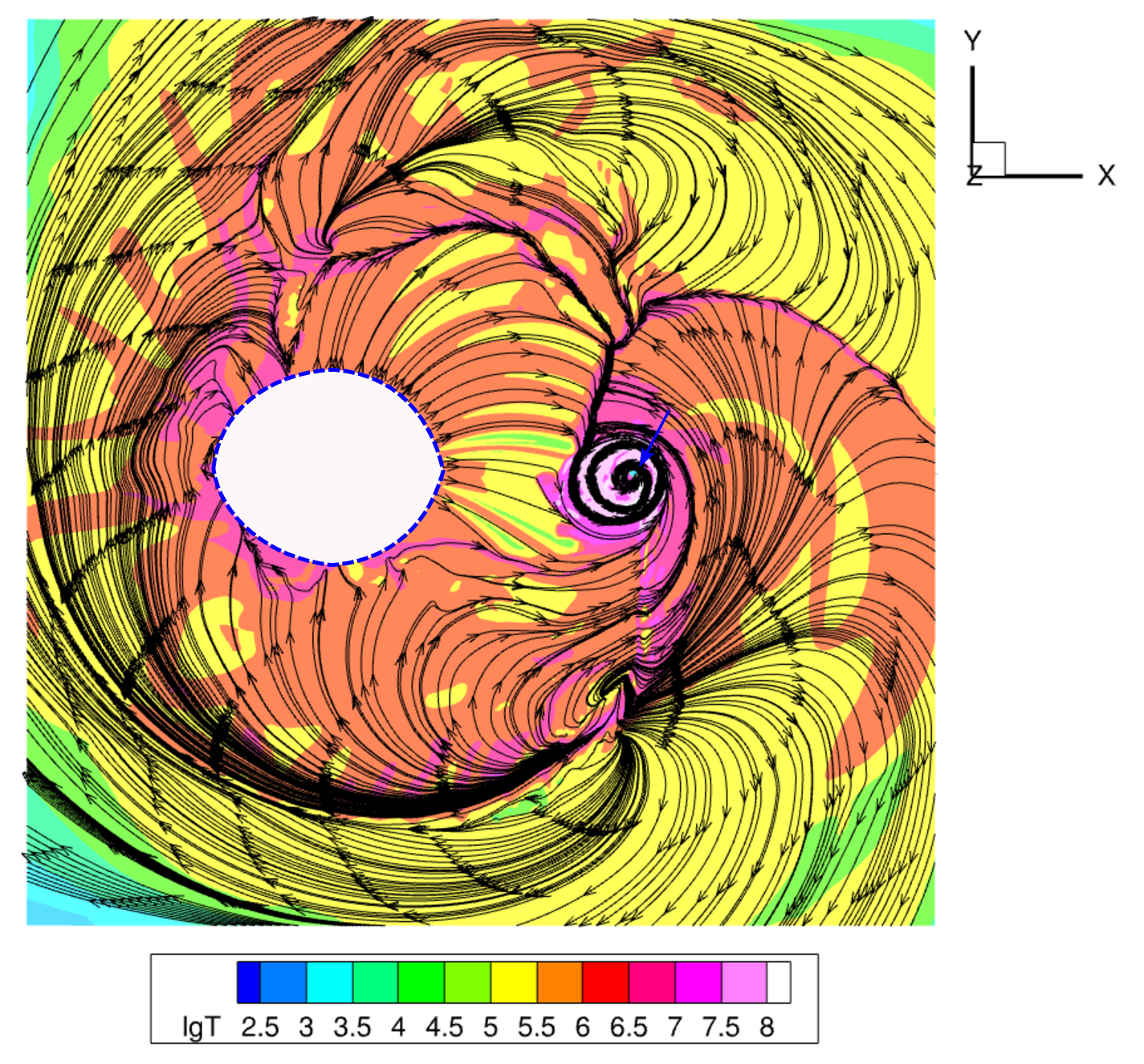}
	\includegraphics[width=0.95\textwidth]{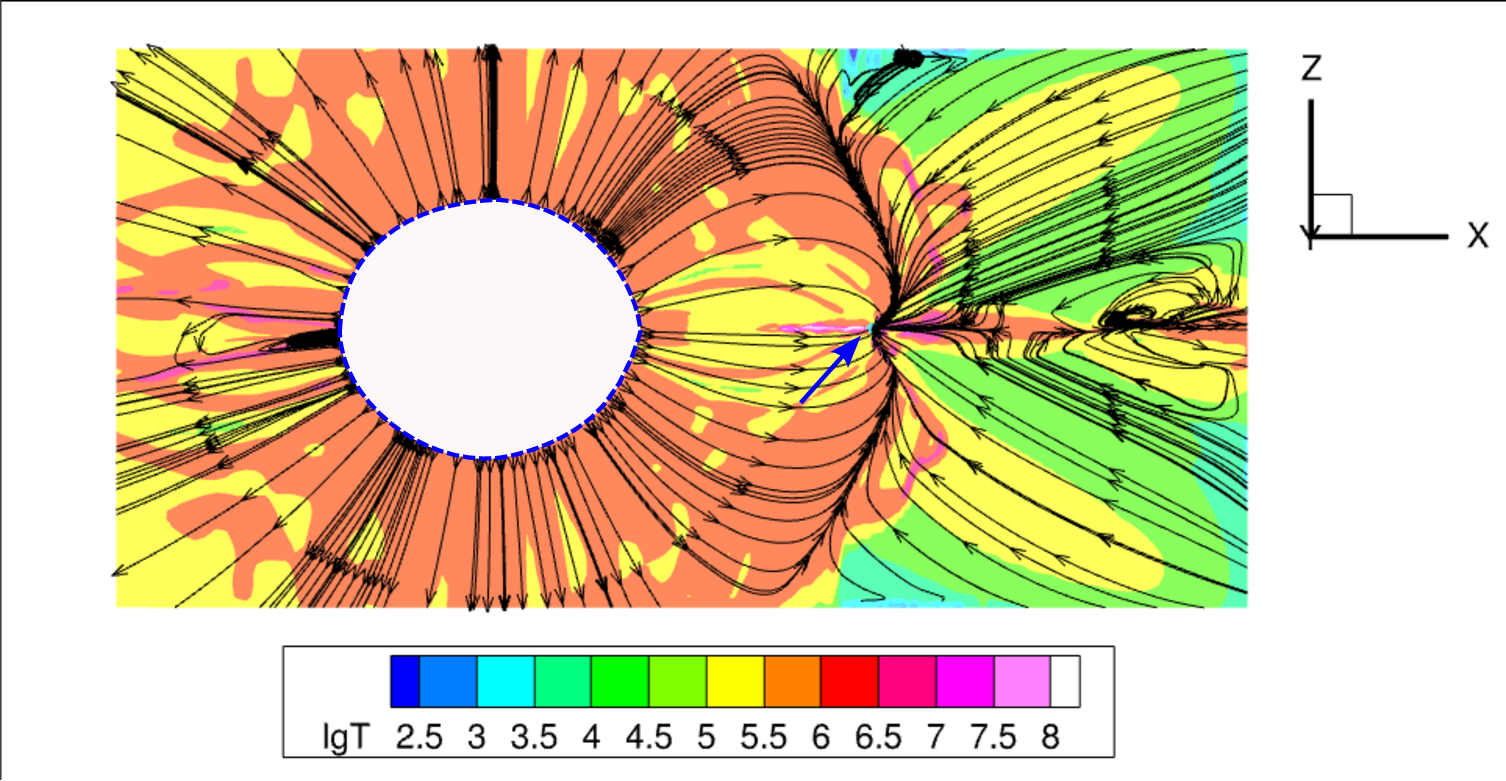}
	\caption{Same as in \ref{fig-10}, but for a donor wind temperature $T_{\text{d}} = 3 \times 10^6$ K.}
	\label{fig-12}
\end{figure}

\begin{figure}[H]
	\centering
	\includegraphics[width=0.95\textwidth]{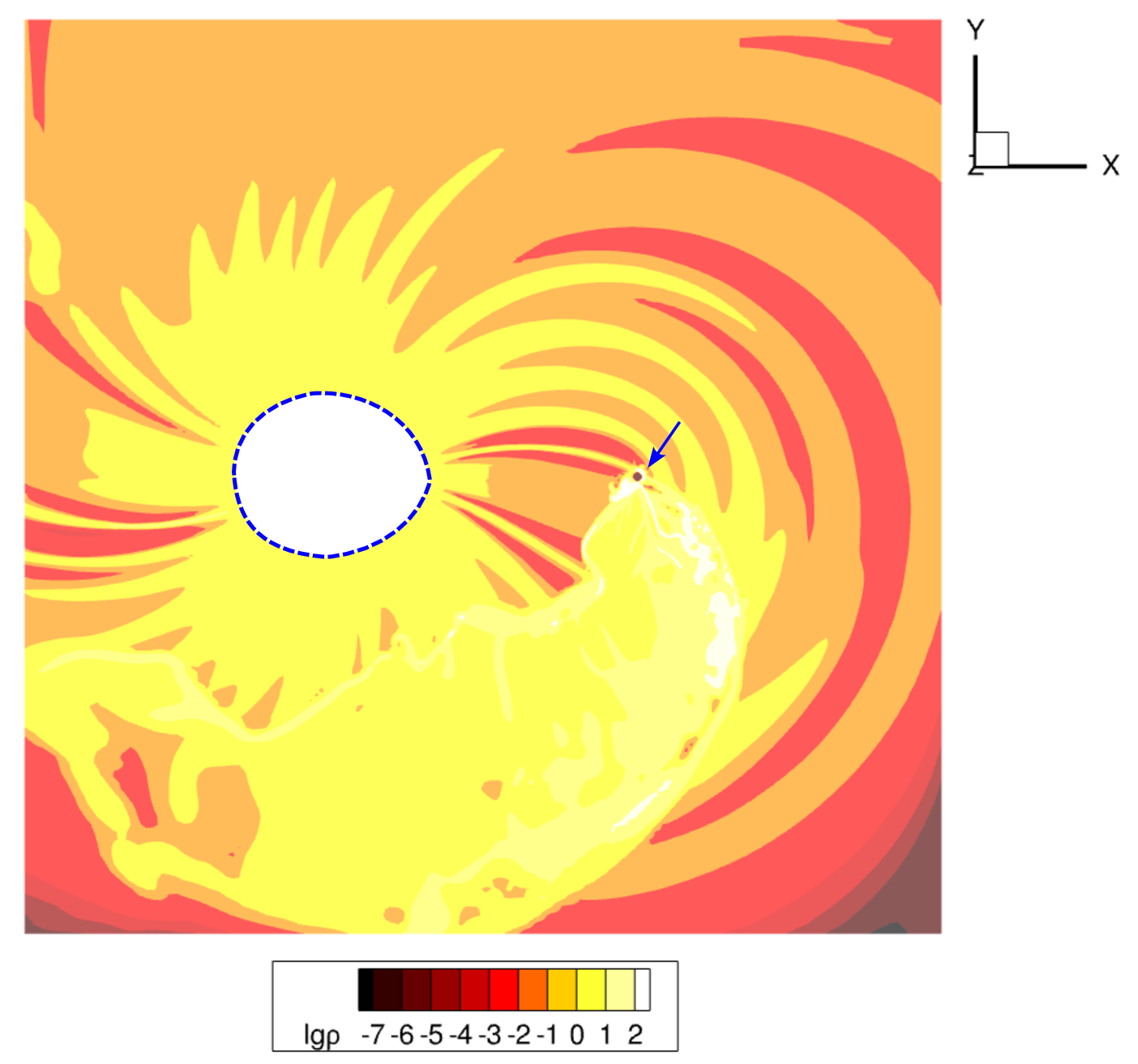}
	\includegraphics[width=0.95\textwidth]{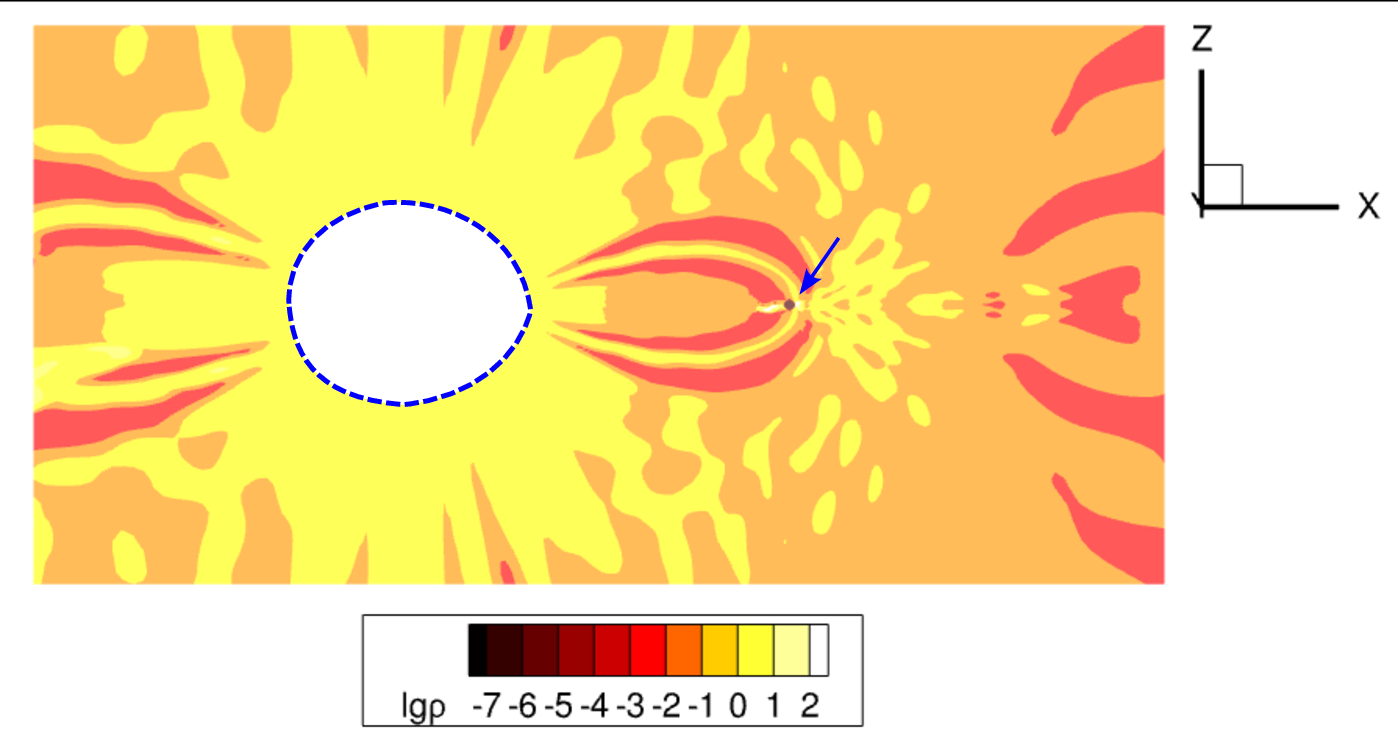}
	\caption{Same as in Fig. \ref{fig-9}, but for a donor wind temperature $T_{\text{d}} = 5.5 \times 10^6$ K (with initial wind speed $v_{\text{w}} = 370$ km/s).}
	\label{fig-13}
\end{figure}

\begin{figure}[H]
	\centering
	\includegraphics[width=0.95\textwidth]{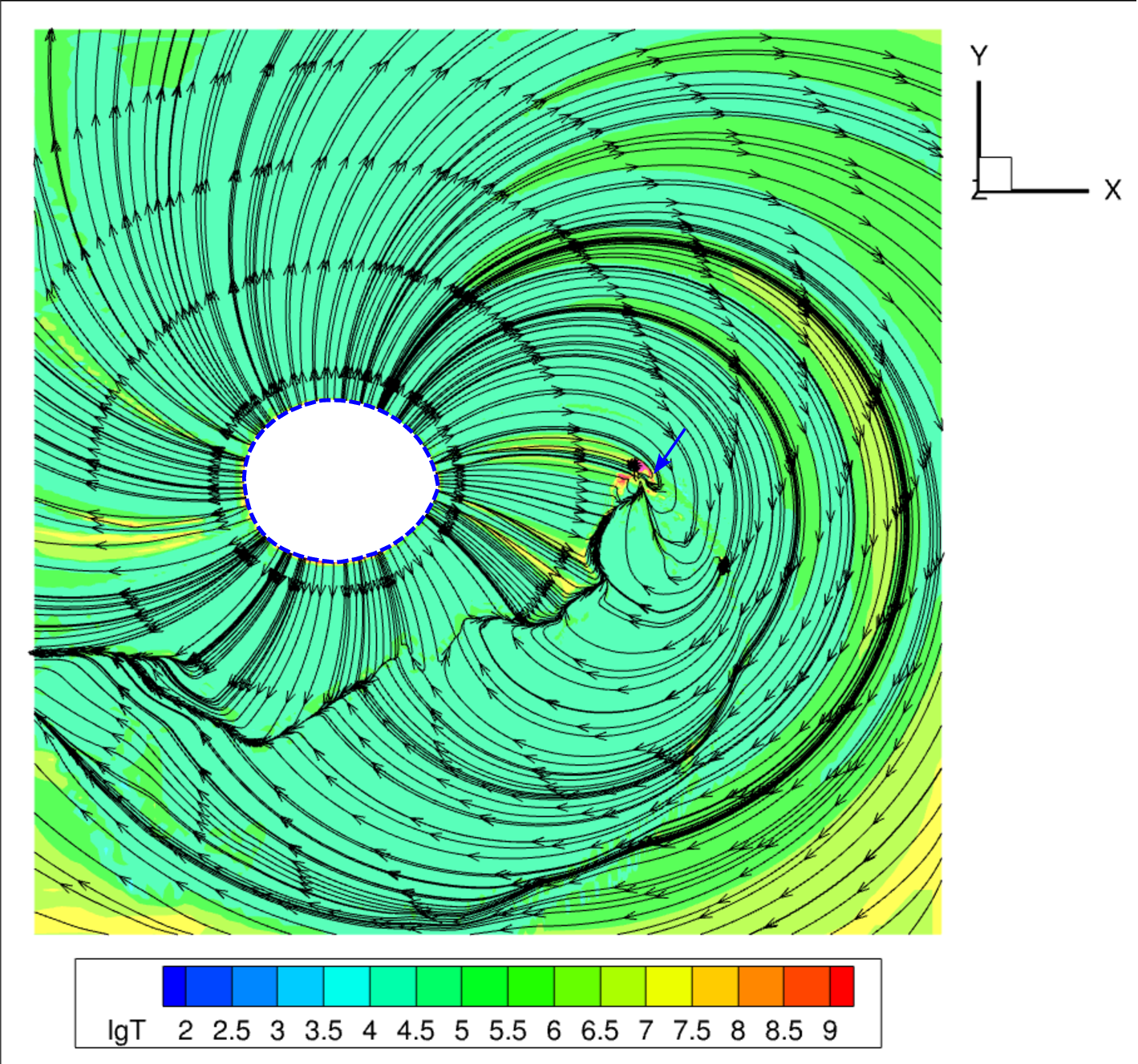}
	\includegraphics[width=0.95\textwidth]{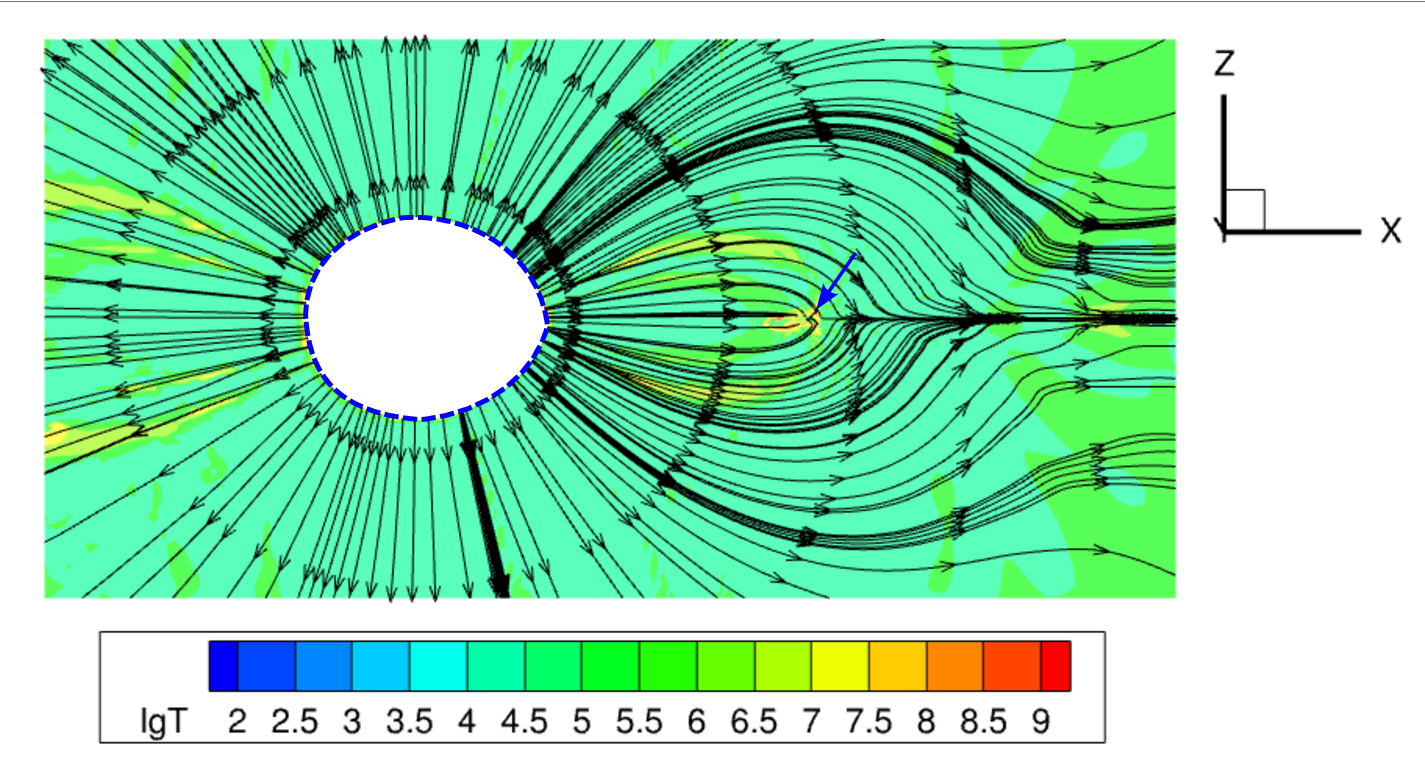}
	\caption{Same as in Fig. \ref{fig-10}, but for a donor wind temperature $T_{\text{d}} = 5.5 \times 10^6$ K.}
	\label{fig-14}
\end{figure}

\begin{figure}[H]
	\centering
	 \includegraphics[width=0.95\textwidth]{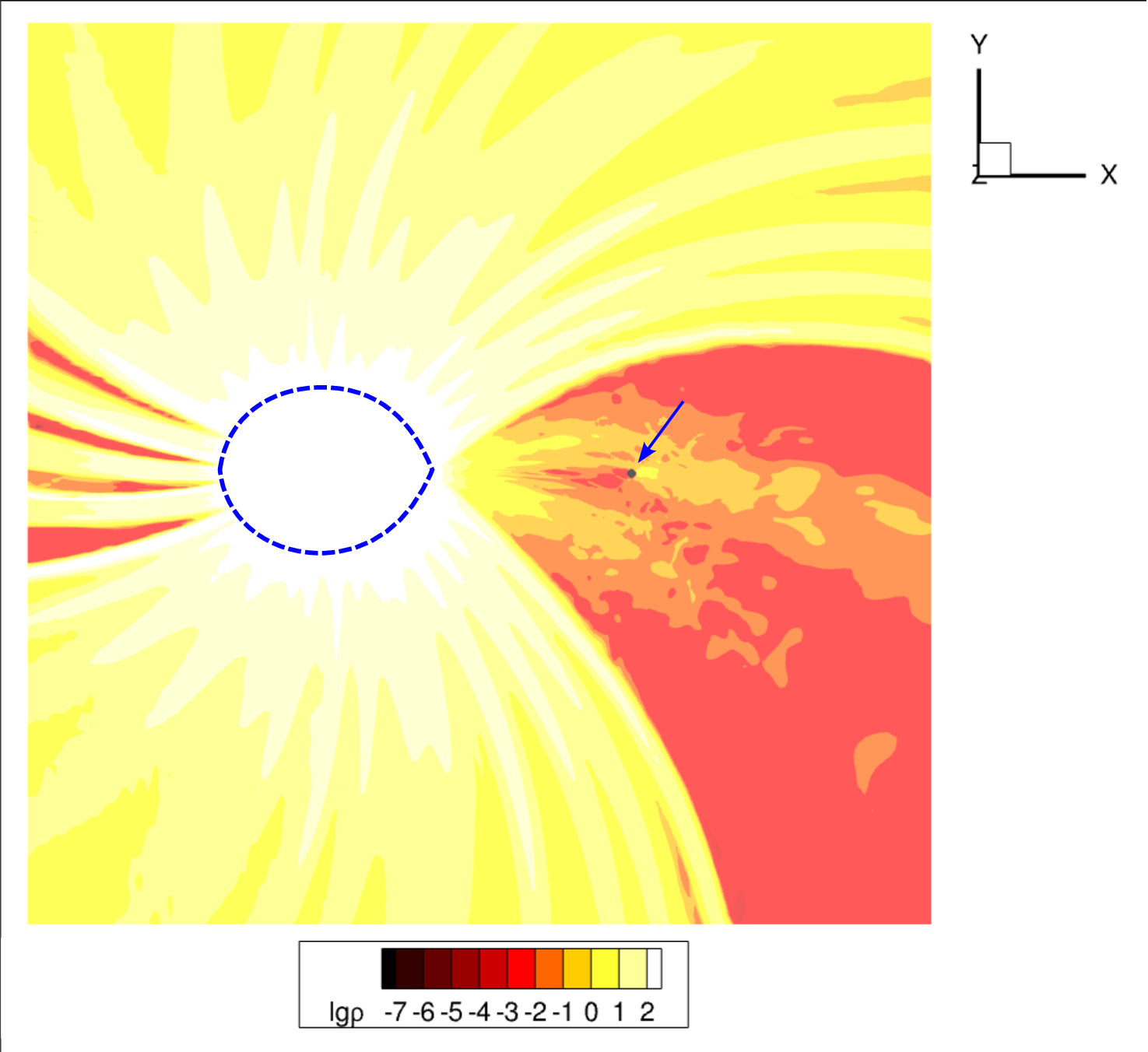}
	\includegraphics[width=0.95\textwidth]{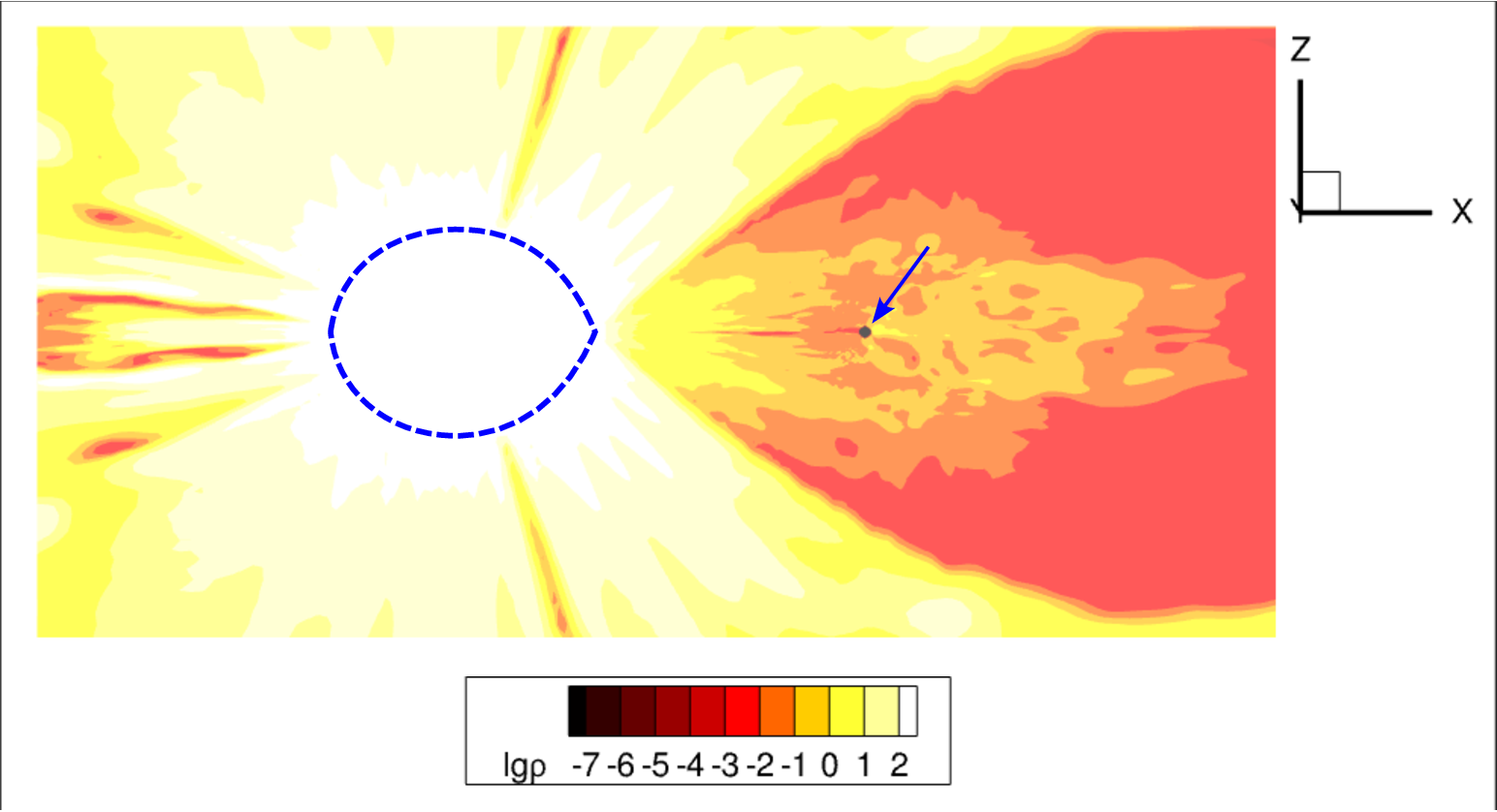}
	\caption{Same as in Fig. \ref{fig-9}, but for a donor wind temperature $T_{\text{d}} = 3 \times 10^7$ K (with initial wind speed $v_{\text{w}} = 860$ km/s).}
	\label{fig-15}
\end{figure}

\begin{figure}[H]
	\centering
	\includegraphics[width=0.95\textwidth]{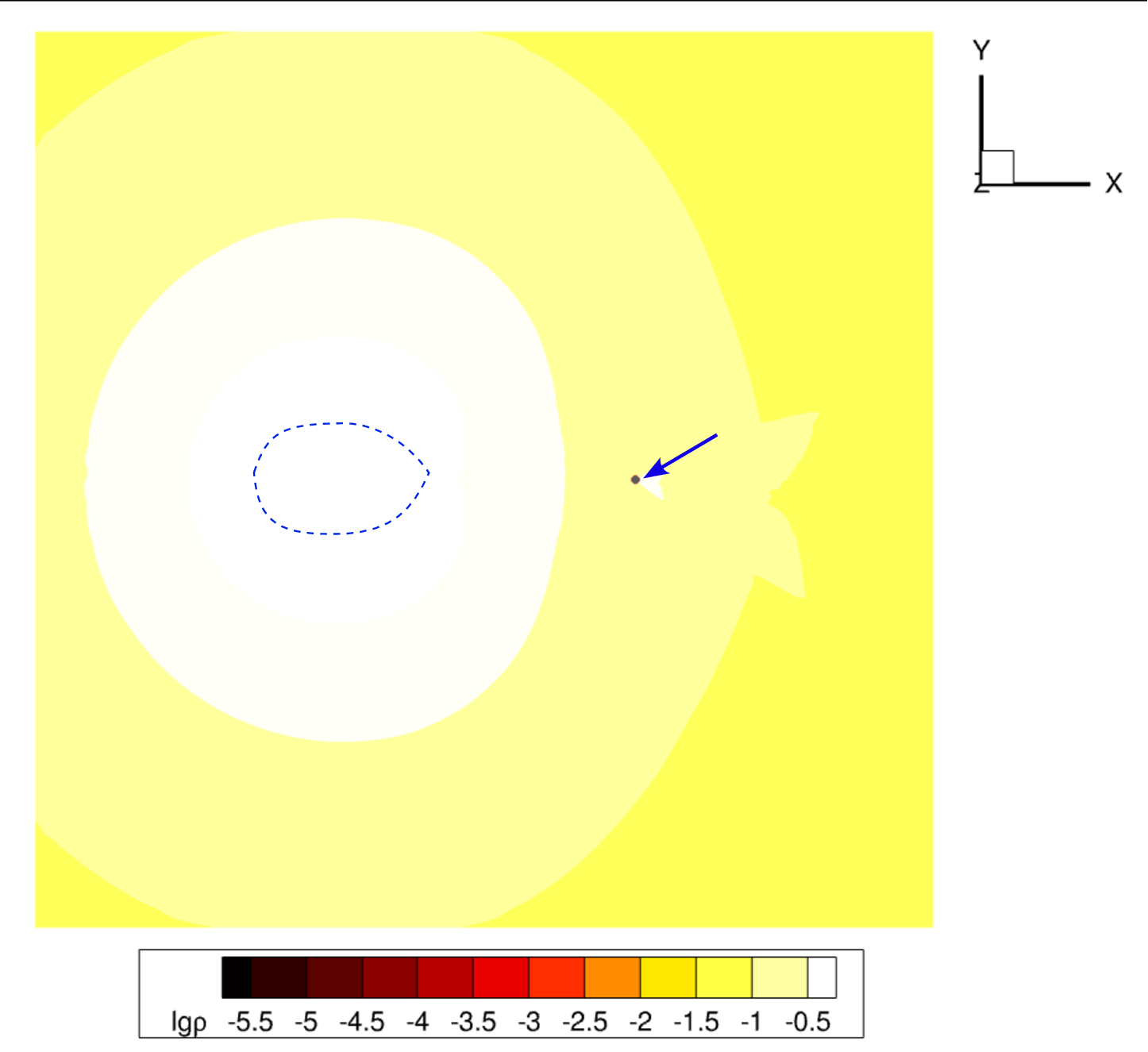}
	\includegraphics[width=0.95\textwidth]{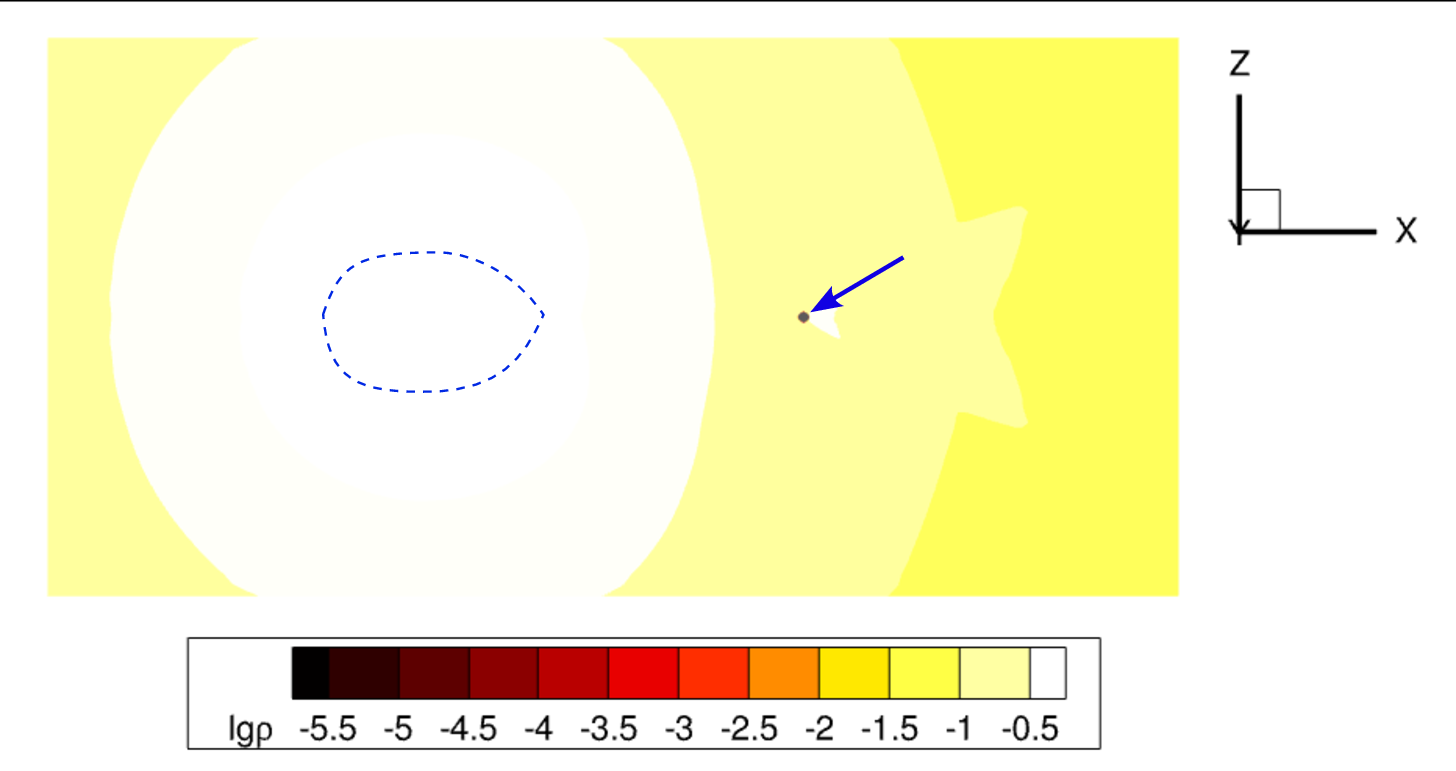}
	\caption{Same as in Fig. \ref{fig-9}, but for a donor wind temperature $T_{\text{d}} = 3 \times 10^8$ K (with initial wind speed $v_{\text{w}} = 2700$ km/s).}
	\label{fig-17}
\end{figure}

Figures  \ref{fig-9}, \ref{fig-11}, \ref{fig-13}, \ref{fig-15} and \ref{fig-17} show the density distribution in units of $\rho(\rm{L}_1)$, while Figs.\ref{fig-10}, \ref{fig-12}, \ref{fig-14}, \ref{fig-16} and \ref{fig-18} show the temperature distribution for the models listed in Table \ref{tab-2}. To visually display the wind structure, the simulation results are shown in two mutually perpendicular cross-sectional planes passing through the $x$ axis: the $XY$ plane coincides with the binary system’s orbital plane, and the orthogonal frontal plane $XZ$ passes through the $z$ axis of the system’s orbital rotation. In the temperature distributions, black lines with arrows in the $XY$ plane indicate the direction of the wind’s velocity vector in the computational domain, while in the $XZ$ plane they show the projection of the velocity vector onto this plane. These figures make it possible to analyze the patterns in the formation of the main elements of the flow structure.

Figures  \ref{fig-9} and \ref{fig-10} present the density and temperature distributions for Model 1. Since these distributions correspond to the quasi-stationary state of the binary system, they show the stable, formed elements of the flow structure. First of all, this is the main outflow of donor atmosphere material from the vicinity of the Lagrange points $\rm{L}_1$ and $\rm{L}_3$. In the first case, the gas flow forms an accretion disk around the white dwarf, provided the accretor does not have a significant magnetic field. The donor’s material leaving the vicinity of point $\rm{L}_3$ becomes, due to the system’s orbital rotation, the source of the general circumstellar envelope. The pattern of wind emission over all regions of the Roche lobe surface, except the Lagrange points $\rm{L}_1$ and $\rm{L}_3$, differs between the orbital and frontal planes. At the first point, this process is prevented by a dense gas
envelope formed as the wind material fills the computational domain. 
\begin{figure}[H]
	\centering
	\includegraphics[width=0.95\textwidth]{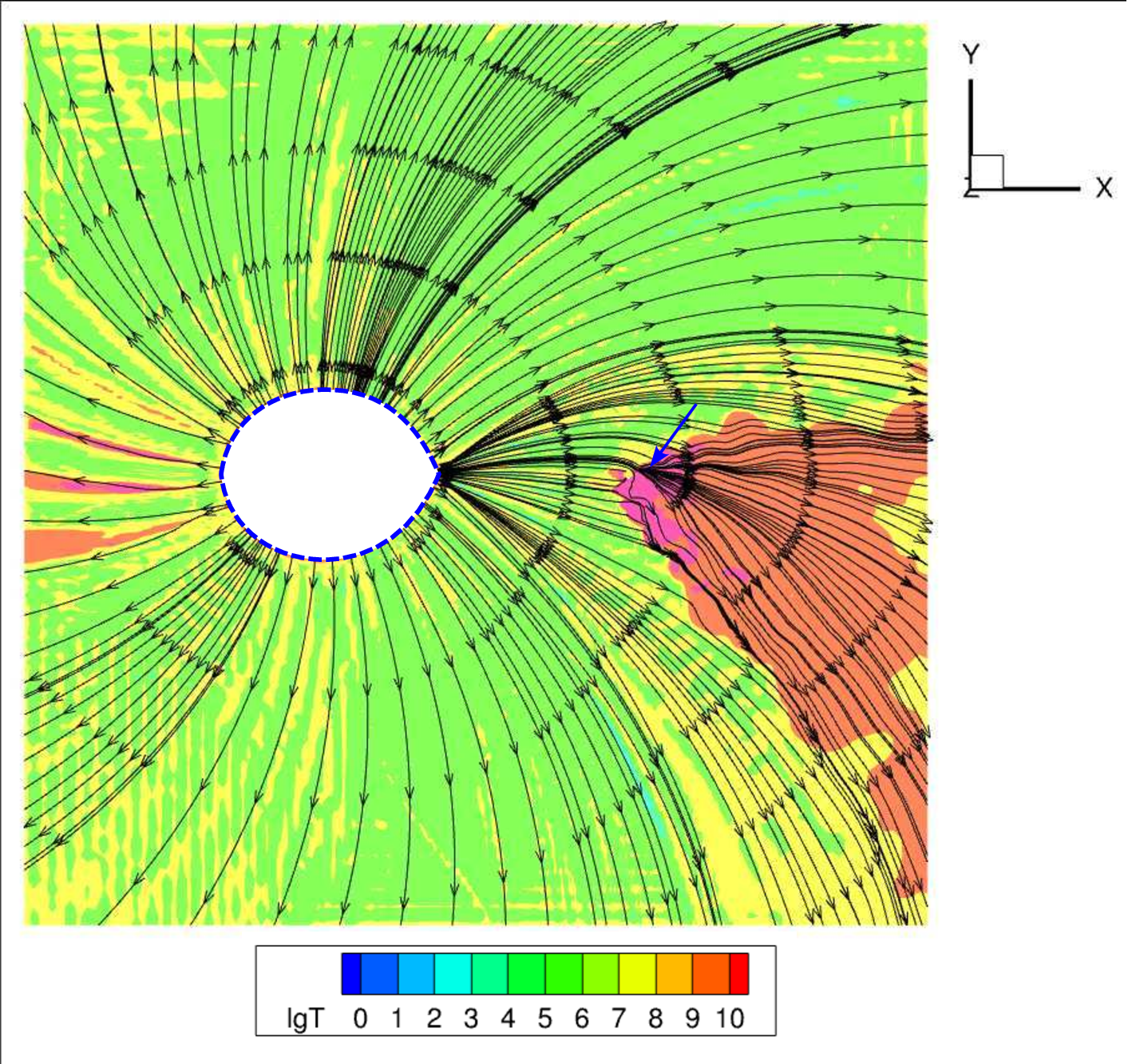}
	\includegraphics[width=0.95\textwidth]{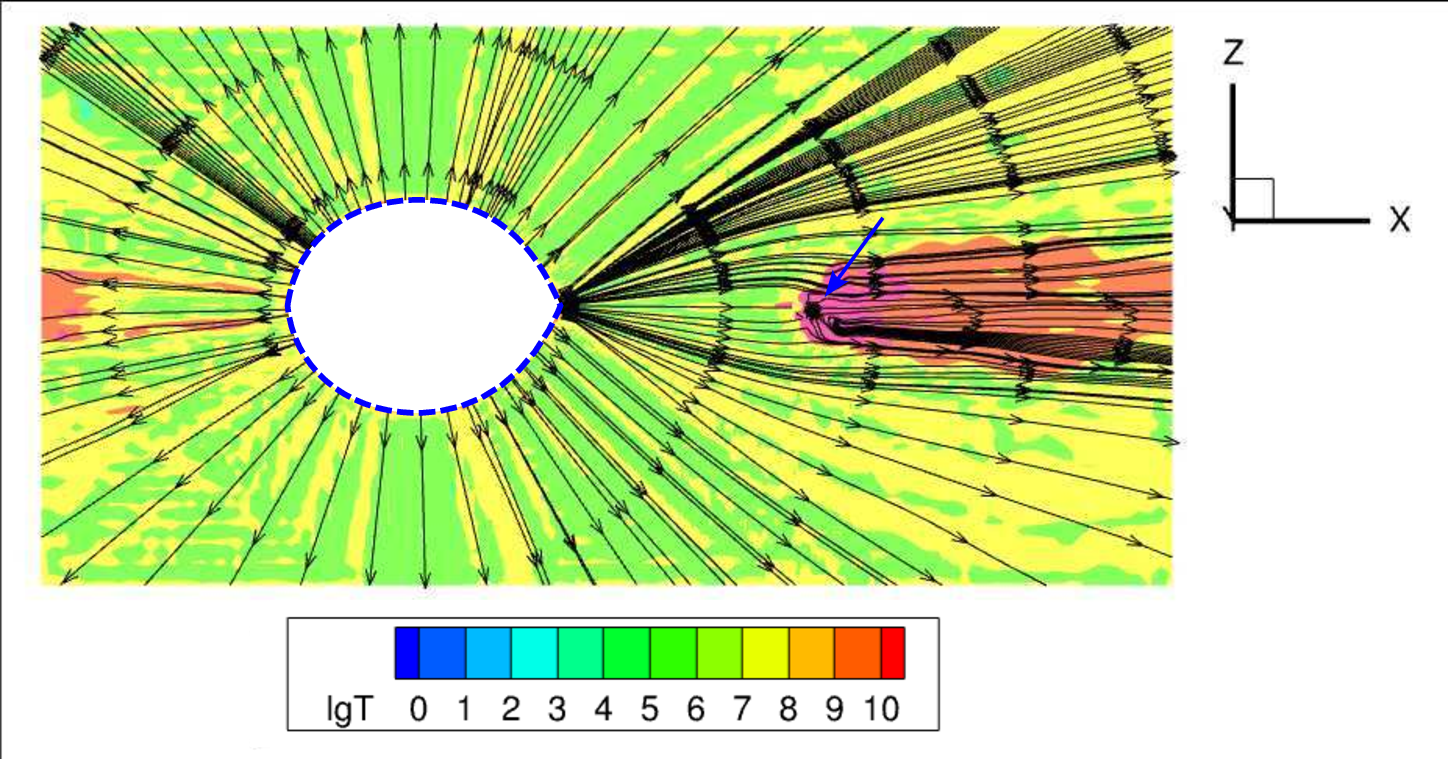}
	\caption{Same as in Fig. \ref{fig-10}, but for a donor wind temperature $T_{\text{d}} = 3 \times 10^7$ K.}
	\label{fig-16}
\end{figure}
\begin{figure}[H]
	\centering
	\includegraphics[width=0.97\textwidth]{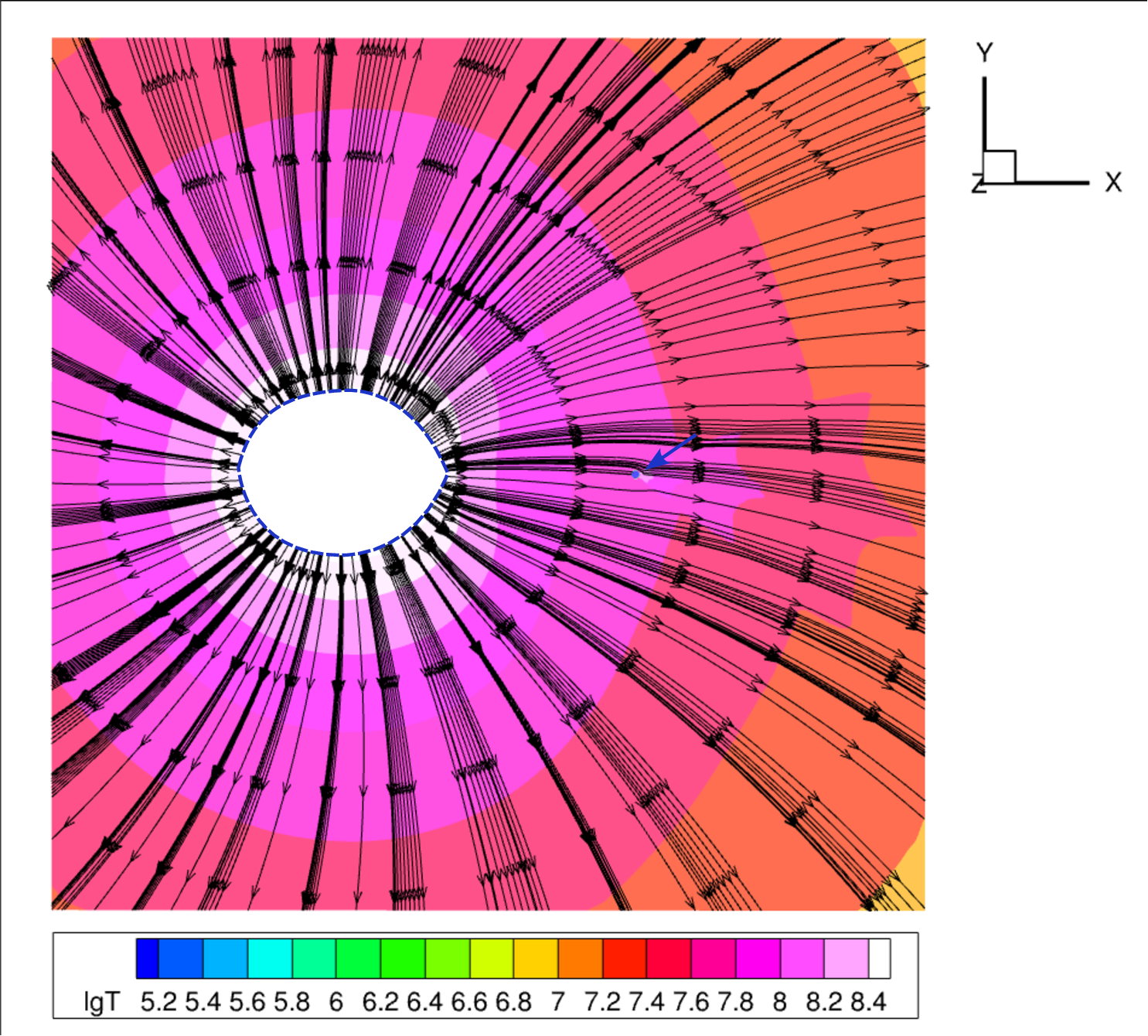}
	\includegraphics[width=0.97\textwidth]{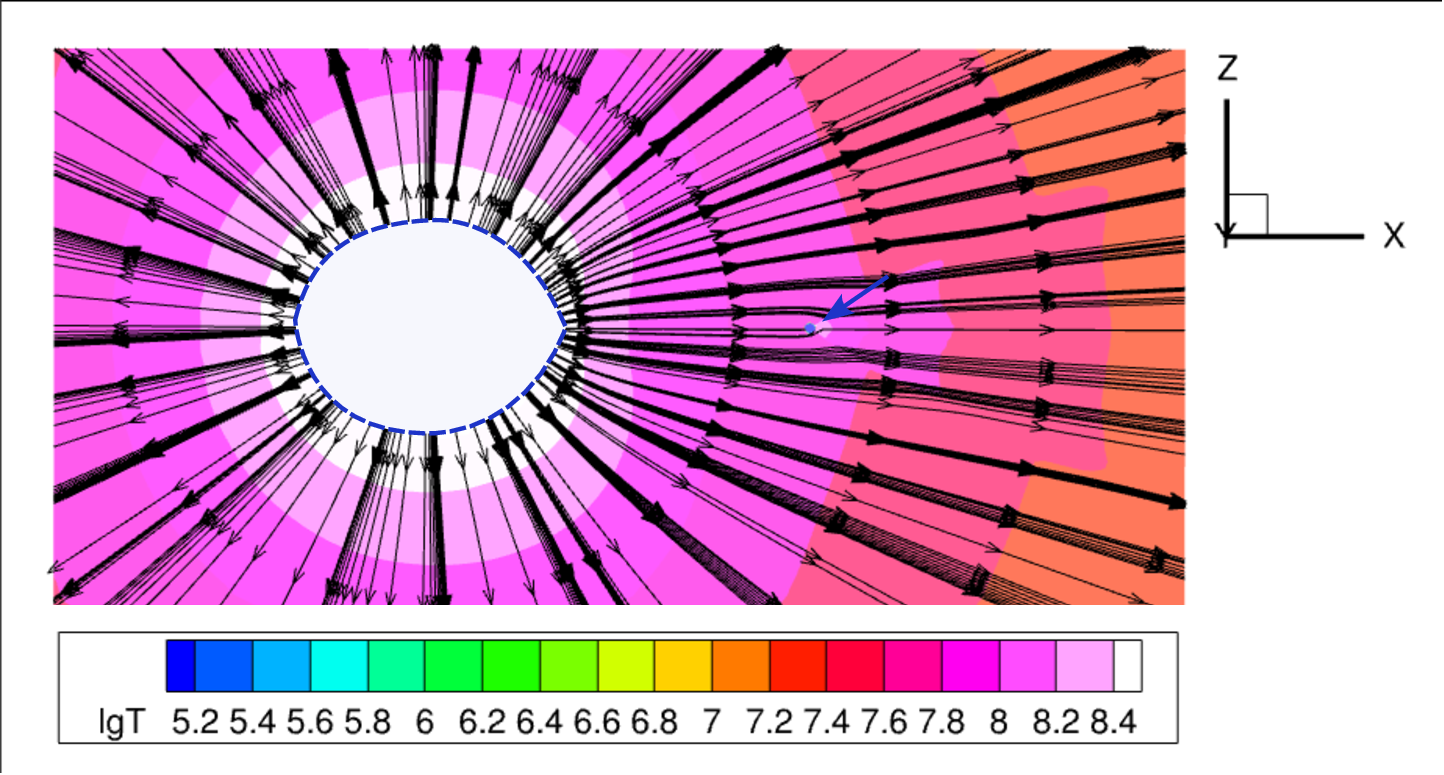}
	\caption{Same as in Fig. \ref{fig-10}, but for a donor wind temperature $T_{\text{d}} = 3 \times 10^8$ K.}
	\label{fig-18}
\end{figure}

By contrast, in the orbital plane, there is an almost uniform wind generation from the entire Roche lobe surface of the donor, although the main gas flow serves as the accretion disk’s feeding source.

For Model 1, the main wind material motion occurs in the orbital plane, while in the $XZ$ plane the computational domain is filled with the general circumstellar envelope. As follows from the temperature distribution (Fig. \ref{fig-10}), wind material falls onto the white dwarf’s surface not only from the inner region of the accretion disk but also from the general circumstellar envelope. This causes an approximately two orders of magnitude increase in the accretion disk halo temperature. Material flowing out from the outer disk edges forms a cone-like structure. The top panel of the figure clearly shows that in the orbital plane, the stream from the vicinity of the Lagrange point $\rm{L}_1$ in the accretion disk area splits into two arms: the first curls in the direction of the disk’s rotation and, colliding with the oncoming wind, is deflected into the region of the Lagrange point $\rm{L}_2$, from where it leaves the computational domain; the second flow is captured by the gravitational potential of the Lagrange
point $\rm{L}_4$ and is deflected back toward the donor, but upon reaching the donor’s far side, it collides with the outflowing wind in the region of the Lagrange point $\rm{L}_3$, which also helps expel material from the computational domain.

With an order-of-magnitude increase in wind temperature (Figs. \ref{fig-11} and \ref{fig-13}) in Model 2, the wind flow structure changes significantly. Wind emission from the donor now occurs over the entire surface of its Roche lobe. The accretion disk is fed by wind flows in the orbital plane, and additional donor material falls
onto the white dwarf through two narrow channels in the vicinity of its geographic poles.

\begin{figure}[H]
	\centering
	\includegraphics[width=0.7\textwidth]{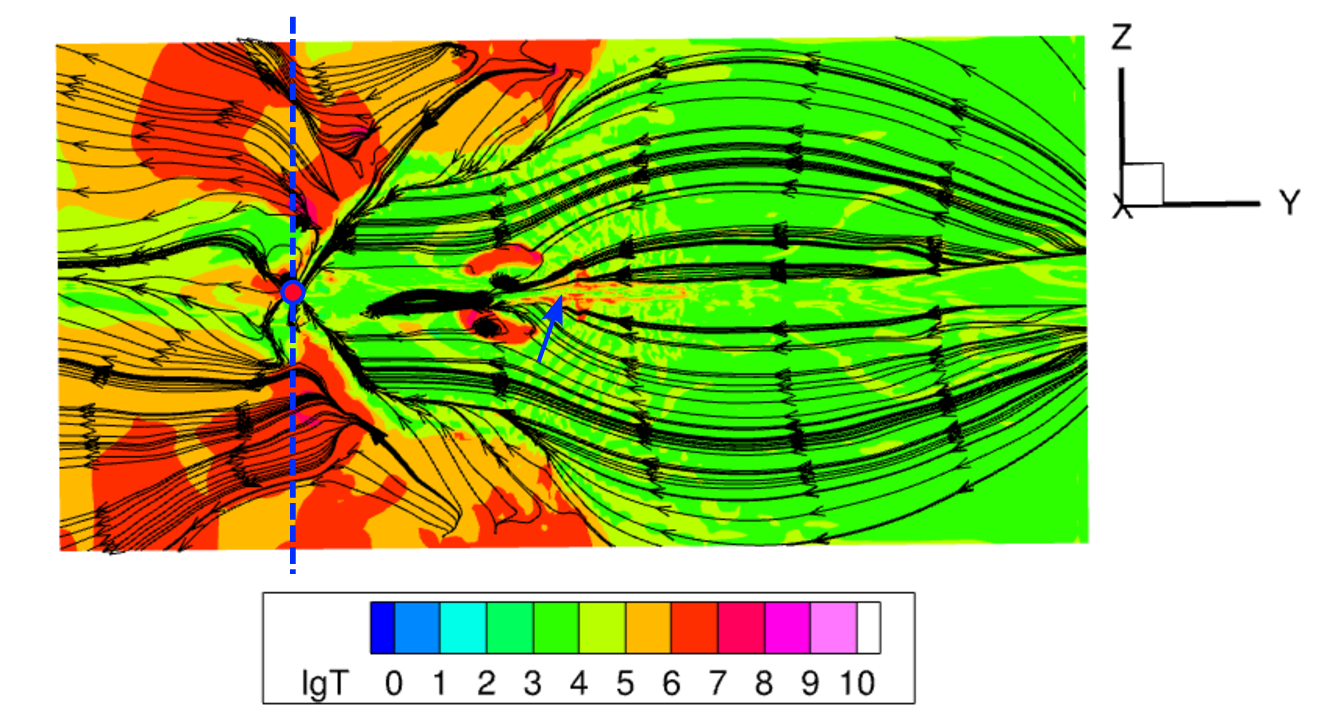}
	\includegraphics[width=0.7\textwidth]{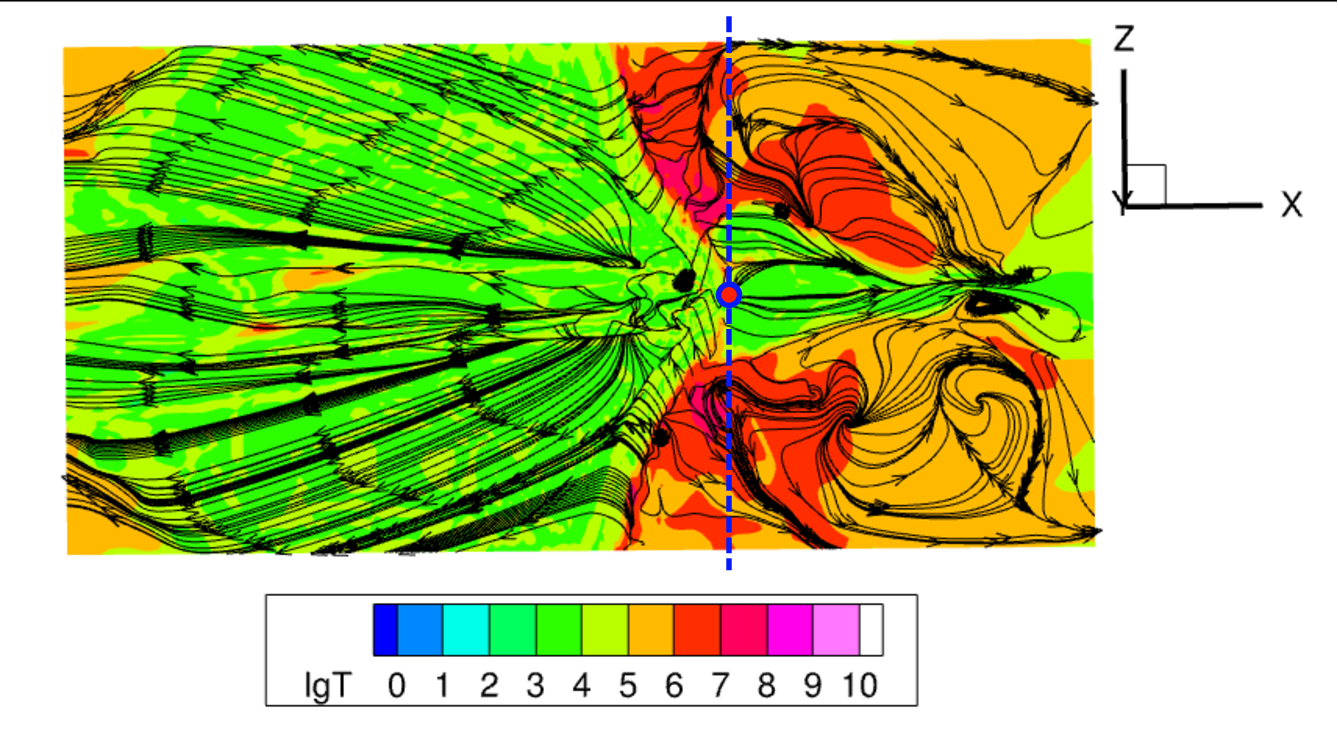}
	\includegraphics[width=0.7\textwidth]{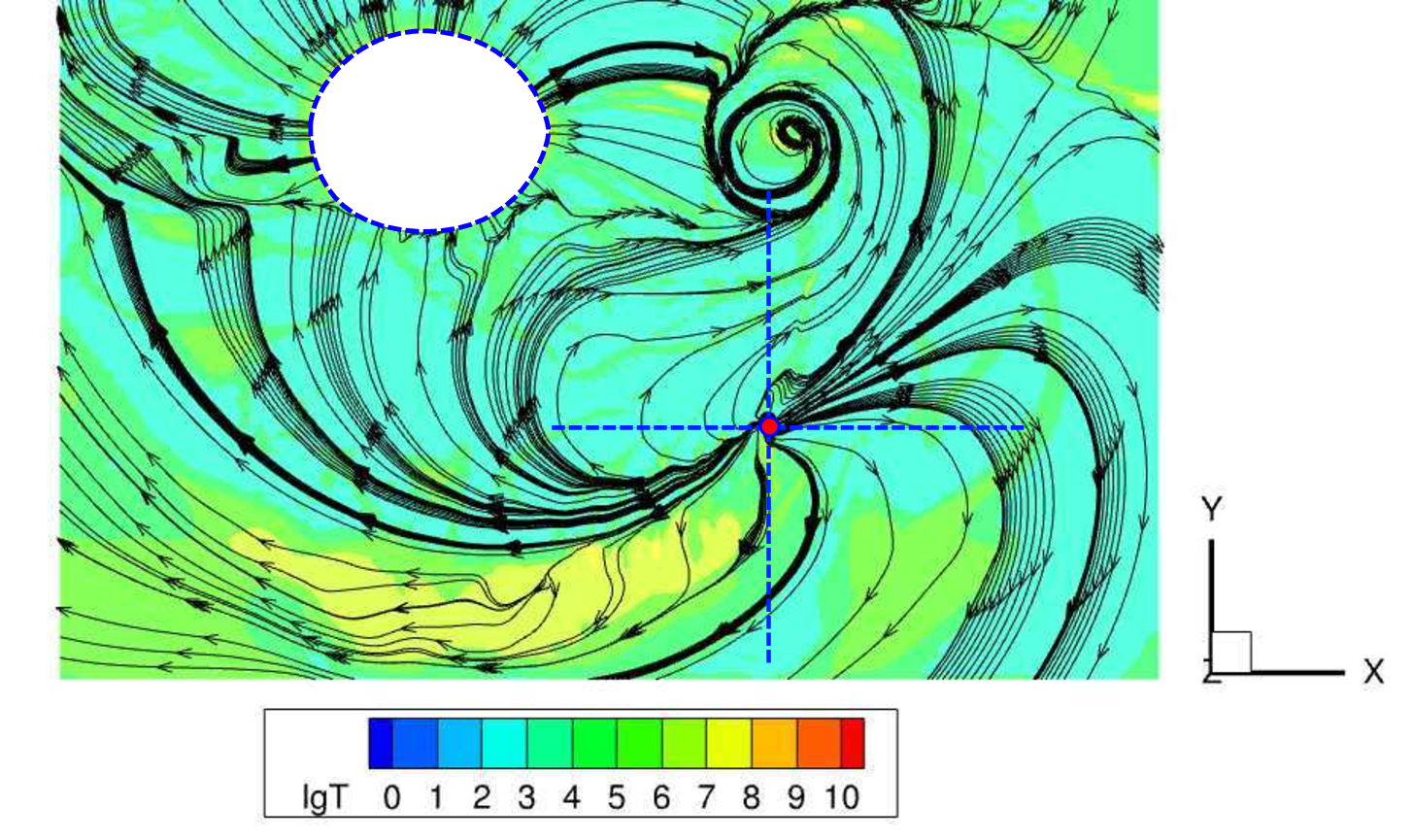}
	\caption{Same as in Fig. \ref{fig-12}, but showing wind motion near the Lagrange point $\rm{L}_4$ (marked by a red dot). The \textit{upper} panel shows a plane parallel to $YZ$ with an offset along the axis $x = 0.08$; the \textit{middle} panel --- a plane parallel to $XZ$, with an offset of $y = -0.08$. Both planes intersect along the blue dashed line passing through the Lagrange point $\rm{L}_4$. The accretion disk plane is indicated by an arrow. The \textit{lower} panel shows a fragment of the system’s orbital plane $XY$, with the positions of the above planes marked by blue dashed lines.}
	\label{fig-19}
\end{figure}

A special role in the wind flow for Model 2 is played by the Lagrange point $\rm{L}_4$. In this point, gas flows in the $XZ$ plane from the donor’s near-polar region, forming a powerful vortex that twists the material within the orbital plane. The movement of gas in this zone is shown in more detail in Fig. \ref{fig-19}. Gas exiting the vortex is distributed along two channels: part of it is carried by the orbital motion of the binary system and returns back to the donor, while the other part feeds the outer regions of the accretion disk. As in Model 1, this latter flow collides with the incoming wind and forms a spiral arm behind the accretor. Some of the gas in this arm leaves the computational domain near the Lagrange point $\rm{L}_2$, while the rest contributes to forming the general circumstellar wind structure
shaped by the system’s orbital rotation.

It should be noted that Figs. \ref{fig-9} -- \ref{fig-12} illustrate the quasi-conservative type of detached system. The wind flowing from the donor’s surface facing the accretor is captured by the latter’s gravitational field and forms an accretion disk around it. Wind from the donor’s far side mainly enters the region of the general circumstellar envelope. Part of it leaves the system through the vicinity of the Lagrange point $\rm{L}_2$. Due to the orbital motion of the binary, the material in the envelope comes under the gravitational influence of the accretor. As a result, the figures show the wind stream colliding with the outer edge of the accretion disk. In
this process, some wind material becomes entrained in the disk’s rotation, while the more distant streams head toward the outer boundary of the computational domain.

Further increase in the donor wind temperature to $3 \times 10^7$ K (Model 4) and $3 \times 10^8$ K (Model 5) also leads to significant changes in the wind flow geometry (Figs. \ref{fig-15} -- \ref{fig-18}). The binary system becomes open (Fig. \ref{fig-2}). These models are characterized by the absence of an accretion disk, while the donor’s wind material is distributed spherically symmetrically along spiral trajectories throughout the computational domain and then leaves it. However, for the temperature $3 \times 10^7$ K, a characteristic feature of the flow is the cone-shaped tail behind the accretor. As shown by the color scale in the figure, the flow density in the inner part of the cone drops by about three orders of magnitude, while its temperature rises by two orders of magnitude. In the orbital plane, due to the binary’s rotation, the morphology of this structure is defined by the motion of part of the wind near the accretor.

The formation of such cone-like structures is also observed in other binary systems. For example, the interaction of the massive main-sequence component’s stellar wind with the relativistic wind of a pulsar in a close binary system was studied within a three-dimensional hydrodynamic model \cite{Bosch2015}. The results showed the formation of a shock wave ahead of the pulsar and significant turbulence in the stellar wind flow behind it. The turbulent <<tail>> was twisted by the
orbital motion into a spiral. Model \cite{Bosch2015} covers a flow
region tens of times larger than the size of the binary system. The study of the flow structure over one orbital period revealed the absence of an accretion disk around the neutron star due to its wind.

The formation of an accretion disk around an isolated black hole was studied in three-dimensional simulations in \cite{Tripathi2024}. This study showed that the formation of temporary disks under initially symmetric flow conditions is only possible when this symmetry is broken by turbulence in the tail induced by the black hole’s gravity.

The interaction of the donor’s stellar wind with the accretor in detached binaries is a complex process. In describing this interaction with a simple analytical model, we aimed to demonstrate its dependence on wind velocity (Eq. \ref{eq-13}). It turned out that this dependence
leads to two interaction regimes. At high angular momentum of the material captured by the accretor, an accretion-decretion disk forms around it.

At low angular momentum, the donor wind material flows past the accretor and leaves the system. The existence of these two interaction regimes supports the presence of two types of binary systems --- quasi-conservative and open --- with the transition between them determined by the wind velocity.

In this study, we performed numerical calculations for an estimated critical wind speed of 370 km/s (see Table \ref{tab-2}), corresponding to a wind temperature of $5.5 \times 10^6$ K (Model 3). According to formula (Eq. \ref{eq-13}) and condition (Eq. \ref{eq-14}), it can be found that $v_{\text{orb}} = 300$ km/s, with the corresponding ratio $\frac{v_{\text{w}}}{v_{\text{orb}}} = 1.26$. Condition (Eq. \ref{eq-14}) is satisfied for $\frac{v_{\text{w}}}{v_{\text{orb}}} \approx 2$. Thus, in our analytical model, the critical wind speed at which an accretion disk around a degenerate white dwarf can still exist lies in the range $v_{\text{w}} = (1-2) v_{\text{orb}}$.

The solution obtained for wind speeds above the critical speed is shown in Figs. \ref{fig-13} and \ref{fig-14}. Compared to Model 2, the flow structure changes significantly: a nearly twofold increase in temperature leads to the disappearance of the accretion disk and the formation of an enveloping cone. The cone axis (Fig. \ref{fig-13}) is almost perpendicular to the intercomponent line (the $x$ axis), which is caused by the significant influence of the accretor’s orbital velocity. The collision of the donor wind with the cone’s material forms a region of increased density along its surface. However, there is no noticeable heating of this zone, which indicates the merging of two flows rather than their collision.

Model 5 with a wind temperature of $3 \times 10^8 $ K (Fig. \ref{fig-17} and \ref{fig-18}) demonstrates a fully open type of system. None of the previously described donor wind flow elements form here, except for a cone-shaped tail behind the accretor as the wind streams past it. The density distribution shown in Fig. \ref{fig-17} reveals only a short conical gas flow disturbed by the accretor. In
fact, in this case the wind appears as a spherically symmetric gas flow whose temperature decreases with distance from the donor. As can be seen in the figure, the structure of this gas is centrally symmetric.

Fig. \ref{fig-19} shows the motion of wind material near the Lagrange point $\rm{L}_4$ for a wind temperature $T_{\text{d}} = 3 \times 10^6$ K (Fig. \ref{fig-12}). Due to the binary’s rotation, the stellar wind flows move along different trajectories depending on the location of their emission zones on the donor. The Lagrange points $\rm{L}_4$ and $\rm{L}_5$ play an important role in this motion. Wind flowing from the
donor’s equatorial regions coils into spiral arms in the orbital plane $XY$. After being deflected by point $\rm{L}_5$ these flows move perpendicularly to the binary’s intercomponent line (the $x$ axis) (Fig. \ref{fig-19}, upper panel).        
\begin{figure}[H]
	\begin{minipage}[h]{0.48\linewidth}
	\centering{\includegraphics[width=0.9\textwidth]{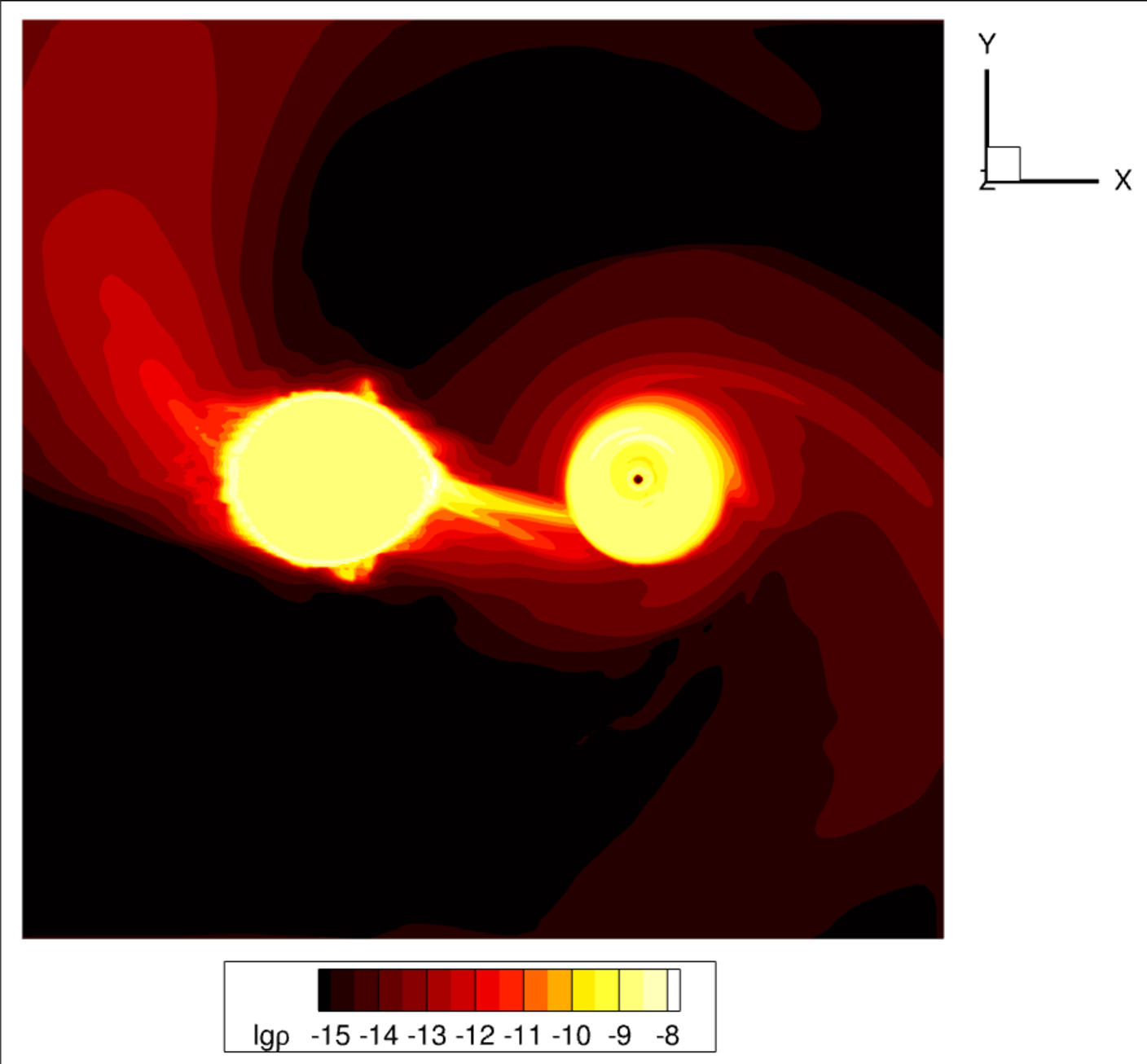} \\ $T_{\text{c}} = 3 \times 10^5 K$}
	\end{minipage}
	\hfill
	\begin{minipage}[h]{0.48\linewidth}
	\centering{\includegraphics[width=0.9\textwidth]{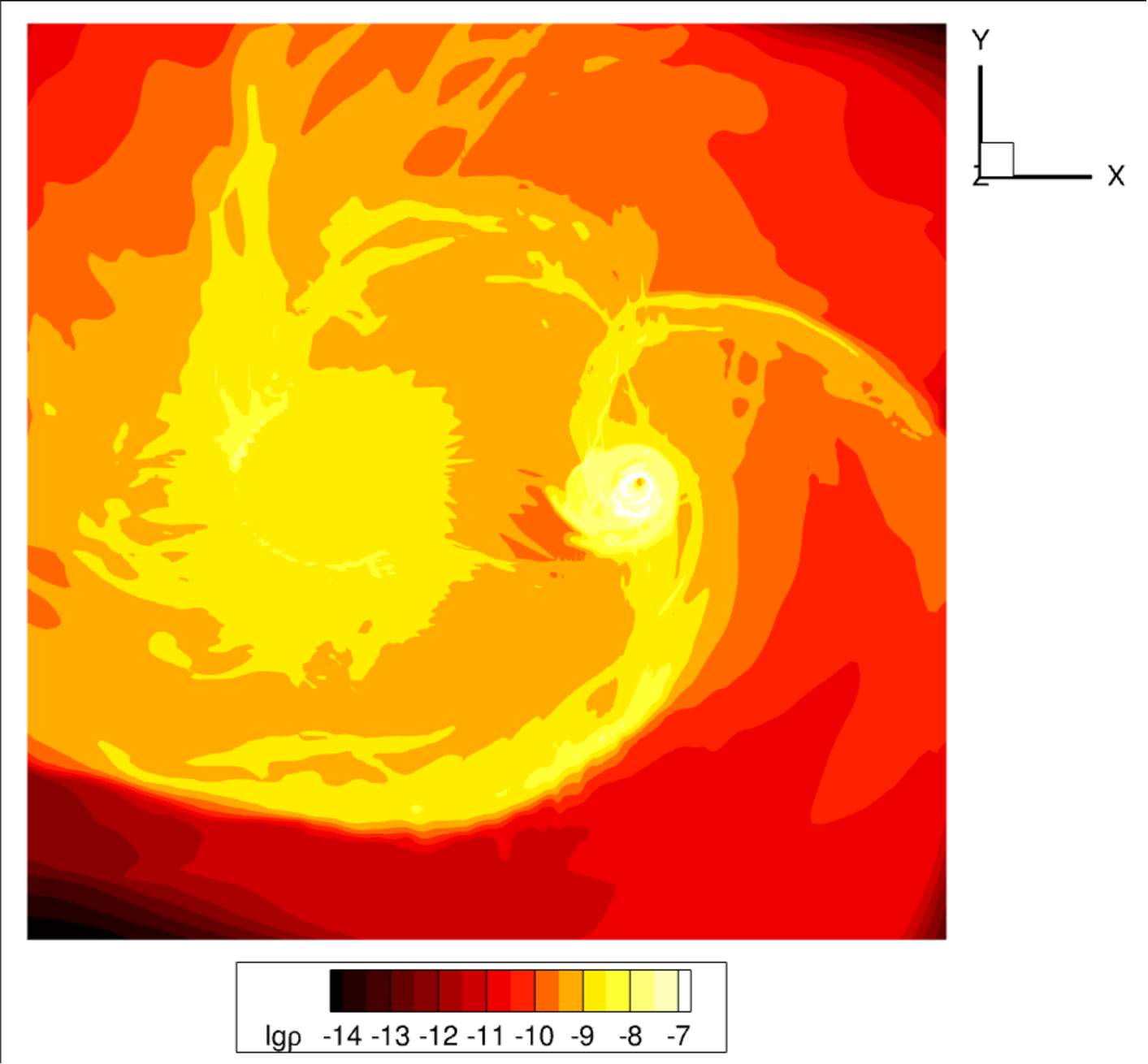} \\ $T_{\text{c}} = 3 \times 10^6 K$}
    \end{minipage}
    \hfill
    \begin{minipage}[h]{0.48\linewidth}
	\centering{\includegraphics[width=0.9\textwidth]{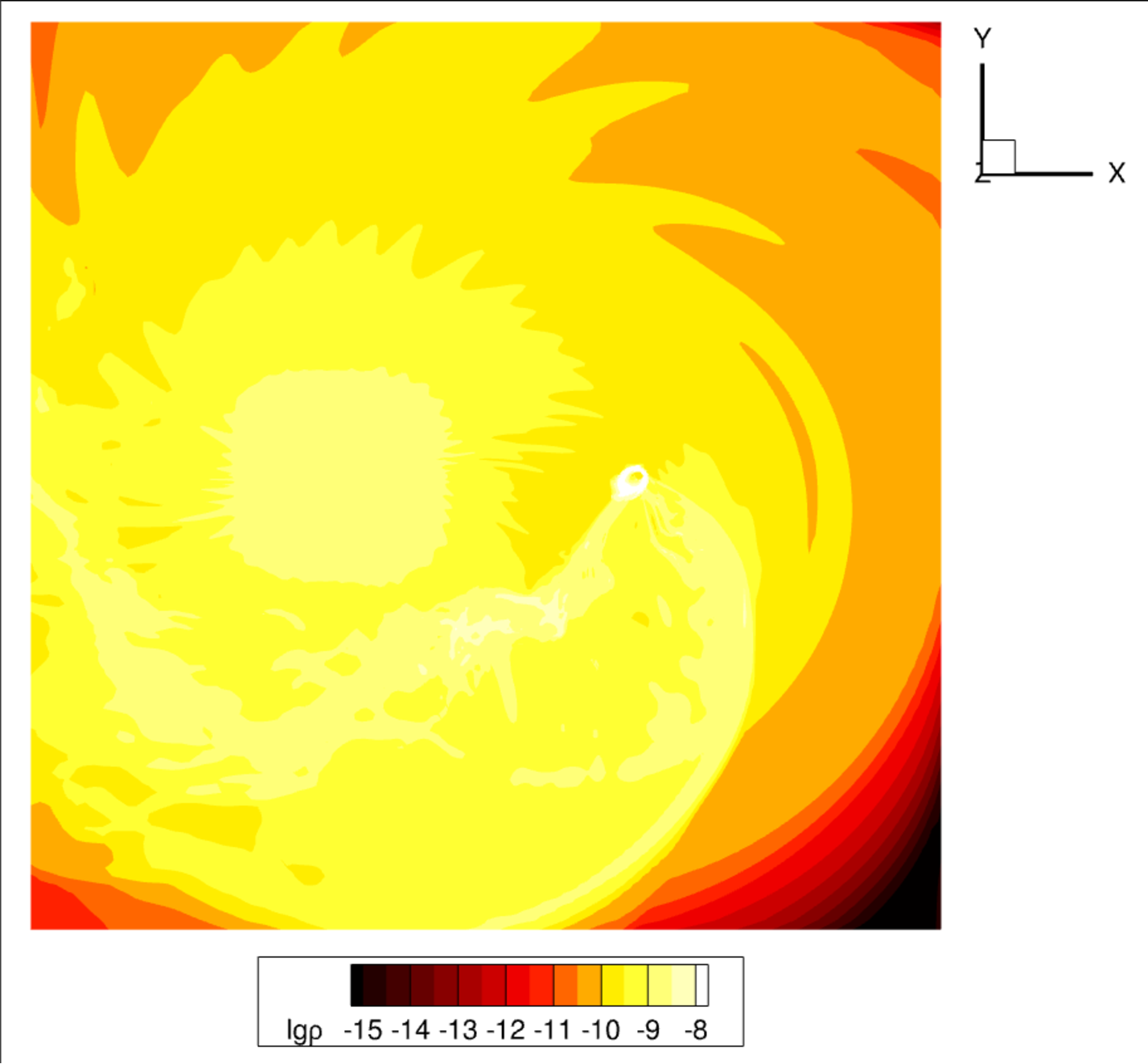} \\ $T_{\text{c}} = 5.5 \times 10^6 K$}
    \end{minipage}
    \hfill
    \begin{minipage}[h]{0.48\linewidth}
	\centering{\includegraphics[width=0.9\textwidth]{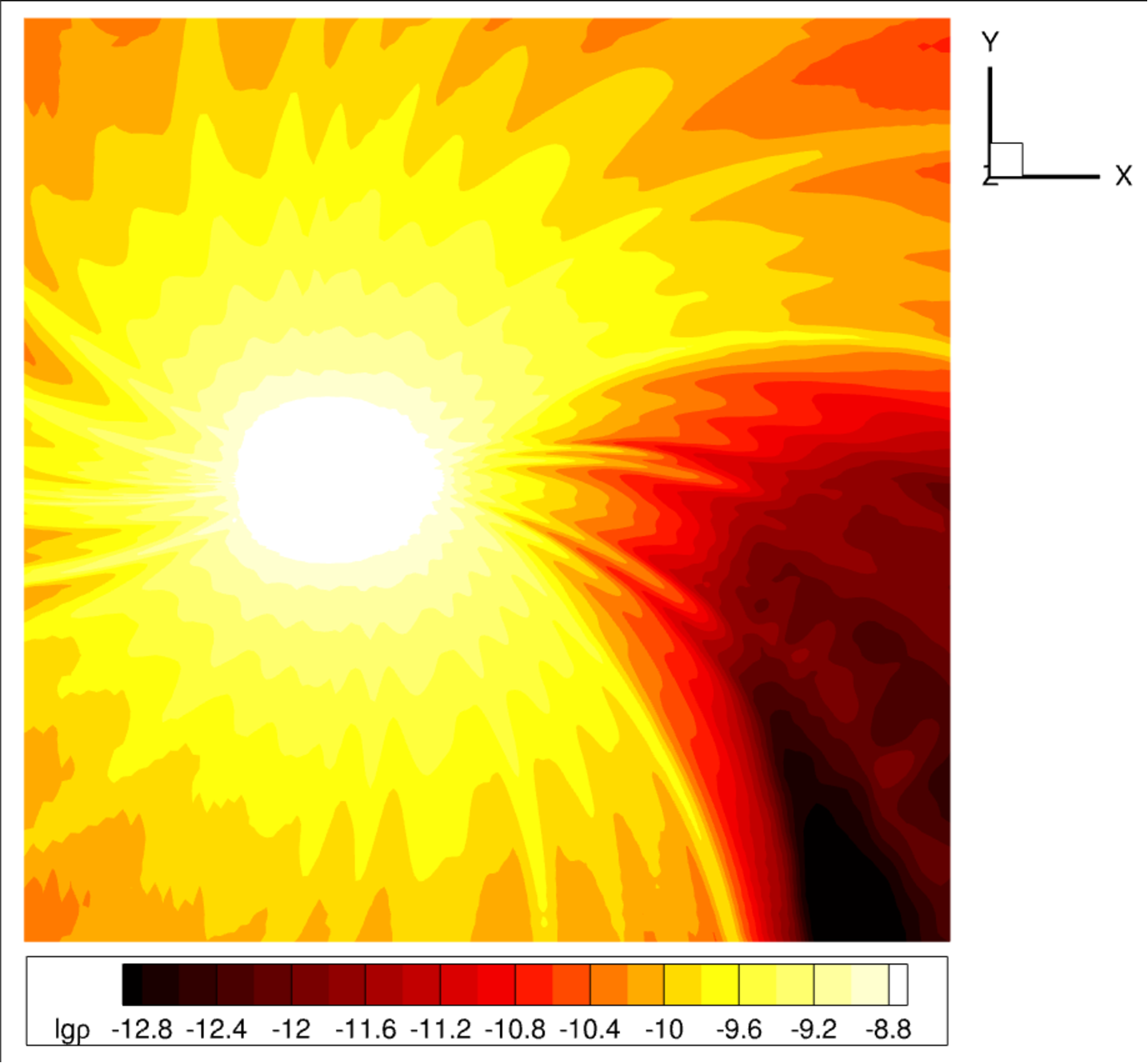} \\ $T_{\text{c}} = 3 \times 10^7 K$}
    \end{minipage}
    \hfill
    \begin{minipage}[h]{0.48\linewidth}
	\centering{\includegraphics[width=0.9\textwidth]{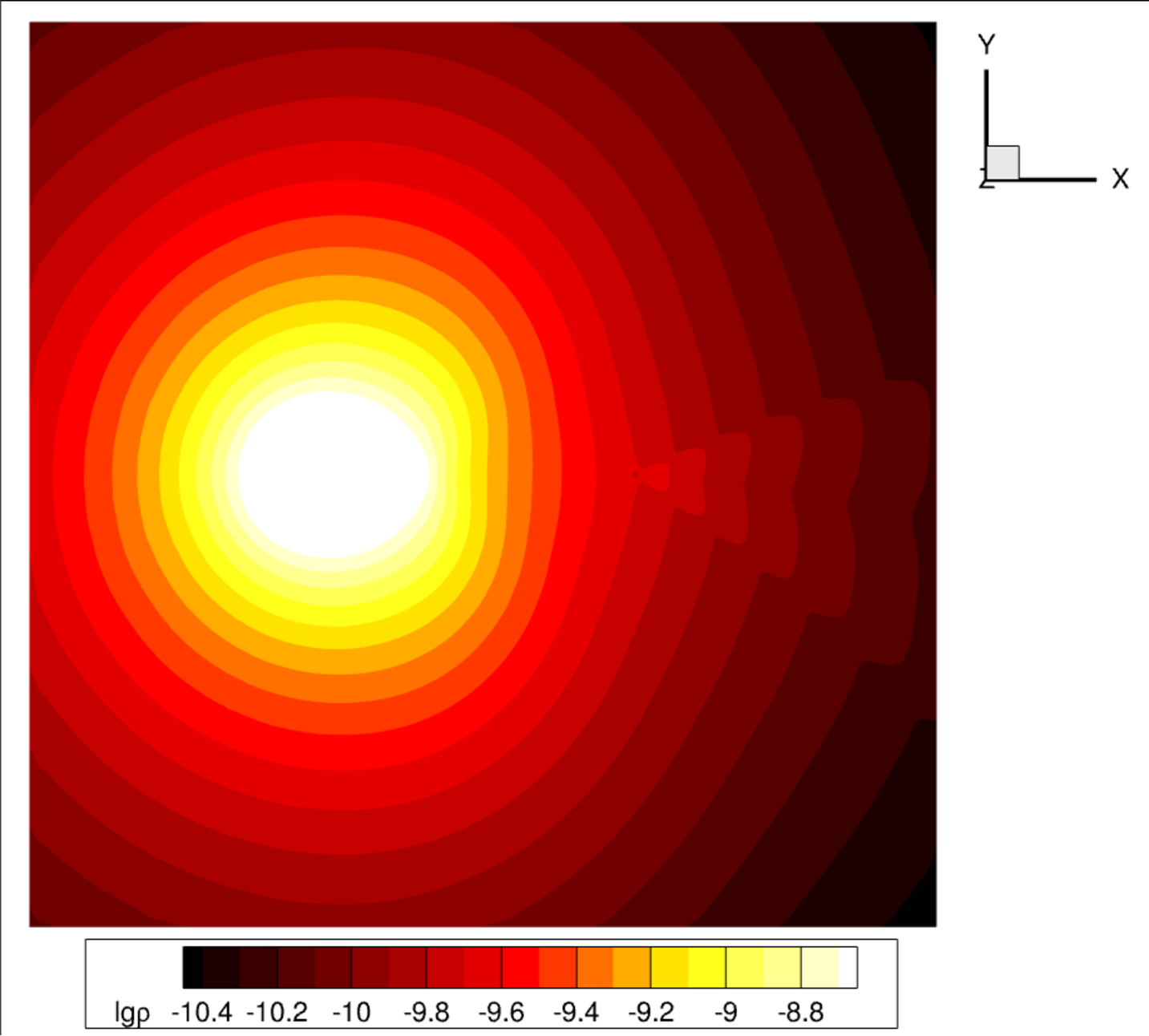} \\ $T_{\text{c}} = 3 \times 10^8 K$}
    \end{minipage}
	\caption{Integral density distribution for all models in a plane parallel to $XY$, with an offset of $z = 0.75$.}
	\label{fig-20}
\end{figure}

Wind streams originating from the donor’s near-polar regions mostly exit the computational domain freely, but as seen in Fig. \ref{fig-12} (bottom panel), a small fraction is deflected toward the orbital plane. As a result, near point $\rm{L}_4$ (marked by a red dot in the figure), the two flows mix, leading to the formation of a vortex whose temperature rises due to gas compression. The further behavior of the wind gas is illustrated in the middle panel of Fig. \ref{fig-19}, which shows the flow in the $XZ$ plane. The left part of the panel corresponds to wind streams moving toward the donor.

Same as in the orbital plane, these flows, driven by the binary’s rotation and pressured by newly generated wind, are directed toward the vicinity of the Lagrange point $\rm{L}_3$ and leave the computational domain. The temperature distribution in the figure shows that these streams cool by an order of magnitude; however, collisions with the fresh, hotter wind cause local heating.

The wind pattern to the right of point $\rm{L}_4$ on this panel of Fig. \ref{fig-19} demonstrates the complex motion of the wind. Heated gas in the upper and lower parts of the vortex, under the influence of the binary’s rotation, is expelled toward the outer boundary of the
computational domain, while the central region of the vortex, by contrast, cools to $10^5$ K, and its material contributes to feeding the accretion disk. The combined movement of these flows leads to the formation of additional turbulent zones symmetrically positioned relative to the disk plane.

Fig. \ref{fig-20} presents the surface density distributions for our models. This figure makes it possible to see the characteristic elements of gas flow structure in close binary systems and how their shape depends on the initial wind speed.

A notable feature of Figs. \ref{fig-12} and \ref{fig-20} for the model
with temperature $T_{\text{c}} = 3 \times 10^6$ K is the formation of a
dense, spiral-shaped compression of donor wind gas around the accretor. At other initial donor corona temperatures (Fig. \ref{fig-12}), this compression disappears. A comparative study of the X-ray binary Vela X-1 light curve at energies $2-4$ keV and $4-10$ keV revealed the existence of a similar gas compression --- a tail near the
neutron star of this system \cite{Abalo2024}. The emergence of such a compression may occur for the same reason as the tail formation in the Sco X-1 model considered in this work: the focusing of part of the donor’s stellar wind flow by the accretor’s gravity when the wind
velocity and the accretor’s orbital velocity are comparable in magnitude.

Equation (Eq. \ref{eq-13}) estimates the critical wind velocity separating the two solutions for the interaction of a compact object with the stellar wind of a close companion that does not fill its Roche lobe. Numerical models confirmed the presence of two regimes. Three-dimensional models with $T_{\text{c}} \leq 3 \times 10^6$ K show
that part of the wind material is captured by the accretor, forming a disk around it. Models with $T_{\text{c}} \ge 3 \times 10^6$ K exclude disk formation around the white dwarf. It is important to emphasize the weak dependence of the critical velocity on the accretor’s size. For an accreting stellar-mass black hole, the critical velocity derived from (Eq. \ref{eq-13}) would be several times higher than that for a degenerate white dwarf.

\section{CONCLUSIONS}

In recent years, three-dimensional modeling of the stellar wind dynamics in close binary systems has become increasingly popular. Several studies have been dedicated to the kinematics of the stellar wind in symbiotic binaries where a supergiant donor is close to
filling its Roche lobe, but does not fill it completely \cite{Lee2022,Malfait2024}. Despite differences in numerical approaches, the resulting gas flow patterns are in good agreement
with each other and with the model presented in this study. Further development of these models will involve incorporating energy transfer physics and the physics of maintaining both intrinsic and X-ray-induced heating of the donor’s stellar wind and the accretion disk. Advancing models of emission-line profiles of gaseous envelopes in interacting binaries will allow researchers to study the kinematics of these envelopes. In addition, it is already clear that the interaction of the accretor’s stellar wind with the donor’s wind (corona) must be specifically investigated and modeled.

When comparing the presented models of donor wind gas flows in close binaries, it is important to note that the accretor’s disk in X-ray binaries may itself have a strong wind (corona) \cite{Tomaru2023}. This can make the system’s X-ray activity recurrent. Initially, the donor wind material accumulates in the disk (torus) around the accretor in the absence of active accretion onto the compact object. Once the disk (torus) accumulates a critical mass, active migration begins and the disk develops its own wind (corona). This wind can almost completely block the disk’s ability to capture donor
material during the active X-ray phase. Depletion of the accretion disk then resumes the accretion of donor wind gas. A similar phenomenon is observed in recurrent cataclysmic systems  \cite{Jordan2024}.

In this study, we proposed a new approach to classifying binaries with components possessing stellar winds that do not fill their Roche lobes. The current division of binary systems into detached and semi-detached is linked to the process of mass transfer through the vicinity of the inner Lagrange point. A key condition for mass transfer is that the donor fills its Roche lobe, creating in point $\rm{L}_1$ a pressure of its atmospheric gas that is not balanced by gravity, leading to outflow.

In the case of detached systems, both components do not fill their Roche lobes, and so they are generally assumed to lack active mass exchange. However, the presence of a stellar wind from the companion changes this picture. The presented modeling shows that an X-ray binary is embedded in a rarefied envelope created by the donor’s stellar wind. This naturally raises the question of whether it can be detected. Hot gas primarily emits in the lines of highly ionized heavy elements from the iron group. Light elements in the hot plasma are fully ionized in the X-ray range. Let us compare the accretor’s emission intensity, as it accumulates wind material, $L_{\text{x}}$ (Eq. \ref{eq-25}), with the envelope’s cooling rate, $L_{\text{R}}$ (Eq. \ref{eq-23}). The emission lines corresponding to the cooling of the hot wind gas are narrow: $\frac{\Delta x}{x} \simeq \frac{v_{\text{w}}}{c} \simeq 10^{-3}$. This makes the disk’s emission intensity comparable to that of these lines, offering a potential opportunity to detect emission lines from the rarefied envelope in which such binary systems are embedded. These emission lines should be
most pronounced in systems with low-mass accretors that have the most intense stellar winds, possibly of induced origin.

For completeness, we should note the drawbacks of our simplified model. Mainly, it is the physics of the stellar wind, which in this model is defined by the initial gas velocity at the donor’s surface. In reality, the physics of heating and cooling coronal gas is complex.
The terminal velocity is acquired by the coronal gas at a distance of several donor radii. Furthermore, the heating of coronal gas depends on the latitude of the region on the donor’s surface, and for a corona
induced by X-ray irradiation from the accreting component, it is asymmetrical. Clearly, these factors could significantly complicate the gas flow pattern in close binaries. Extending the simplified model discussed here will require addressing these simplifications.

It should be noted that in dense galactic nuclei, the donor can be an accreting SMBH with a strong accretion disk wind. The special, and so far unexplored, fate of stars located near active galactic nuclei is of particular interest. Observations have shown that gas accretion from an SMBH disk produces a disk wind with an intensity $\dot M_{\text{w}} = 10^{-2} - 10^2 M_\odot$/year and speed of $\sim 10^3$ km/s for a typical SMBH mass of $10^5 - 10^7 M_\odot$/year \cite{Baron2019}. A large number of stars in the nucleus of a galaxy hosting an active SMBH or quasar find themselves immersed in this wind field, acquiring the ability to accrete its material. As a rule, neutron stars and black holes near the active nucleus can accrete wind material from the nucleus, giving rise to various interesting phenomena and objects. Thus, stars near an active galactic nucleus form peculiar binary-like systems --- once again divided into interacting and non-interacting. A simple estimate shows that for a neutron star to achieve Eddington luminosity in a nucleus wind with a rate of $10^2 M_\odot$/year and speed of $10^3$ km/s, it must lie within $\sim 10^{-3}$ pc of the SMBH. Given that there are many neutron stars and stellar black holes in the nucleus, we can conclude that compact stars accreting the nuclear wind could significantly contribute to its X-ray luminosity.

The modeling of the donor’s stellar wind performed in this study is based on a formal approach to the physics of its formation --- by fixing the initial temperature of its material. Compact accretors (neutron
stars and black holes) in close binaries, by accreting wind gas, become powerful sources of X-ray radiation. The physics of the donor wind’s interaction with the accretor’s radiation and the possible wind from the bright accretion disk is not considered in the presented model. Nevertheless, developing a three-dimensional model of stellar wind gas dynamics in X-ray binaries requires accounting for the physical processes occurring in such interactions.

Below is a brief summary of the main conclusions of this study.
\begin{enumerate}
\item 
We classified detached binaries with components possessing stellar winds. Analytical estimates were provided for the conditions of mass exchange in close binaries and for the limits on wind velocity
required to form an accretion--decretion disk around the accretor.

\item 
Using several initial gas velocities (temperatures) of the stellar wind, we studied, within a 3D gas-dynamical model, the process of reaching a steady-state solution for a binary system with component parameters matching those of the X-ray binary Sco X-1.

\item
Numerical 3D models with initial donor corona
gas temperatures below $\sim 4 \times 10^6$ K demonstrated the
formation of accretion–decretion disks around the
accretor, while the model with $T_{\text{c}} = 3 \times 10^6$ K also
showed the formation of tidal gas tails around the system.

\item
Numerical 3D models with initial donor corona gas temperatures above $\sim 4 \times 10^6$ K showed the stellar wind gas flowing around the accretor without forming an accretion disk around the compact companion. The reason for the change in accretion regime as wind velocity increases is the reduction of the angular momentum of material falling onto the accretor below the threshold required for disk formation.

\end{enumerate}

\section*{ACKNOWLEDGMENTS}

The authors express their gratitude to A.M. Cherepashchuk for discussions on X-ray binaries, A.G. Zhilkin for providing the MHD code, and A.V. Fedorova for discussions of the results. This study was carried out using the equipment of the Joint Supercomputer Center of the Russian Academy of Sciences\footnote[1]{https://www.jscc.ru/}. The study was performed as part of the project <<Exoplanets>>.

\section*{FUNDING}

This work was supported by ongoing institutional funding. No additional grants to carry out or direct this particular research were obtained.

\section*{CONFLICT OF INTEREST}
The authors of this work declare that they have no conflicts of interest.


\thebibliography{99}

\bibitem{Masevich1988}
A. G. Masevich, A. V. Tutukov, {\it{Stellar Evolution: Theory and Observations}} (Nauka, Moscow, 1988) [in Russian]. %1

\bibitem{Cherepaschuk2013} 
A. M. Cherepashchuk, {\it{Close Binary Stars}} (Fizmatlit, Moscow, 2013), Vols. 1, 2 [in Russian].%2

\bibitem{BZB2013}
D. V. Bisikalo, A. G. Zhilkin, and A. A. Boyarchuk, {\it{Gas
Dynamics of Close Binary Stars}} (Fizmatlit, Moscow,
2013) [in Russian]. %3

\bibitem{Tutukov1992} 
A. Tutukov, L. Yungelson, I. Iben, ApJ. \textbf{386}, 197 (1992).%4

\bibitem{Lipunov1994} 
V. Lipunov, K. Postnov, M. Prokhorov, ApJ. \textbf{423}, 121 (1994).%5

\bibitem{Oskinova2018} 
L. Oskinova, T. Bulik, A. Gomez-Moran, A\&A \textbf{613}, 10 (2018). %6

\bibitem{Tutukov1976}
A. Tutukov, L. Yungelson, ApJ. \textbf{12}, 342 (1976).%7

\bibitem{MacLeod2024} 
M. MacLeod et al., astro-ph/2409.13332 (2024). %8

\bibitem{Hill2002} 
G. Hill, A. Moffat, N. St-Louis, MNRAS, \textbf{335}, 1069 (2002). %9

\bibitem{Romeo2007} 
G. Romeo, A. Okazaki, M. Orellana, S. Owocki, A\&A, \textbf{474}, 15 (2007). %10

\bibitem{Parkin2008} 
E. Parkin, J. Puttard, MNRAS, \textbf{388}, 1047 (2008). %11

\bibitem{Bisikalo2006} 
D. Bisikalo, A. Boyarchuk, E. Kilpio, N. Tomov et al., ARep, \textbf{50}, 722 (2006). %12

\bibitem{Bisikalo1994} 
D. Bisikalo, A. Boyarchuk, O. Kuznetsov, ARep, \textbf{38}, 494 (1994). %13

\bibitem{Skopal2015} 
A. Skopal, Z. Carikova, A\&A, \textbf{573}, 5 (2015). %14

\bibitem{Taani2022} 
A Taani, S. Karino, L. Song, C. Zhang, S. Chaty, PASA, \textbf{39}, 40 (2022). %15

\bibitem{Pittard2006} 
J. Pittard, S. Dougherty, MNRAS, \textbf{372}, 801 (2006). %16

\bibitem{Pittard2010} 
J. Pittard, E. Parkin, MNRAS, \textbf{403}, 1657 (2010). %17

\bibitem{Cherepaschuk2020}
A. Tutukov, A. Cherepaschuk, UFN, \textbf{190}, 225 (2020).%18
	
\bibitem{Tutukov2023}
A. Tutukov, ARep., \textbf{67}, 867 (2023). %19

\bibitem{Ransome2024}
C. Ransome, V. Villar,  astro-ph/2409.10596 (2024). %20

\bibitem{Tutukov1969}
A. V. Tutukov, Nauch. Inform. Astron. Sov. Akad. Nauk SSSR \textbf{11}, 62 (1969). %21

\bibitem{Iben1984} 
I. Iben, A. Tutukov, ApJ., \textbf{284}, 719 (1984). %22

\bibitem{Iben1997} 
I. Iben, A. Tutukov, A. Fedorova, ApJ., \textbf{486}, 955 (1997). %23

\bibitem{Baron2019} 
D. Baron, H. Netzer, MNRAS, \textbf{486}, 4290 (2019). %24

\bibitem{Shaposhnikov2024}
I. Shaposhnikov, A. Cherepashchuk, A. Dodin, K. Postnov, A\&A, \textbf{683}, L17 (2024). %25

\bibitem{Rochowicz1995}
K. Rochowicz, A. Niedzielski, A\&A, \textbf{45}, 307 (1995). %26

\bibitem{Karino2021} 
S. Karino, MNRAS, \textbf{507}, 1002 (2021). %27

\bibitem{Tutukov2007} 
A. Tutukov, A. Fedorova,  ARep., \textbf{51}, 847 (2007). %28

\bibitem{Harmanec2002} 
P. Harmanec, D. Bisikalo, A. Boyarchuk, O. Kuznetsov, A\&A,  \textbf{396}, 937 (2002). %29

\bibitem{Niwano2024} 
M. Niwano, M. Fausnaugh, R. Lau, K. De et al., astro-ph/2409.09581 (2024). %30

\bibitem{Krticka2001} 
J. Krticka, J. Kubat, A\&A,  \textbf{377}, 175 (2001). %31

\bibitem{Vink2018} 
J. Vink, A\&A,  \textbf{615}, A119 (2018). %32

\bibitem{Shoda2021} 
M. Shoda, B. Chandran, S. Cranmer, ApJ., \textbf{915}, 52 (2021). %33

\bibitem{Skumanich1972} 
A. Skumanich, C. Smythe, E. Frazier, 
Bulletin of the American Astronomical Society, \textbf{4}, 391 (1972). %34

\bibitem{Tutukov1978} 
A. Tutukov, A\&A, \textbf{70}, 57 (1978). %35

\bibitem{Davidson1973}
K. Davidson, J. Ostriker, ApJ., \textbf{179}, 585 (1973). %36

\bibitem{Bu2024} 
D. Bu, F. Yuan, astro-ph/2409.01664 (2024). %37

\bibitem{Boetcher2024}
E. Boetcher, E. Hodges-Kluck, astro-ph/2408.15327 (2024). %38

\bibitem{Loose1982}
K. Loose, E. Krugel, A. Tutukov, A\&A, \textbf{105}, 342 (1982). %39

\bibitem{Gupta2024}
K. Gupta et al., astro-ph/2409.12239 (2024). %40

\bibitem{Kimbrell2024}
S. Kimbrell, A. Reines, astro-ph/2409.10630 (2024). %41

\bibitem{Semenov2003}
D. Semenov et al., A\&A, \textbf{410}, 611 (2003). %42

\bibitem{Kholtygin2002}
A. Kholtygin, V. Bratsev, V. Ochkur, Ap., \textbf{45}, 32 (2002). %43

\bibitem{Cranmer1999}
S. Cranmer, G. Field, J. Kohl, AJ., \textbf{518}, 937 (1999). %44

\bibitem{Antokhin2022}
I. Antokhin, A. Cherepaschuk, E. Antokhina, A. Tatarnikov, ApJ., \textbf{926}, 123 (2022). %45

\bibitem{Fortin2024}
F. Fortin et al., A\&A, \textbf{684}, A124 (2024). %46

\bibitem{Folini2000}
D. Folini, R. Walder, Aph\&SS, \textbf{274}, 189 (2000). %47

\bibitem{Madura2013}
T. Madura, T. Gull, A. Okazaki, MNRAS, \textbf{436}, 3820 (2013). %48

\bibitem{Mitsumoto2005}
M. Mitsumoto et al., ARep., \textbf{49}, 884 (2005). %49

\bibitem{Saladino2019}
M. Saladino, O. Pols, C. Abate, A\&A, \textbf{626}, 68 (2019). %50

\bibitem{Gawryszczak2003}
A. Gawryszczak, J. Mikolajewska, M. Razyczka, ASPC, \textbf{303}, 418 (2003). %51

\bibitem{Campana2018} 
S. Campana, T. Di. Salvo, ASSL, \textbf{457}, 149 (2018). %52

\bibitem{Karino2021}
S. Karino, MNRAS, \textbf{507}, 1002 (2021). %53

\bibitem{Cherepaschuk2022}
A. Cherepaschuk, T. Khruzina, A. Bogomazov, ARep., \textbf{66}, 348 (2022). %54

\bibitem{Zhilkin2019}
A. G. Zhilkin, A. V. Sobolev, D. V. Bisikalo, and M. M. Gabdeev, Astron. Rep. \textbf{63}, 751 (2019). %55

\bibitem{Balsara1998} 
D. Balsara, ApJ. Supl. Ser., \textbf{116}, 119 (1998). %56

\bibitem{Tutukov2016} 
A. Tutukov, A. Fedorova, ARep., \textbf{60}, 116 (2016). %57

\bibitem{Brahic1974} 
A. Brahic, IAU Symp., \textbf{62}, 83 (1974). %58

\bibitem{Chatterjee1981} 
T. Chatterjee, Ap\&SS, \textbf{76}, 491 (1981). %59

\bibitem{Ipatov2000} 
S. Ipatov, J. Henrard, SSR, \textbf{34}, 61 (2000). %60

\bibitem{Bosch2015} 
V. Bosch-Ramon, M. Barkov, M. Perucho, A\&A, \textbf{577}, A89 (2015). %61

\bibitem{Tripathi2024} 
P. Tripathi, I. Chattopadhyay, R. Joshi, astro-ph/2412.04815 (2024). %62

\bibitem{Lee2022} 
Y-M. Lee, Y. Kim, H-W. Lee, AstJ., \textbf{931}, 142, (2022). %63

\bibitem{Malfait2024} 
J. Malfait et al., A\&A, \textbf{691}, 84, (2024). %64

\bibitem{Tomaru2023} 
R. Tomaru, Ch. Done, H. Odaka, MNRAS, \textbf{527}, 7047, (2023). %65

\bibitem{Jordan2024} 
I. Jordan, D. Wehner, R. Kuiper, A\&A, \textbf{689}, 354, (2024). %66

\bibitem{Abalo2024} 
L. Abalo et al., astro-ph/2412.21456 (2024). %67
\end{document}